\documentclass{esub2acm}

\usepackage{theorem}
\usepackage{latexsym}
\usepackage{amsfonts}
\usepackage{amssymb}
\usepackage{amsmath}
\usepackage{float}
\usepackage{color}

\newenvironment{example}{\medskip \noindent{\em Example:}\em}{\medskip}

\floatstyle{plain}
\floatname{algorithm}{\footnotesize Alg.}
\newfloat{algorithm}{htbp}{1op}

\renewcommand{\citeN}[1]{\shortciteN{#1}}

\newcommand{\citetwoN}[2]{\shortciteANP{#1} [\citeyearNP{#1,#2}]}

\newcommand{\caja}[1]{\fbox{\hspace{-0.5mm}$\mathsf{#1}$\hspace{-0.5mm}}}

\newcommand{\dd}{\mathinner{.\,.}}

\begin{document}

\title{Indexing Highly Repetitive String Collections}

\author{%
Gonzalo Navarro \\
University of Chile, Chile 
}


\begin{abstract}
Two decades ago, a breakthrough in indexing string collections made it
possible to represent them within their compressed space while at the same
time offering indexed search functionalities. As this new technology
permeated through applications like bioinformatics, the string collections
experienced a growth that outperforms Moore's Law and challenges our ability 
to handle them even in compressed form. It turns out, fortunately, that 
many of these rapidly growing string collections are highly repetitive, 
so that their information content is orders of magnitude lower than their
plain size. The statistical compression methods used for classical collections,
however, are blind to this repetitiveness, and therefore a new set of techniques
has been
developed in order to properly exploit it. The resulting indexes form a new
generation of data structures able to handle the huge repetitive string
collections that we are facing.

In this survey we cover the algorithmic developments that have led to these
data structures. We describe the distinct compression paradigms that have been
used to exploit repetitiveness,
the fundamental algorithmic ideas that form the base of all the existing
indexes, and the various structures that have been proposed, 
comparing them both in theoretical and practical aspects. We conclude with
the current challenges in this fascinating field.
\end{abstract}



\begin{bottomstuff}

Funded by ANID Basal Funds FB0001, Millennium Science Initiative Program - 
Code ICN17\_002, and Fondecyt Grant 1-200038, Chile.
\begin{authinfo}
\address{%
Gonzalo Navarro, Center for Biotechnology and Bioengineering (CeBiB) and
Millennium Institute for Foundational Research on Data (IMFD), Department of 
Computer Science, University of Chile, Beauchef 851, Santiago, Chile,
{\tt gnavarro@dcc.uchile.cl}.}
\end{authinfo}
\permission

\end{bottomstuff}

\markboth{G. Navarro}{Indexing Highly Repetitive String Collections}

\maketitle

%
%

\tableofcontents

\newpage

\section{Introduction}

Our increasing capacity for gathering and exploiting all sorts of data around us
is shaping modern society into ways that were unthinkable a couple of decades 
ago. In bioinformatics, we have stepped in 20 years from sequencing the first 
human genome to completing projects for sequencing 100,000 genomes%
\footnote{\texttt{https://www.genomicsengland.co.uk/about-genomics-england/the-100000-genomes-project}}. Just storing such a collection requires about
70 terabytes, but a common data analysis tool like a suffix tree \cite{Apo85}
would require
5.5 petabytes. In astronomy, telescope networks generating terabytes per hour 
are around the corner\footnote{\texttt{https://www.nature.com/articles/d41586-018-01838-0}}. The web is estimated to have 60 billion pages, with a 
total size of about 4 petabytes counting just text content\footnote{\texttt{https://www.worldwidewebsize.com and https://www.keycdn.com/support/ the-growth-of-web-page-size}, see the average HTML size.}. Estimations of the yearly amount of data generated in the world 
are around 1.5 exabytes\footnote{\texttt{http://groups.ischool.berkeley.edu/archive/how-much-info/how-much-info.pdf}}.

Together with the immense opportunities brought by the data in all sorts of
areas,
we are faced to the immense challenge of efficiently storing, processing, and 
analyzing such volumes of data. Approaches such as parallel and distributed
computing, secondary memory and streaming algorithms reduce time, but still
pay a price proportional to the \textit{data size} in terms of amount of 
computation, storage requirement, network use, energy consumption, and/or 
sheer hardware. This is problematic because the growth rate of the data has 
already surpassed Moore's law 
in areas like bioinformatics and astronomy \cite{Plos15}. Worse, these methods
must access the data in secondary memory, which is much slower than the main
memory. Therefore, not only we have to cope with orders-of-magnitude larger
data volumes, but we must operate on orders-of-magnitude slower storage devices.

A promising way to curb this growth is to focus on how much \textit{actual 
information} is carried by those data volumes. It turns out that many of 
the applications where the data is growing the fastest feature large degrees of 
\textit{repetitiveness} in the data, that is, most of the content in each
element is equal to content of other elements. For example, let us focus on
sequence and text data. Genome repositories typically store many genomes of 
the same species. Two human genomes differ by about 0.1\% \cite{PHDR00}, and 
Lempel-Ziv-like compression \cite{LZ76} on such repositories
report compression ratios (i.e., compressed divided by uncompressed space)
around 1\% \cite{FLCB11}. A versioned document collection like Wikipedia stored
10 terabytes by 2015, and it reported over 20 versions per article, with the
versions (i.e., near-repetitions) growing faster than original articles, and
1\% Lempel-Ziv compression ratios\footnote{\texttt{https://en.wikipedia.org/wiki/Wikipedia:Size\_of\_Wikipedia}}. A versioned software repository like GitHub 
stored over 20 terabytes in 2016 and it also reported over 20 versions per 
project\footnote{\texttt{https://blog.sourced.tech/post/tab\_vs\_spaces and http://blog.coderstats.net/github/ 2013/event-types}}. Degrees of 40\%--80\% of 
duplication have been observed in tweets \cite{TAHHG13}, emails \cite{EO06}, 
web pages \cite{Hen06}, and general software repositories \cite{KG05} as well.

These sample numbers show that we can aim at 100-fold reductions in the data 
representation size by using appropriate compression methods on highly 
repetitive data sets. Such a reduction would allow us handling much larger data
volumes in main memory, which is considerably faster. Even in cases where the 
reduced data still does not fit in main memory, we can expect a 100-fold 
reduction in storage, network, hardware, and/or energy costs. 

Just compressing the data, however, is not sufficient to reach this goal,
because we still need to decompress it in order to carry out any processing on
it. For the case of text documents, a number of ``version control'' systems like
CVS\footnote{\tt https://savannah.nongnu.org/projects/cvs},
SVN\footnote{\tt https://subversion.apache.org}, and
Git\footnote{\tt https://git-scm.com},
support particular types of repetitive collections, namely {\em versioned} ones,
where documents follow a controlled structure (typically linear or hierarchical)
and the systems can track which document is a variant of which. Those systems
do a good job in reducing space while supporting extraction of any version
of any document, mainly by storing the set of ``edits'' that distinguish each
document from a close version that is stored in plain form.

Still, just extraction of whole documents is not sufficient. In order to
process the data efficiently, we need {\em data structures} built on top it.
What is needed is a more ambitious
concept, a \textit{compressed data structure} \cite{Nav16}. Such a data
structure aims not only at representing the data within space close to its
actual information content, but also, within that space, at efficiently
supporting direct access, queries, analysis, and manipulation of the data
without ever decompressing it. This is in sharp contrast with classical data
structures, which add (sometimes very significant) extra space on top of the
raw data (e.g., the suffix trees already mentioned typically use 80 times the
space of a compacted genome).

Compressed data structures are now 30 years old \cite{Jac89}, have made their 
way into applications and companies \cite{Nav16}, and include mature libraries%
\footnote{\texttt{https://github.com/simongog/sdsl-lite}}. Most compressed data
structures, however, build on statistical compression \cite{CT06}, 
which is blind to repetitiveness \cite{KN13}, and therefore fail to get even
close to the compression ratios we have given for highly repetitive scenarios.
The development of compressed data structures aimed at highly repetitive data 
is much more recent, and builds on variants of \textit{dictionary compression}
\cite{CT06,SS82}. 

In this survey we focus on {\em pattern matching on string collections}, one
of the most fundamental problems that arise when extracting information from 
text data. The problem consists in, given a collection of strings, build a 
data structure (called an {\em index}) so that, later, given a short query 
string (the pattern), we efficiently locate the places where the pattern occurs
in the string collection. Indexed pattern matching is at the core of areas like 
Information Retrieval \cite{BCC10,BYRN11},
Data Mining \cite{Liu07,LBNRLB09,Sil10}, 
Bioinformatics \cite{Gus97,Ohl13,MBCT15},
Multimedia Retrieval \cite{TWV05,SHYT10}, 
and others.

The need of compressed data structures for pattern matching was recognized
a couple of decades ago \cite{FM00,GV00}, and there are nowadays mature and
successful indexes \cite{NM06,Gro11} that have made their way to applications;
see for example bioinformatic software like {\tt Bowtie}\footnote{\tt
http://bowtie-bio.sourceforge.net}, {\tt BWA}\footnote{\tt
http://bio-bwa.sourceforge.net}, or {\tt Soap2}\footnote{\tt
http://soap.genomics.org.cn}. Just like the general compressed data structures,
however,
these indexes build on statistical compression, and therefore do not exploit
the high degree of repetitiveness that arise in many applications. 

This challenge was explicitly recognized almost a decade later \cite{MNSV08}, 
in the same bioinformatic context. After about another decade,
the specific challenges of text indexing
on highly repetitive string collections have become apparent, but also there
has been significant progress and important results have been reached.
These include, for example, searching in optimal time within 
dictionary-compressed space, and developing new and more robust 
measures of compressibility.

Our aim in this survey is to give an exhaustive, yet friendly, coverage of the
discoveries in this area. We start with the fascinating issue of 
how to best measure compressibility via repetitiveness, just like the entropy
of \citeN{Sha48} is the right concept to measure compressibility via
frequency skews. Section~\ref{sec:measures} covers a number of repetitiveness
measures, from ad-hoc ones like the size of a Lempel-Ziv parse \cite{LZ76}
to the most recent and abstract ones based on string attractors and string
complexity \cite{KP18,KNP20}, and the relations between them. 
Section~\ref{sec:access} explores the problem of giving direct access to any
part of a string that is compressed using some of those measures, which 
distinguishes a compressed data structure from sheer compression: some measures
enable compression but apparently not direct access. The more
ambitious topic of developing indexes whose size is bounded in terms of some 
of those measures is developed next, building on parsings in 
Section~\ref{sec:parsed-indexing} and on string suffixes in
Section~\ref{sec:suffix-indexing}. The progress on other problems related to 
pattern matching is briefly covered in Section~\ref{sec:others}. 
Finally, Section~\ref{sec:concl} discusses the challenges that remain open.

\section{Notation and Basic Concepts}
\label{sec:def}

We assume basic knowledge on algorithms, data structures, and algorithm analysis.
In this section we define some fundamental concepts on strings, preceded by
a few more general concepts and notation remarks.

\paragraph*{Computation model}
We use the RAM model of computation, where we assume the programs run
on a random-access memory where words of $w = \Theta(\log n)$ bits are accessed
and manipulated in constant time, where $n$ is the input size. All the typical 
arithmetic and logical operations on the machine words are carried out in 
constant time, including multiplication and bit operations.

\paragraph*{Complexities}
We will use big-$O$ notation for the time complexities, and in many cases for
the space complexities as well. Space complexities are measured in amount of
computer words, that is, $O(X)$ space means $O(X \log n)$ bits. By 
$\mathrm{poly}\,x$ we mean any polynomial in $x$, that is, $x^{O(1)}$, and
$\mathrm{polylog}\,x$ denotes $\mathrm{poly}\,(\log x)$.
Logarithms will be to the base 2 by default. Within big-$O$ complexities,
$\log x$ must be understood as $\lceil \log (2+x) \rceil$, to avoid border 
cases.

\subsection{Strings}

A {\em string} $S=S[1\dd n]$ is a sequence of {\em symbols} drawn from a set 
$\Sigma$ called the {\em alphabet}. We will assume $\Sigma = 
\{1,2,\ldots,\sigma\}$. The {\em length} of $S[1\dd n]$ is $n$, also denoted 
$|S|$. We use $S[i]$ to denote the $i$-th symbol of $S$ and $S[i\dd j] = S[i]
\cdots S[j]$ to denote a {\em substring} of $S$. If $i > j$, then
$S[i\dd j]=\varepsilon$, the empty string. A {\em prefix} of $S$ is a substring
of the form $S[1\dd j]$ and a {\em suffix} is a substring of the form 
$S[i\dd n]=S[i\dd]$. With $SS'$ we denote the {\em concatenation} of the strings $S$ 
and $S'$, that is, the symbols of $S'$ are appended after those of $S$. 
Sometimes we identify a single symbol with a string of length 1, so that
$aS$ and $Sa$, with $a\in\Sigma$, denote concatenations as well. 
In general, string
collections will be viewed as a single string that concatenates them all, with
a suitable separator symbol among them.

The {\em lexicographic order} among strings is defined as in a dictionary.
Let $a,b \in \Sigma$ and let $S$ and $S'$ be strings. Then $aS \le bS'$ if
$a<b$, or if $a=b$ and $S \le S'$; and $\varepsilon \le S$ for every $S$.

For technical convenience, we will often assume that strings $S[1\dd n]$
are terminated with a special symbol $S[n]=\$$, which does not appear elsewhere
in $S$ nor in $\Sigma$. We assume that $\$$ is smaller than every symbol
in $\Sigma$ to be consistent with the lexicographic order.
The string
$S[1\dd n]$ read backwards is denoted $S^{rev} = S[n] \cdots S[1]$; note that
in this case the terminator does not appear at the end of $S^{rev}$.

\subsection{Pattern Matching}

The {\em indexed pattern matching problem} consists in, given a sequence 
$S[1\dd n]$, build a data structure (called an {\em index}) so that, later, 
given a query string $P[1\dd m]$, one efficiently finds the $occ$ places in 
$S$ where $P$ occurs, that is, one outputs the set 
$Occ = \{ i,~ S[i\dd i+m-1] = P \}$.

With ``efficiently'' we mean that, in an indexed scenario, we expect the search
times to be sublinear in $n$, typically of the 
form $O((\mathrm{poly}\,m+occ)\,\mathrm{polylog}\,n)$.  
The {\em optimal search time}, since we have to read the input and write the
output, is $O(m+occ)$. Since $P$ can be represented in $m\log\sigma$ bits, in
a few cases we will go further and assume that $P$ comes packed into 
$O(m/\log_\sigma n)$ consecutive machine words, in which case the RAM-optimal
time is $O(m/\log_\sigma n + occ)$.

In general we will handle a collection of $\$$-terminated strings, 
$S_1,\ldots,S_d$, but we model the collection by concatenating the strings
into a single one, $S[1\dd n] = S_1 \cdots S_d$, and doing pattern matching
on $S$. 

\subsection{Suffix Trees and Suffix Arrays} \label{sec:stree}

{\em Suffix trees} and {\em suffix arrays} are the most classical pattern
matching indexes. The {\em suffix tree} \cite{Wei73,McC76,Apo85} is a trie
(or digital tree) containing all the suffixes of $S$. That is, every suffix of
$S$ labels a single root-to-leaf path in the suffix tree, and no node has two
distinct children labeled by the same symbol. Further, the unary paths (i.e., 
paths of nodes with a single child) are compressed into single edges labeled
by the concatenation of the contracted edge symbols. Every internal node in the 
suffix tree corresponds to a substring of $S$ that appears more than once, and 
every leaf corresponds to a suffix. The leaves of the suffix tree indicate the 
position of $S$ where their corresponding suffixes start. Since there are $n$ 
suffixes in $S$, there are $n$ leaves in the suffix tree, and since there are 
no nodes with a single child, it has less than $n$ internal nodes. The suffix
tree can then be represented within $O(n)$ space, for example by representing
every string labeling edges with a couple of pointers to an occurrence of the
label in $S$. The suffix tree can also be built in linear (i.e., $O(n)$) time 
\cite{Wei73,McC76,Ukk95,FFM00}. 

The suffix tree is a very popular data structure in stringology and 
bioinformatics \cite{Apo85,CR02,Gus97}, supporting a large number of complex
searches (by using extra information, such as suffix links, that we omit here). 
The most basic search is pattern matching: since all the occurrences of
$P$ in $S$ are prefixes of suffixes of $S$, we find them all by descending
from the root following the successive symbols of $P$. If at some point we 
cannot descend by some $P[i]$, then $P$ does not occur in $S$. Otherwise, we
exhaust the symbols of $P$ at some suffix tree node $v$ or in the middle of
some edge leading to $v$. We then say that $v$ is the {\em locus} of $P$: every
leaf descending from $v$ is a suffix starting with $P$. If the children $v_1,
\ldots,v_k$ of every suffix tree node $v$ are stored with perfect hashing (the
keys being the first symbols of the strings labeling the edges $(v,v_i)$), then
we reach the locus node in time $O(m)$. Further, since the suffix tree has no
unary paths, the $occ$ leaves with the occurrences of $P$ are traversed from
$v$ in time $O(occ)$. In total, the suffix tree supports pattern matching in
optimal time $O(m+occ)$. With more sophisticated structures, it supports 
RAM-optimal time search, $O(m/\log_\sigma n + occ)$ \cite{NN17}.

A convenient way to regard the suffix tree is as the {\em Patricia tree} 
\cite{Mor68} of all the suffixes of $S$. The Patricia tree, also known as
{\em blind trie} \cite{FG99} (their technical differences are not important
here) is a trie where we compact the unary paths and retain only the first
symbol and the length of the string labeling each edge. In this case we use
the first symbols to choose the appropriate child, and simply trust that the
omitted symbols match $P$. When arriving at the potential locus $v$ of $P$,
we jump to any leaf, where a potential occurrence $S[i\dd i+m-1]$ of $P$ is 
pointed, and compare $P$ with $S[i\dd i+m-1]$. If they match, then $v$ is the
correct locus of $P$ and all its leaves match $P$; otherwise $P$ does not 
occur in $S$. A pointer from each node $v$ to a leaf descending from it is 
needed in order to maintain the verification within the optimal search time.

The suffix array \cite{MM93} of $S[1\dd n]$ is the array $A[1\dd n]$ of the 
positions of the suffixes of $S$ in lexicographic order. If the children of 
the suffix tree nodes are lexicographically ordered by their first symbol,
then the suffix array corresponds to the leaves of the suffix tree in 
left-to-right order.
The suffix array can be built directly, without building the suffix tree, in 
linear time \cite{KSPP05,KA05,KSB06}.

All the suffixes starting with $P$ form a range in the suffix array 
$A[sp\dd ep]$. We can find the range with binary search in time $O(m\log n)$,
by comparing $P$ with the 
strings $S[A[i]\dd A[i]+m-1]$, so as to find the smallest and largest suffixes
that start with $P$. The search time can be reduced to $O(m+\log n)$ by using
further data structures \cite{MM93}.

\begin{figure}[t]
\begin{center}
\includegraphics[width=12cm]{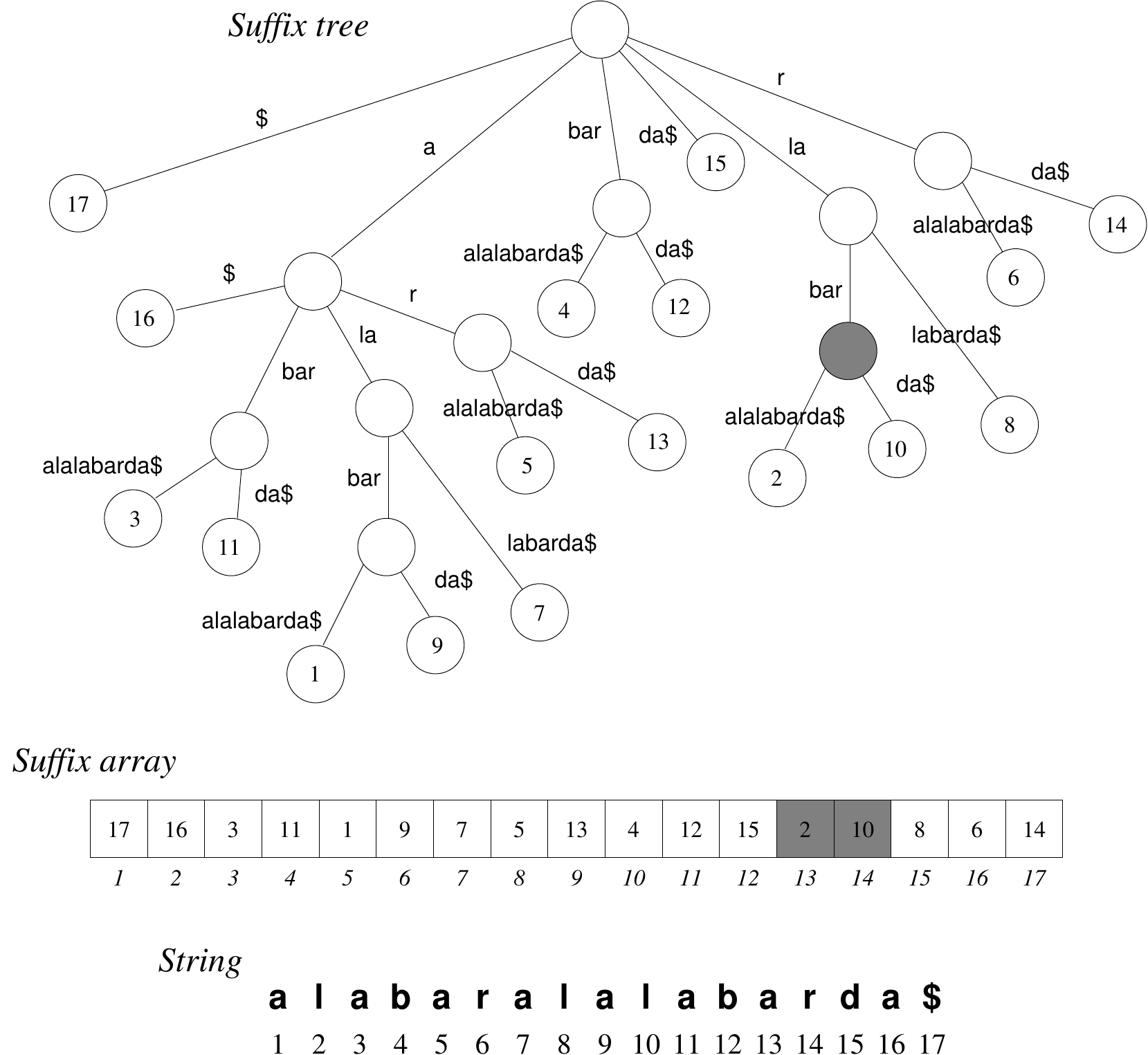}
\end{center}
\caption{The suffix tree and suffix array of the string
$S=\mathsf{alabaralalabarda\$}$.
The suffix tree leaves indicate the positions where the corresponding suffixes 
start, and those positions, collected left to right, form the suffix array. The locus suffix
tree node, and the suffix array interval, for $P=\mathsf{lab}$, are grayed.}
\label{fig:stree}
\end{figure}
 
\begin{example}
Figure~\ref{fig:stree} shows the suffix tree and array of the string
$S=\mathsf{alabaralalabarda\$}$. The search for $P=\mathsf{lab}$ in the
suffix tree leads to the grayed locus node: the search in fact falls in
the middle of the edge from the parent to the locus node. The two leaves
descending from the locus contain the positions $2$ and $10$, which is where
$P$ occurs in $S$. In the suffix array, we find with binary search the interval
$A[13\dd 14]$, where the answers lie.
\end{example}

\subsection{Karp-Rabin Fingerprints} \label{sec:kr}

\citeN{KR87} proposed a technique to compute a {\em signature} or {\em
fingerprint} of a string via hashing, in a way that enables (non-indexed)
string matching in $O(n)$ average time. The signature $\kappa(Q)$ of a string
$Q[1\dd q]$ is defined as
\[ \kappa(Q) ~~=~~ \left( \sum_{i=1}^q Q[i] \cdot b^{i-1} \right) \!\!\!\mod p,
\]
where $b$ is an integer and $p$ a prime number. It is not hard to devise the
arithmetic operations to compute the signatures of composed and decomposed
strings, that is, compute $\kappa(Q\cdot Q')$ from $\kappa(Q)$ and
$\kappa(Q')$, or $\kappa(Q)$ from $\kappa(Q\cdot Q')$ and $\kappa(Q')$, or
$\kappa(Q')$ from $\kappa(Q\cdot Q')$ and $\kappa(Q)$ (possibly storing some 
precomputed exponents together with the signatures).

By appropriately choosing $b$
and $p$, the probability of two substrings having the same fingerprint is very
low. Further, in $O(n\log n)$ expected time, we can find a function $\kappa$
that ensures that no two different substrings of $S[1\dd n]$ have the same fingerprint
\cite{bille2014time}: they build fingerprints $\kappa'$ that are collision-free
only over substrings of lengths that are powers of two, and then define
$\kappa(Q) = \langle \kappa'(Q[1\dd 2^{\lfloor \log_2 q \rfloor}]),
\kappa'(Q[q-2^{\lfloor \log_2 q \rfloor}+1\dd q]) \rangle$.

%

\section{Compressors and Measures of Repetitiveness} \label{sec:measures}

In statistical compression, where the goal is to exploit frequency skew, the
so-called {\em statistical entropy} defined by \citeN{Sha48} offers a measure 
of compressibility that is both optimal and reachable. While statistical
entropy is defined for infinite sources, it can be adapted to individual 
strings. The resulting measure for individual strings, called {\em empirical 
entropy} \cite{CT06}, turns out to be a reachable lower bound (save for 
lower-order terms) to the space a semistatic statistical compressor can 
achieve on that string.

Statistical entropy, however, does not adequately capture
other sources of compressibility, particularly repetitiveness. In this arena,
concepts are much less clear. Beyond the ideal but uncomputable measure of 
string complexity proposed by \citeN{Kol65}, most popular measures of 
(compressibility by exploiting) repetitiveness are ad-hoc, defined as the 
result of particular compressors, and there is not yet a consensus
measure that is both 
reachable and optimal within a reasonable set of compression techniques. Still, 
many measures work well and have been used as the basis of compressed and 
indexed sequence representations. In this section we describe the most relevant
concepts and measures.

\subsection{The Unsuitability of Statistical Entropy} \label{sec:Hk}

\citeN{Sha48} introduced a measure of compressibility that exploits the 
different probabilities of the symbols emitted by a source. In its simplest 
form, the source is ``memoryless'' and emits each symbol $a \in \Sigma$ with 
a fixed probability $p_a$. The {\em entropy} is then defined as
\[ {\mathcal H}(\{ p_a \}) ~~=~~ \sum_{a \in \Sigma} p_a \log \frac{1}{p_a}.\]
When all the probabilities $p_a$ are equal to $1/\sigma$, the entropy is 
maximal, ${\mathcal H} = \log\sigma$. In general, the entropy decreases as the 
probabilities are more skewed. This kind of entropy is called {\em statistical 
entropy}.

In a more general form, the source may remember the last $k$ symbols emitted,
$C[1\dd k]$ and the probability $p_{a|C}$ of the next symbol $a$ may depend on 
them. The entropy is defined in this case as
\[ {\mathcal H}(\{ p_{a,C} \}) ~~=~~ \sum_{C \in \Sigma^k} p_C \sum_{a \in \Sigma} p_{a|C} \log \frac{1}{p_{a|C}},\]
where $p_C$ is the global probabilty of the source emitting $C$.

Other more general kinds of sources are considered, including those that have 
``infinite'' memory of all the previous symbols emitted. \citeN{Sha48} shows 
that any encoder of a random source of symbols with entropy $\mathcal H$ must
emit, on average, no less than $\mathcal H$ bits per symbol. The measure is 
also reachable:
arithmetic coding \cite{WNC87} compresses $n$ symbols from such a source into
$n\mathcal{H}+2$ bits.

Shannon's entropy can also be used to measure the entropy of a {\em finite
individual} sequence $S[1\dd n]$. The idea is to {\em assume} that the only 
source of compressibility of the sequence are the different frequencies of 
its symbols. If we take the frequencies as independent, the result is the 
{\em zeroth order empirical entropy} of $S$:
\[ {\mathcal H}_0(S) ~~=~~ \sum_{a\in\Sigma} \frac{n_a}{n}\log\frac{n}{n_a},\]
where $n_a$ is the number of times $a$ occurs in $S$ and we assume $0\log 0 =0$.
This is exactly the Shannon's entropy of a memoryless source with probabilities
$p_a = n_a/n$, that is, we use the relative frequencies of the symbols in $S$
as an estimate of the probabilities of a hypothetical source that generated
$S$ (indeed, the most likely source). The string $S$ can then be encoded in
$n{\mathcal H}_0(S)+2$ bits with arithmetic coding based on the symbol 
frequencies.

If we assume, instead, that the symbols in $S$ are better predicted by knowing
their $k$ preceding symbols, then we can use the {\em $k$th order empirical
entropy} of $S$ to measure its compressibility:
\[ {\mathcal H}_k(S) ~~=~~  \sum_{C\in\Sigma^k} \frac{n_C}{n} \cdot \mathcal{H}_0(S_C),\]
where $S_C$ is the sequence of the symbols following substring $C$ in $S$ and 
$n_C = |S_C|$. Note that, if $S$ has a unique \$-terminator, $n_C$ is
also the number of times $C$ occurs in $S$,\footnote{Except if $C$ corresponds 
to the last $k$ symbols of $S$, but this does not affect the measure because 
this substring contains $\$$, so it is unique and $n_C = 0$.} and the measure 
corresponds to the Shannon entropy of a source with memory $k$. Once again, an
arithmetic coder encodes $S$ into $n {\mathcal H}_k(S)+2$ bits.

At this point, it is valid to wonder what disallows us to take $k=n$, so that
$n_C = n_S = 1$ if $C=S$ and $n_C = 0$ for the other strings of length $k=n$,
and therefore $\mathcal{H}_n(S)=0$. We could then encode $S$ with arithmetic
coding into 2 bits!

The trick is that the encoding sizes we have given assume that the decoder
knows the distribution, that is, the probabilities $p_a$ or $p_{a|C}$ or, in 
the case of empirical entropies, the frequencies $n_a$ and $n_C$. This may
be reasonable when analyzing the average bit rate to encode a source that
emits infinite sequences of symbols, but not when we consider actual compression
ratios of finite sequences.

Transmitting the symbol frequencies (called the {\em model}) to the decoder
(or, equivalently, storing it together with the compressed string) in plain form
requires $\sigma \log n$ bits for the zeroth order entropy, and 
$\sigma^{k+1}\log n$ bits for the $k$th order entropy. With this simple
model encoding, we cannot hope to achieve compression for $k \ge 
\log_\sigma n$, because encoding the model then takes more space than the
uncompressed string. In fact, this is not far from the best that can be done: 
\citeN{Gag06} shows that, for about that value of $k$, $n{\mathcal H}_k(S)$
sometimes falls below Kolmogorov's complexity, and thus there is no hope of encoding $S$
within that size. In his words, ``$k$th-order empirical entropy stops being
a reasonable complexity metric for almost all strings''.

With the restriction $k \le \log_\sigma n$, consider now that we concatenate
two identical strings, $S\cdot S$. All the relative symbol frequencies in 
$S\cdot S$ are identical to those in $S$, except for the $k-1$ substrings 
$C$ that cover the concatenation point; therefore we can expect that 
${\mathcal H}_k(S\cdot S) \approx {\mathcal H}_k(S)$. Indeed, it can be shown 
that ${\mathcal H}_k(S\cdot S)
\ge {\mathcal H}_k(S)$ \cite[Lem.~2.6]{KN13}. That is, the empirical entropy is
insensitive to the repetitiveness, and any compressor reaching the empirical
entropy will compress $S\cdot S$ to about twice the space it uses to compress 
$S$. Instead, being aware of repetitivenss allows us to compress $S$ in any form
and then somehow state that a second copy of $S$ follows.

This explains why the compressed indexes based on statistical entropy
\cite{NM06} are not suitable for indexing highly repetitive string collections.
In those, the space reduction that can be obtained by exploiting the 
repetitiveness is much more significant than what can be obtained by exploiting
skewed frequencies. 

{\em Dictionary methods} \cite{SS82}, based on representing $S$ as the 
concatenation of strings from a set (a ``dictionary''), generally obtained from
$S$ itself, are more adequate to exploit repetitiveness: a small dictionary of
distinct substrings of $S$
should suffice if $S$ is highly repetitive. Though many dictionary methods can 
be shown to converge to Shannon's entropy, our focus here is their ability to
capture repetitiveness. In the sequel we cover various such methods, not only
as compression methods but also as ways to measure repetitiveness.

We refer the reader to \citeN{CT06} for a deeper discussion of the concepts
of Shannon entropy and its relation to dictionary methods.

\subsection{Lempel-Ziv Compression: Measures $z$, $z_{no}$, and $z_{end}$} \label{sec:lz}

\citeN{LZ76} proposed a technique to measure the ``complexity'' of individual
strings based on their repetitiveness (in our case, complexity can be 
interpreted as incompressibility). The compressor LZ77 
\cite{ZL77} and many other variants \cite{BCW90} that derived from this measure
have become very popular; they are behind compression software like {\tt zip}, 
{\tt p7zip}, {\tt gzip}, {\tt arj}, etc.

\subsubsection{The compressor}
The original Lempel-Ziv method {\em parses} (i.e., partitions) $S[1\dd n]$ into
{\em phrases} (i.e., substrings) as follows, starting from $i \leftarrow 1$:
\begin{enumerate}
	\item Find the shortest prefix $S[i\dd j]$ of $S[i\dd n]$ that does not
occur in $S$ starting before position $i$.
	\item The next phrase is then $S[i\dd j]$. 
	\item Set $i \leftarrow j+1$. If $i \le n$, continue forming phrases.
\end{enumerate}

This greedy parsing method can be proved to be optimal (i.e., producing the 
least number of phrases) among all {\em left-to-right} parses (i.e., those where
phrases must have an occurrence starting to their left) \cite[Thm.~1]{LZ76}. 

A compressor can be obtained by
encoding each phrase as a triplet: If $S[i\dd j]$ is the next phrase, then
$S[i\dd j-1]$ occurs somewhere to the left of $i$ in $S$. Let $S[i'\dd j']$ be 
one such occurrence (called the {\em source} of the phrase), that is, $i' < i$. 
The next triplet is then $\langle i',j-i,S[j] \rangle$. When $j-i=0$, any
empty substring can be the source, and it is customary to assume $i'=0$, so
that the triplet is $\langle 0,0,S[j] \rangle$. 

\begin{example}
The string $S=\mathsf{alabaralalabarda\$}$ is parsed as 
$\mathsf{a|l|ab|ar|alal|abard|a\$}$, where we use the vertical bar to 
separate the phrases. A possible triplet encoding is
$\langle 0,0,\mathsf{a} \rangle
 \langle 0,0,\mathsf{l} \rangle
 \langle 1,1,\mathsf{b} \rangle
 \langle 1,1,\mathsf{r} \rangle
 \langle 1,3,\mathsf{l} \rangle
 \langle 3,4,\mathsf{d} \rangle
 \langle 11,1,\mathsf{\$} \rangle$.
\end{example}

From the triplets, we easily recover $S$ by starting with an empty string $S$
and, for each new triplet $\langle p,\ell,c \rangle$, appending 
$S[p\dd p+\ell-1]$ and then $c$.%
\footnote{Since the source may overlap the formed phrase, the copy of 
$S[p\dd p+\ell-1]$ to the end of $S$ must be done left to right: consider 
recovering $S=\mathsf{a}^{n-1}\mathsf{\$}$ from the encoding 
$\langle 0,0,\mathsf{a} \rangle \langle 1,n-2,\mathsf{\$} \rangle$.}

This extremely fast decompression is one of the reasons of the popularity of 
Lempel-Ziv compression. Another reason is that, though not as easily as 
decompression, it is also possible to carry out the compression
(i.e., the parsing) in $O(n)$ time \cite{RPE81,SS82}.
Recently, there has been a lot of research on doing the parsing within little 
extra space, see for example \citeN{KKP16}, \citeN{FIKS18}, and references 
therein.

\subsubsection{The measure} \label{sec:zmeasure}
For this survey we will use a slightly different variant of the Lempel-Ziv 
parsing,
which is less popular for compression but more coherent with other measures of 
repetitiveness, and simplifies indexing. The parsing into phrases is redefined 
as follows, also starting with $i \leftarrow 1$.

\begin{enumerate}
	\item Find the longest prefix $S[i\dd j]$ of $S[i\dd n]$ that 
occurs in $S$ starting before position $i$.
	\item If $j \ge i$, that is, $S[i\dd j]$ is nonempty, then the next
phrase is $S[i\dd j]$, and we set $i \leftarrow j+1$.
	\item Otherwise, the next phrase is the explicit symbol $S[i]$, which 
has not appeared before, and we set $i \leftarrow i+1$.
	\item If $i \le n$, continue forming phrases.
\end{enumerate}

We define the Lempel-Ziv measure of $S[1\dd n]$ as the number $z=z(S)$ of 
phrases into which $S$ is parsed by this procedure.

\begin{figure}[t]
\begin{center}
\includegraphics[width=8cm]{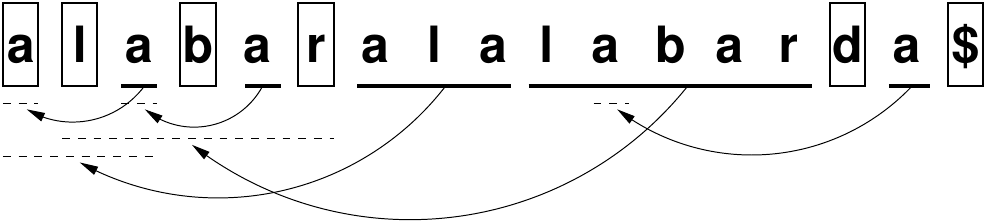}
\end{center}
\caption{Lempel-Ziv parse of $S=\mathsf{alabaralalabarda\$}$. Each phrase is
either an underlined string, which appears before, or a boxed symbol. The 
arrows go from each underlined string to one of its
occurrences to the left (which is underlined with a dashed line).}
\label{fig:lzparse}
\end{figure}

\begin{example}
Figure~\ref{fig:lzparse} shows how the string $S=\mathsf{alabaralalabarda\$}$
is parsed into $z(S)=11$ phrases, $\mathsf{\caja{a}|\caja{l}|a|
\caja{b}|a|\caja{r}|ala|labar|\caja{d}|a|\caja{\$}}$,
with the explicit symbols boxed.
\end{example}

The two parsing variants are closely related. If the original variant forms the 
phrase $S[i\dd j]$ with $j>i$, then $S[i\dd j-1]$ is the longest prefix of
$S[i\dd n]$ that appears starting to the left of $i$, so this new variant will
form the phrase $S[i\dd j-1]$ (if $i<j$) and its next phrase will be either
just $S[j]$ or a longer prefix of $S[j\dd n]$. It is not hard to see that the
compression algorithms for both variants are the same, and that the greedy 
parsing is also optimal for this variant. It follows that $z' \le z \le 2z'$,
where $z'$ is the number of phrases created with the original method. Thus,
$z$ and $z'$ are equivalent in asymptotic terms. In particular, the triplet
encoding we described shows that one can encode $S$ within $O(z\log n)$ bits 
(or $O(z)$ words), which makes $z$ a reachable compressibility measure.

\subsubsection{Weaker variants}
\citeN{SS82} use a slightly weaker Lempel-Ziv parse, where the source
$S[i'\dd j']$ of $S[i\dd j]$ must be completely contained in 
$S[1\dd i-1]$. That is, it must hold that $j' < i$, not just $i' < i$. 
The same greedy parsing described, using this stricter condition, also 
yields the least number of phrases \cite[Thm.~10 with $p=1$]{SS82}.
The phrase encoding and decompression proceed in exactly the same way, and
linear-time parsing is also possible \cite{CIKRW12}.
The number of phrases obtained in this case will be called $z_{no}$, with $no$
standing for ``no overlap'' between phrases and their sources. 

This parsing simplifies, for example, direct pattern matching on the 
compressed string \cite{GKPR96,FT98} or creating context-free grammars from the 
Lempel-Ziv parse \cite{Ryt03}.\footnote{A Lempel-Ziv based index also claims
to need this restricted parsing \cite{KN13}, but in fact they can handle the
original parsing with no changes.} It comes with a price, however. Not only
$z_{no}(S) \ge z(S)$ holds for every string $S$ because the greedy parsings
are optimal, but also
$z_{no}$ can be $\Theta(\log n)$ times larger than $z$, for example
on the string $S = \mathsf{a}^{n-1}\mathsf{\$}$, where $z=3$ (with parsing 
$\caja{a}|\mathsf{a}^{n-2}| \mathsf{\caja{\$}}$) and $z_{no} = \Theta(\log n)$ 
(with parsing $\mathsf{\caja{a}|\mathsf{a}|\mathsf{a}^2|\mathsf{a}^4|
\mathsf{a}^8|}\cdots$).

An even more restricted variant called LZ-End was proposed by \citeN{KN13}.
Adapted to the definition we use in Section~\ref{sec:zmeasure}, it corresponds 
to forcing the previous occurrence of $S[i\dd j]$ to end at a phrase
boundary. The number of phrases produced by this parsing, $z_e$, clearly
satisfies $z_e \ge z_{no}$, but it is unknown if there are string families 
where $z_{no} = o(z_e)$; only cases where they differ by a constant factor 
have been devised \cite{KN13,IMFNIT21}. As for upper bounds, \citeN{KS22} 
recently proved that $z_e = O(z\log^2(n/z))$.

The LZ-End parsing was defined in order to speed up substring extraction in
some convenient cases \cite{KN13}; recently it has been shown that the 
restriction suffices to provide extraction of arbitrary substrings in 
polylogarithmic time \cite{KS22}.
Further, the LZ-End parsing can be computed in linear time \cite{KK17,KK17esa}.

In this survey we are more interested in the strength of the 
repetitiveness measures, so we rather define $z_{end}$ as the {\em smallest} 
parsing that can be obtained so that every phrase is either an individual
symbol or occurs earlier ending at a phrase boundary. Since the greedy
parsing defined by \citeN{KN13} is not necessarily optimal, there is no
obvious method to compute the measure $z_{no} \le z_{end} \le z_e$. It inherits,
however, the other good properties of the LZ-End parsing and its optimality
will allow us bound it better in terms of other measures.

\subsubsection{Evaluation}
Apart from fast compression and decompression, a reason for the popularity of 
Lempel-Ziv compression is that all of its variants converge to the statistical 
entropy \cite{LZ76}, even on individual strings \cite{KM00}, though statistical
methods converge faster (i.e., their sublinear extra space over the empirical 
entropy of $S[1\dd n]$ is a slower-growing function of $n$).
In particular, it holds that $z_e \log n = O(n\mathcal{H}_k)$ \cite{KN13}, so 
the space $O(z_e)$ is at worst $\Theta(n\log\sigma)$
bits, which is proportional to the plain size of $S$.

More important for us is that Lempel-Ziv captures repetitiveness. In our 
preceding example of the string $S\cdot S$, we have $z(S\cdot S) \le z(S)+1$,
$z_{no}(S\cdot S) \le z_{no}(S)+1$, and $z_{end}(S\cdot S) \le z_{end}(S)+1$
(i.e., we need at most one extra phrase to capture the second copy of $S$);
this could not be the case with $z_e$.

Despite the success of Lempel-Ziv compression and its frequent use as a gold 
standard to quantify repetitiveness, the measure $z$ (and, analogously, $z_{no}$ and
$z_{end}$) has 
some pitfalls:
\begin{itemize}
	\item It is asymmetric, that is, $z(S)$ may differ from $z(S^{rev})$.
For example, removing the terminator $\mathsf{\$}$ to avoid complications,
$\mathsf{alabaralalabarda}$ is parsed into $z=10$ phrases, whereas its reverse
$\mathsf{adrabalalarabala}$ requires only $z=9$.
	\item It is monotonic when removing suffixes, but not prefixes, that 
is, $z(S')$ can be larger than $z(S \cdot S')$. For example, 
$\mathsf{aaabaaabaaa}$ is parsed
into $z=4$ phrases, $\mathsf{\caja{a}|aa|\caja{b}|aaabaaa}$, but 
$\mathsf{aabaaabaaa}$ needs $z=5$, $\mathsf{\caja{a}|a|\caja{b}|aa|abaaa}$.
	\item Although it is the optimal size of a left-to-right parse, $z$ is 
arguably not optimal within a broader class of plausible
compressed representations. One can represent $S$ using fewer phrases by 
allowing their sources to occur also to their right in $S$, 
as we show with the next measure.
\end{itemize}

\subsection{Bidirectional Macro Schemes: Measure $b$}

\citeN{SS82} proposed an extension of Lempel-Ziv parsing that allows sources
to be to the left or to the right of their corresponding phrases, as long as 
every symbol can
eventually be decoded by following the dependencies between phrases and sources.
They called such a parse a ``bidirectional macro scheme''. Analogously to the
Lempel-Ziv parsing variant we are using, a phrase is either a substring that 
appears elsewhere, or an explicit symbol. 

The dependencies between sources and
phrases can be expressed through a function $f$, such that $f(i)=j$ if position
$S[i]$ is to be obtained from $S[j]$; we set $f(i)=0$ if $S[i]$ is an explicit
symbol. Otherwise, if $S[i\dd j]$ is a copied phrase, then $f(i+t)=f(i)+t$
for all $0 \le t \le j-i$, that is, $S[f(i)\dd f(j)]$ is the source of 
$S[i\dd j]$. The bidirectional macro scheme is valid if, for each $1\le i\le n$,
there is a $k > 0$ such that $f^k(i)=0$, that is, every position is eventually
decoded by repeatedly looking for the sources.

We call $b = b(S)$ the minimum number of phrases of a bidirectional macro 
scheme for $S[1\dd n]$. It obviously holds that $b(S) \le z(S)$ for every
string $S$ because Lempel-Ziv is just one possible bidirectional macro scheme.

\begin{figure}[t]
\begin{center}
\includegraphics[width=8cm]{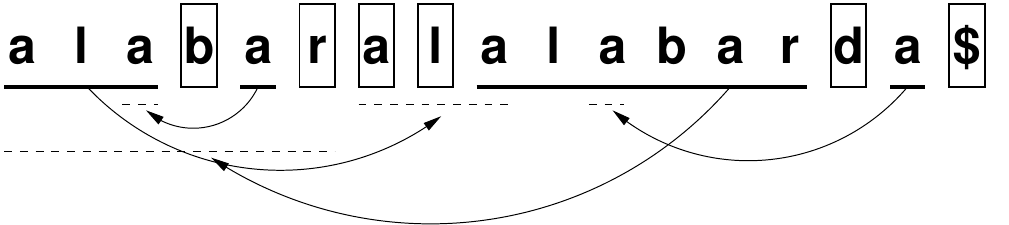}
\end{center}
\caption{A bidirectional macro scheme for $S=\mathsf{alabaralalabarda\$}$, 
with the conventions of Figure~\ref{fig:lzparse}.}
\label{fig:bms}
\end{figure}

\begin{example}
Figure~\ref{fig:bms} shows a bidirectional macro scheme for 
$S = \mathsf{alabaralalabarda\$}$ formed by $b=10$ phrases, 
$S=\mathsf{ala|\caja{b}|a|\caja{r}|\caja{a}|\caja{l}|alabar|\caja{d}|a|\caja{\$}}$ (we had $z=11$ for the same string,
see Figure~\ref{fig:lzparse}). 
It has function $f[1\dd n] = \langle 7,8,9,0,3,0,0,0,1,2,3,4,5,6,$ $0,11,0\rangle$.
One can see that every symbol can eventually be 
obtained from the explicit ones. For example, following the arrows in the
figure (or, similarly, iterating the function $f$),
we can obtain $S[13]=S[5]=S[3]=S[9]=S[1]=S[7]=\mathsf{a}$. 
\end{example}

Just as for Lempel-Ziv, we can compress $S$ to $O(b)$ space by encoding the
source of each of the $b$ phrases. 
It is not hard to recover $S[1\dd n]$ in $O(n)$ time from the 
encoded phrases, because when we traverse $t$ positions until finding an 
explicit symbol in time $O(t)$, we discover the contents of all those $t$  
positions. Instead, finding the smallest bidirectional macro scheme is 
NP-hard \cite{Gal82}. 
This is probably the reason that made this technique less popular, although 
some attempts exist to approach it heuristically \cite{NT19,RCNF20}.

As said, it always holds that $b \le z$. \citeN{GNP18latin} (corrected in
\citeN{NPO19}) showed that $z = O(b\log(n/b))$ for every string family,
and that this bound is tight: in the Fibonacci words, where $F_1 = 
\mathsf{b}$, $F_2 = \mathsf{a}$, and $F_k = F_{k-1} \cdot F_{k-2}$ for
$k>2$, it holds that $b=O(1)$ and $z = \Theta(\log n)$.

Measure $b$ is the smallest of those we study that is {\em reachable}, that is,
we can compress $S[1\dd n]$ to $O(b)$ words. It is also symmetric, 
unlike $z$: $b(S) = b(S^{rev})$. 
Still, $b$ is not monotonic: there are strings $S$ and $S'$ where
$b(S) > b(S \cdot S')$.\footnote{Paolo Ferragina and Francesco Tosoni
(personal communication) show that $b(\mathsf{aabaaaabaa})=5$ but 
$b(\mathsf{aabaaaabaaa})=4$.}

\subsection{Grammar Compression: Measures $g$ and $g_{rl}$}
\label{sec:grammar}

\citeN{KY00} introduced a compression technique based on context-free
grammars (the idea can be traced back to \citeN{Rub76}). Given $S[1\dd n]$, 
we find a grammar that generates only the
string $S$, and use it as a compressed representation. The size of a 
grammar is the sum of the lengths of the right-hand sides of the rules
(we avoid empty right-hand sides). 

\begin{figure}[t]
\begin{center}
\includegraphics[width=8cm]{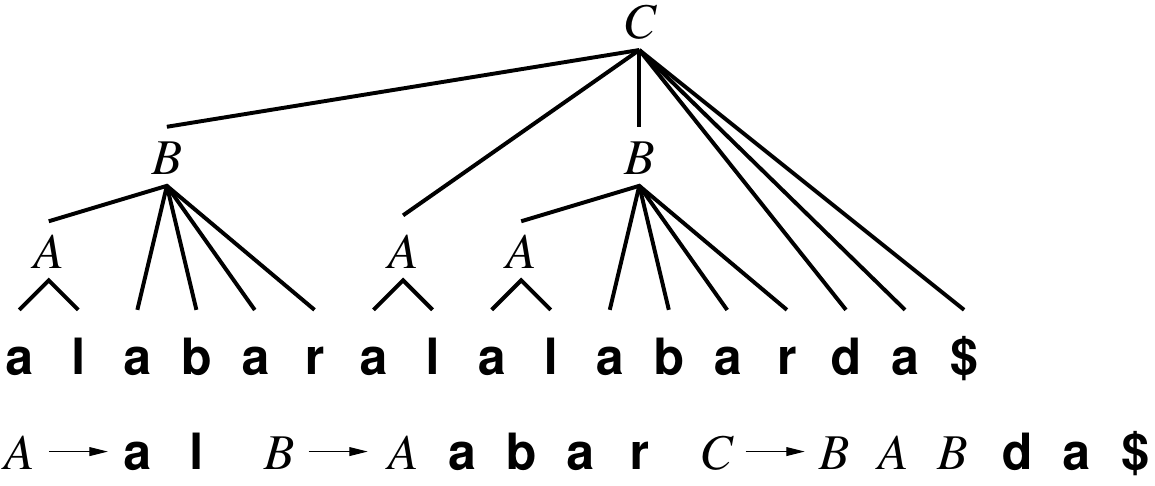}
\end{center}
\caption{A context-free grammar generating the string
$S=\mathsf{alabaralalabarda\$}$. The rules of the grammar are on the
bottom. On the top we show the parse tree.}
\label{fig:g}
\end{figure}

\begin{example}
Figure~\ref{fig:g} shows a context-free grammar that generates only the string
$S=\mathsf{alabaralalabarda\$}$. The grammar has three rules, $A \rightarrow
\mathsf{al}$, $B \rightarrow A \mathsf{abar}$, and the initial rule
$C \rightarrow BAB\mathsf{da\$}$. The sum of the lengths of the right-hand sides
of the rules is $13$, the grammar size. 
\end{example}

Note that, in a grammar that generates
only one string, there is exactly one rule $A \rightarrow X_1 \cdots X_k$ per 
nonterminal $A$, where each $X_i$ is a terminal or a nonterminal (if there is
more than one rule per nonterminal, these must be redundant and we can leave
only one).

Figure~\ref{fig:g} also displays the {\em parse tree} of the grammar: an ordinal
labeled tree where the root is labeled with the initial symbol, the leaves
are labeled with the terminals that spell out $S$, and each internal node
is labeled with a nonterminal $A$: if $A \rightarrow X_1\cdots X_k$, then
the node has $k$ children labeled, left to right, $X_1,\ldots,X_k$.

\citeN{KY00} prove that grammars that satisfy a few reasonable rules reach the 
$k$th order entropy of a source, and the same holds for the empirical entropy 
of individual strings \cite{ON18}. Their size is always $O(n/\log_\sigma n)$.

Grammar compression is interesting for us because repetitive strings should
have small grammars. Our associated measure of repetitiveness is then the size
$g=g(S)$ of the smallest grammar that generates only $S$. 
It is known that $z_{no} \le g=O(z_{no}\log(n/z_{no}))$ 
\cite{CLLPPRSS02,Ryt03,CLLPPSS05},
and even $g = O(z\log(n/z))$ \cite[Lem.~8]{Gaw11}. 
The same construction of \citeN{CLLPPSS05} (their Lemma 9) shows
that $z_{end} \le g$, since the parsing they obtain indeed satisfies the LZ-End
rules.

Finding such a smallest grammar is NP-complete, however \cite{SS82,CLLPPSS05}. 
This has not made grammar compression unpopular, because several
efficient constructions yield grammars of size $O(z_{no}\log(n/z_{no}))$
and even $O(z\log(n/z))$ \cite{Ryt03,CLLPPSS05,Sak05,Jez15,Jez16}. Further,
heuristics like RePair \cite{LM00} or Sequitur \cite{NMWM04} perform extremely 
well and are preferred in practice.

Since it always holds that $z_{end} \le g$, a natural question is why grammar
compression is interesting. One important reason is that grammars allow for
direct access to the compressed string in logarithmic time, as we will describe
in Section~\ref{sec:access-grammars}. 
For now, a simple version illustrates its power.
A grammar construction algorithm produces {\em balanced} grammars if the height
of their parse tree is $O(\log n)$ when built on strings of length $n$. On a
balanced grammar for $S[1\dd n]$, with constant-size rules, it is very easy to 
extract any symbol $S[i]$ by virtually traversing the parse tree, if one stores
the lengths of the string represented by each nonterminal. The first grammar
constructions built from a Lempel-Ziv parse \cite{CLLPPRSS02,Ryt03} had these
properties, and thus they were the first structures of size
$O(z_{no}\log(n/z_{no}))$ with access time $O(\log n)$.

A little more notation on grammars will be useful. We call $exp(A)$ the string
of terminals to which nonterminal $A$ expands, and $|A| = |exp(A)|$. To 
simplify matters, we forbid rules of right-hand side length $0$ or $1$. An 
important concept will be the {\em grammar tree} (cf.\ partial parse-tree
\cite{Ryt03}), which is obtained by pruning
the parse tree: for each nonterminal $A$, only one internal node labeled $A$
is retained; all the others are converted to leaves by pruning their subtree.
Since the grammar tree will have the $k$ children of each unique nonterminal 
$A \rightarrow X_1 \cdots X_k$, plus the root, its total number of nodes is
$g+1$ for a grammar of size $g$. 

\begin{figure}[t]
\begin{center}
\includegraphics[width=8cm]{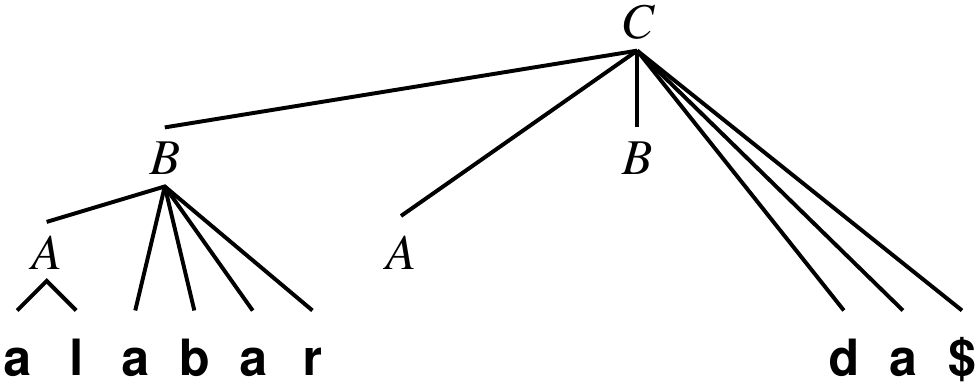}
\end{center}
\caption{The grammar tree of the grammar of Figure~\ref{fig:g}.}
\label{fig:gt}
\end{figure}

With the grammar tree we can easily see that $z_{end} \le g$, for
example. Consider a grammar tree where only leftmost occurrence of every 
nonterminal is an internal node. The string $S$ is then cut into at most $g$ 
substrings, each covered by a leaf of the grammar tree. Each leaf is either a 
terminal or a pruned nonterminal. We can then define a 
left-to-right parse with $g$ phrases: the phrase covered by pruned nonterminal 
$A$ points to its copy below the internal node $A$, which is to its left;
terminal leaves become explicit symbols.
Since this is a left-to-right parse where every source ends at a phrase
boundary and $z_{end}$ is the least
size of such a parse, we have $z_{end} \le g$. 

\begin{example}
Figure~\ref{fig:gt} shows the grammar tree of the parse tree of 
Figure~\ref{fig:g}, with $14$ nodes. It induces the left-to-right parse
$S=\mathsf{\caja{a}|\caja{l}|\caja{a}|\caja{b}|\caja{a}|\caja{r}|al|alabar|\caja{d}|\caja{a}|\caja{\$}}$.
\end{example}


\subsubsection{Run-length grammars}

To handle some anomalies that occur when compressing repetitive strings with
grammars, \citeN{NIIBT16} proposed to enrich context-free grammars with
{\em run-length} rules, which are of the form $A \rightarrow X^t$, where $X$
is a terminal or a nonterminal and $t \ge 2$ is an integer. The rule is
equivalent to $A \rightarrow X \cdots X$ with $t$ repetitions of $X$, but it
is assumed to be of size $2$. Grammars that use run-length rules are called
{\em run-length (context-free) grammars}.

We call $g_{rl} = g_{rl}(S)$ the size of the smallest run-length grammar that 
generates $S$. It obviously holds that $g_{rl}(S)\le g(S)$ for every string $S$.
It can also be proved that $z(S) \le g_{rl}(S)$ for every string $S$
\cite{GNP18latin}\footnote{They claim $z \le 2g_{rl}$ because they use a 
different definition of grammar size.}.
An interesting connection with bidirectional macro schemes is that
$g_{rl} = O(b\log(n/b))$ (from where $z=O(b\log(n/b))$ is obtained)
\cite{GNP18latin,NPO19}.
There is no clear dominance between $g_{rl}$ and $z_{no}$, however: On the
string family $S=\mathsf{a}^{n-1} \mathsf{\$}$ we have $g_{rl}=O(1)$ and 
$z_{no}=\Theta(\log n)$ (as well as $g=\Theta(\log n)$), but there exist
string families where $g_{rl} = \Omega(z_{no}\log n/\log\log n)$
\cite{BGGP18} (the weaker result $g = \Omega(z_{no}\log n/\log\log n)$ was 
known before \cite{CLLPPSS05,HLR16}). Further, the string family used by 
\citeN{BGGP18} is parsed into $O(z_{no})$ phrases by
the original LZ-End, thus showing that $g_{rl} = \Omega(z_e\log n/\log\log n)$.

The parse tree of a run-length grammar is the same as if rules $A \rightarrow
X^t$ were written as $A \rightarrow X \cdots X$. The grammar tree, instead,
is modified to ensure it has $g_{rl}+1$ nodes if the grammar is of size
$g_{rl}$: The internal node labeled $A$ has two children, the left one is 
labeled $X$ and the right one is labeled $X^{[t-1]}$. Those special marked
nodes are treated differently in the various indexes and access methods on
run-length grammars.

\subsection{Collage Systems: Measure $c$} \label{sec:collage}

To generalize sequential pattern matching algorithms, \citeN{KMSTSA03} proposed
an extension of run-length grammars called {\em collage systems}. These allow,
in addition, {\em truncation} rules of the form $A \rightarrow B^{[t]}$ and
$A \rightarrow \!^{[t]}B$, which are of size 2 and mean that $exp(A)$ consists 
of the first or last $t$ symbols of $exp(B)$, respectively. Collage systems 
also extend the {\em composition systems} of \citeN{GKPR96}, which lack the
run-length rules.

\begin{example}
A collage system generating $S=\mathsf{alabaralalabarda\$}$, though larger than
our grammar, is $A \rightarrow \mathsf{al}$, $B \rightarrow AA \mathsf{abar}$, 
$B' \rightarrow \!^{[6]}B$, and the initial rule
$C \rightarrow B' B \mathsf{da\$}$.
\end{example}

The size of the smallest collage
system generating a string $S$ is called $c = c(S)$, and thus it obviously holds
that $c(S) \le g_{rl}(S)$ for every string $S$. \citeN{KMSTSA03} also proved
that $c = O(z \log z)$; a better bound $c = O(z)$ was recently proved by
\citeN{NPO19}. This is interesting because it sheds light on what must be added
to grammars in order to make them as powerful as Lempel-Ziv parses.

\citeN{NPO19} also prove lower bounds on $c$ when restricted to what they call
{\em internal} collage systems, where $exp(A)$ must appear in $S$ for every
nonterminal $A$. This avoids collage systems that generate a huge string from
which a small string $S$ is then obtained by truncation. For internal collage
systems it holds that $b = O(c)$, and on Fibonacci words it holds $b=O(1)$
and $c = \Theta(\log n)$. Instead, while the bound $c = O(z)$ also
holds for internal collage systems, it is unknown if there are string
families where $c = o(z)$, even for general collage systems.

\subsection{Burrows-Wheeler Transform: Measure $r$} \label{sec:bwt}

\citeN{BW94} designed a reversible transformation with the goal of making 
strings easier to compress by local methods. The Burrows-Wheeler Transform
(BWT) of $S[1\dd n]$, $S^{bwt}$, is a permutation of $S$ obtained as follows:
\begin{enumerate}
	\item Sort all the suffixes of $S$ lexicographically (as in the suffix 
array).
	\item Collect, in increasing order, the symbol {\em preceding} each
suffix (the symbol preceding the longest suffix is taken to be \$).
\end{enumerate}

\begin{figure}[t]
\begin{center}
\includegraphics[width=9cm]{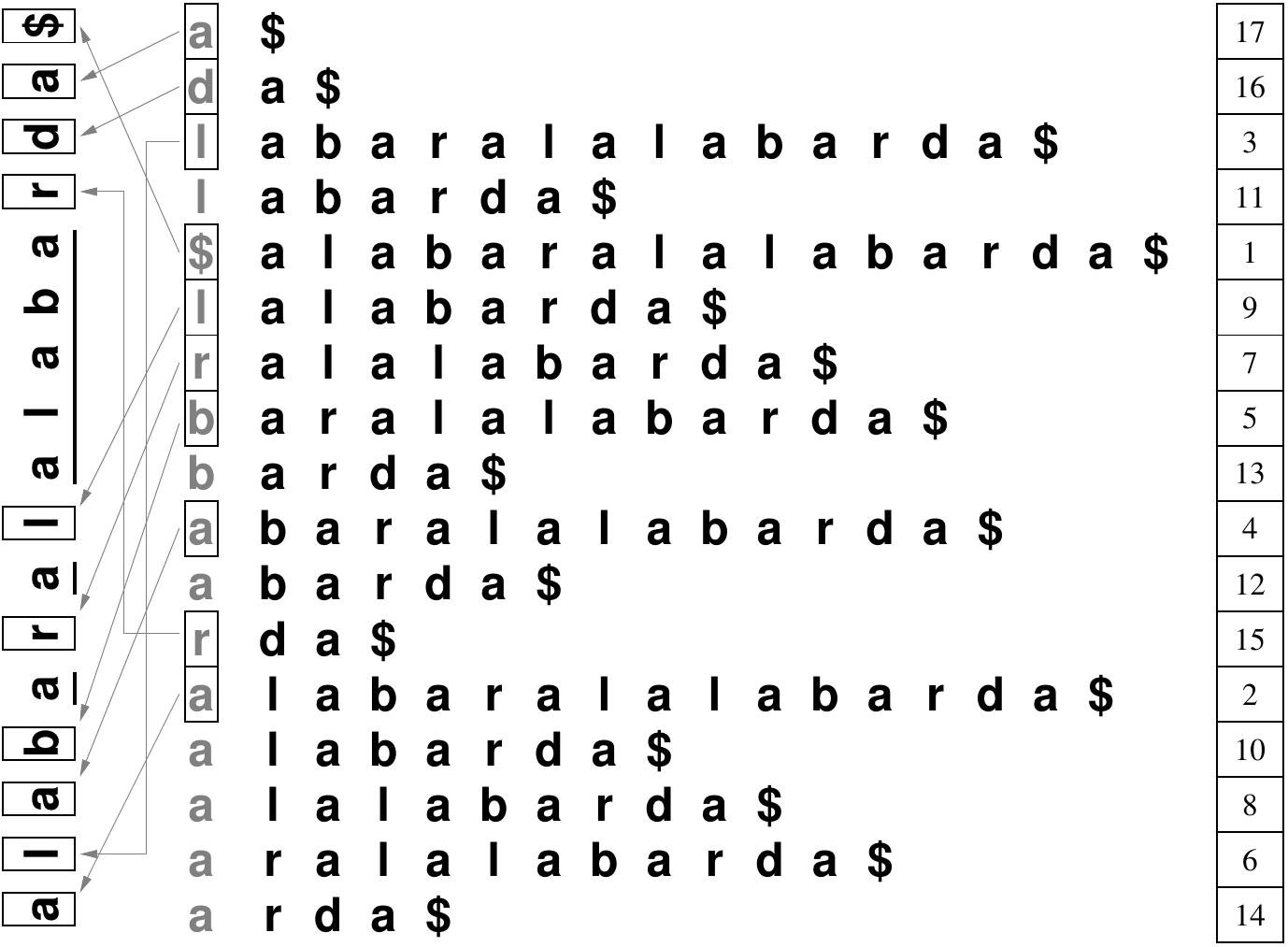}
\end{center}
\caption{The list of suffixes of $S=\mathsf{alabaralalabarda\$}$ in increasing
lexicographic order. The sequence of preceding symbols (in gray) forms the BWT
of $S$, $S^{bwt}=\mathsf{adll\$lrbbaaraaaaa}$. The run heads are boxed.
To their left, we show how they are mapped to $S$ and become the explicit 
symbols of the induced bidirectional macro scheme, using the same conventions 
of Figure~\ref{fig:lzparse}. On the right we show the suffix array of $S$.}
\label{fig:runs}
\end{figure}

\begin{example}
The BWT of $S=\mathsf{alabaralalabarda\$}$ is 
$S^{bwt} = \mathsf{adll\$lrbbaaraaaaa}$, as shown in Figure~\ref{fig:runs}
(ignore the leftmost part of the figure for now).
\end{example}

It turns out that, by properly 
partitioning $S^{bwt}$ and applying zeroth-order compression to each piece,
one obtains $k$th order compression of $S$ \cite{Man01,FGMS05,GKKPP19}. The
reason behind this fact is that the BWT puts together all the suffixes 
starting with the same context $C$ of length $k$, for any $k$. Then, 
encoding the symbols preceding those suffixes is the same as encoding together 
the symbols preceding the occurrences of each context $C$ in $S$. Compare
with the definition of empirical $k$th order entropy we gave in 
Section~\ref{sec:Hk} applied to the reverse of $S$, ${\cal H}_k(S^{rev})$.

The BWT has a strong connection with the suffix array $A$ of $S$; see the
right of Figure~\ref{fig:runs}: it is not hard to see that
\[   S^{bwt}[j] ~~=~~ S[A[j]-1], \]
if we interpret $S[0]=S[n]$.
From the suffix array, which can be built in linear time,
we easily compute the BWT. The BWT is also easily reversed in linear
time \cite{BW94} and even within compact space \cite{KP10,KKP12}.
The connection between the BWT and the suffix array
has been used to implement fast searches on 
$S[1\dd n]$ in $n{\cal H}_k(S) + o(n\log\sigma)$ bits of space \cite{FM05,NM06}; we
see in Section~\ref{sec:suffix-indexing} how this technique is implemented 
when $S$ is highly repetitive.

In addition, the BWT has 
interesting properties on highly repetitive strings, with connections to
the measures we have been studying. Let us define $r=r(S)$ as the number
of equal-symbol {\em runs} in $S^{bwt}$.
Since the BWT is reversible, we can represent $S$ in $O(r)$ space, by
encoding the $r$ symbols and run lengths of $S^{bwt}$. While it is not known
how to provide fast access to any $S[i]$ within this $O(r)$ space, it is 
possible to provide fast pattern searches by emulating the original BWT-based
indexes \cite{MN05,MNSV09,GNP19}.

\begin{example}
The BWT of $S=\mathsf{alabaralalabarda\$}$,
$S^{bwt} = \mathsf{adll\$lrbbaaraaaaa}$, has $r(S)=10$ runs. It can be 
encoded by the symbols and lengths of the runs: 
$(\mathsf{a},1),(\mathsf{d},1),(\mathsf{l},2),$ $(\mathsf{\$},1),(\mathsf{l},1),
(\mathsf{r},1),(\mathsf{b},2),(\mathsf{a},2),(\mathsf{r},1),(\mathsf{a},5)$,
and then we can recover $S$ from $S^{bwt}$.
\end{example}

There is no direct dominance relation between BWT runs and Lempel-Ziv parses or
grammars: on the de Bruijn sequences over binary alphabets,%
\footnote{The de Bruijn sequence of degree $k$ over an alphabet of size $\sigma$
contains all the possible substrings of length $k$ within the minimum possible
length, $\sigma^k+k-1$.}
it holds $r = \Theta(n)$ \cite{BCGPR15} and thus, since
$g = O(n/\log n)$, we have $r = \Omega(g \log n)$. On the even Fibonacci words,
on the other hand, it holds that $r=O(1)$ and $z=\Theta(\log n)$
\cite{Pre16,NPO19}.

Interestingly, a relation of $r$ with bidirectional macro schemes can be proved,
$b = O(r)$ \cite{NPO19}, by noting that the BWT runs induce a bidirectional 
macro scheme of size $2r$: If we map each position $S^{bwt}[j]$ starting a run
to the position $S[i]$ where $S^{bwt}[j]$ occurs, then we define the
phrases as all those explicit positions $i$ plus the (non-explicit) substrings
between those explicit positions. Since that is shown to be a valid
bidirectional scheme, it follows that $b \le 2r$. 

\begin{example}
On $S = \mathsf{alabaralalabarda\$}$, with
$S^{bwt} = \mathsf{adll\$lrbbaaraaaaa}$, the corresponding bidirectional macro
scheme is
$S=\mathsf{\caja{a}|\caja{l}|\caja{a}|\caja{b}|a|\caja{r}|a|\caja{l}|alaba|\caja{r}|\caja{d}|\caja{a}|\caja{\$}}$,
of size $13$. See the leftmost part of Figure~\ref{fig:runs}.
\end{example}

As a final observation, we note that a drawback of $r$ as a repetitiveness
measure is that $r(S)$ and $r(S^{rev})$ may differ by a factor of 
$\Omega(\log n)$ \cite{GILPST20}. Further, $r$ depends on the order of the
alphabet; it is NP-hard to find the alphabet permutation that minimizes $r$ \cite{BGT19}.

\begin{example}
Replacing $\mathsf{a}$ by $\mathsf{e}$ in $S=\mathsf{alabaralalabarda\$}$
we obtain $S'=\mathsf{elebereleleberde\$}$, whose BWT has only $r(S')=8$ runs.
\end{example}

\subsection{Lexicographic Parsing: Measure $v$} \label{sec:lex-parse}

\citeN{NPO19} generalized the Lempel-Ziv parsing into ``ordered'' parsings, 
which are bidirectional macro schemes where each nonexplicit phrase equals
some substring of $S$ that is 
smaller under some criterion (in Lempel-Ziv, the criterion is to start earlier 
in $S$). A particularly interesting case are the so-called {\em lexicographic 
parsings}, where each nonexplicit phrase $S[i\dd j]$ must have a copy 
$S[i'\dd j']$ where the suffix $S[i'\dd]$ is lexicographically smaller than 
$S[i\dd]$. The smallest lexicographic parse of $S$ is called the {\em 
lex-parse} of $S$ and has $v = v(S)$ phrases. It is obtained by
processing $S$ left to right and maximizing the length of each phrase, as for
Lempel-Ziv, that is, $S[i\dd j]$ is the longest prefix of $S[i\dd]$ that
occurs at some $S[i'\dd j']$ where the suffix $S[i'\dd]$ is lexicographically
smaller than $S[i\dd]$. Note that this is the same to say that $S[i'\dd]$ 
is the suffix that lexicographically precedes $S[i\dd]$. 

\begin{figure}[t]
\begin{center}
\includegraphics[width=8cm]{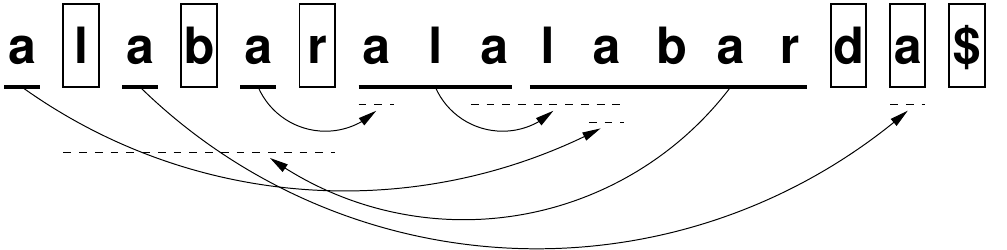}
\end{center}
\caption{The lex-parse of $S=\mathsf{alabaralalabarda\$}$,
with the same conventions of Figure~\ref{fig:lzparse}.}
\label{fig:v}
\end{figure}

\begin{example}
Figure~\ref{fig:v} gives the lex-parse of 
$S=\mathsf{a|\caja{l}|a|\caja{b}|a|\caja{r}|ala|labar|\caja{d}|\caja{a}|\caja{\$}}$,
of size $v(S)=11$. For example, the first phrase is 
$\mathsf{a}$ because the suffix $\mathsf{alabaralalabarda\$}$ shares a prefix
of length $1$ with its lexicographically preceding suffix, $\mathsf{abarda\$}$, 
and the second phrase is the explicit symbol $\mathsf{\caja{l}}$ because it
shares no prefix with its lexicographically preceding suffix, 
$\mathsf{da\$}$. Just like $r$, the value of $v$ depends on the alphabet 
ordering, for example for $S'=\mathsf{elebereleleberde\$}$ we have $v(S')=10$.
\end{example}

Measure $v$ has several interesting characteristics. First, it can be computed
in linear time via the so-called longest common prefix array \cite{KLAAP01}. 
Second, apart from $b = O(v)$ because the lex-parse is a bidirectional macro 
scheme, it holds that $v = O(r)$, because the bidirectional macro scheme 
induced by the runs of $S^{bwt}$ is also lexicographic \cite{NPO19} ($v$ similarly
subsumes other lexicographic parses like {\em lcpcomp} \cite{DFKLS17}). 
Note that,
although $r$ and $z$ are incomparable, $v$ is never asymptotically larger than
$r$. \citeN{NPO19} also connect $v$ with grammars, by showing that $v \le 
g_{rl}$, and therefore $v=O(b\log(n/b))$ and $v \le n{\cal H}_k+o(n\log\sigma)$.
It follows that $r = \Omega(v\log n)$ on binary de Bruijn sequences, where
$r=\Theta(n)$ and $v=O(n/\log n)$. They also show that $v=\Theta(\log n)$
(and thus $v=\Omega(b\log n)$) on the odd Fibonacci words and $r=O(1)$ (and
thus $c = \Omega(r\log n)$) on the even ones. On the other hand, it is unknown
if there are string families where $z=o(v)$.
 
\subsection{Compact Directed Acyclic Word Graphs: Measure $e$} \label{sec:e}

Measure $r$ exposes the regularities that appear in the suffix array of 
repetitive sequences $S$. As seen in Section~\ref{sec:stree}, the suffix array 
corresponds to the leaves of the suffix tree, where each suffix of $S$ labels
a path towards a distinct leaf. A Compact Directed Acyclic Word Graph (CDAWG) 
\cite{BBHMCE87} is obtained by merging all the identical subtrees of the suffix 
tree. The suffix trees of repetitive strings tend to have large isomorphic 
subtrees, which yields small CDAWGs. The number $e$ of nodes plus edges in the 
CDAWG of $S$, is then a 
repetitiveness measure. The CDAWG is also built in linear time \cite{BBHMCE87}.

\begin{figure}[t]
\begin{center}
\includegraphics[width=8cm]{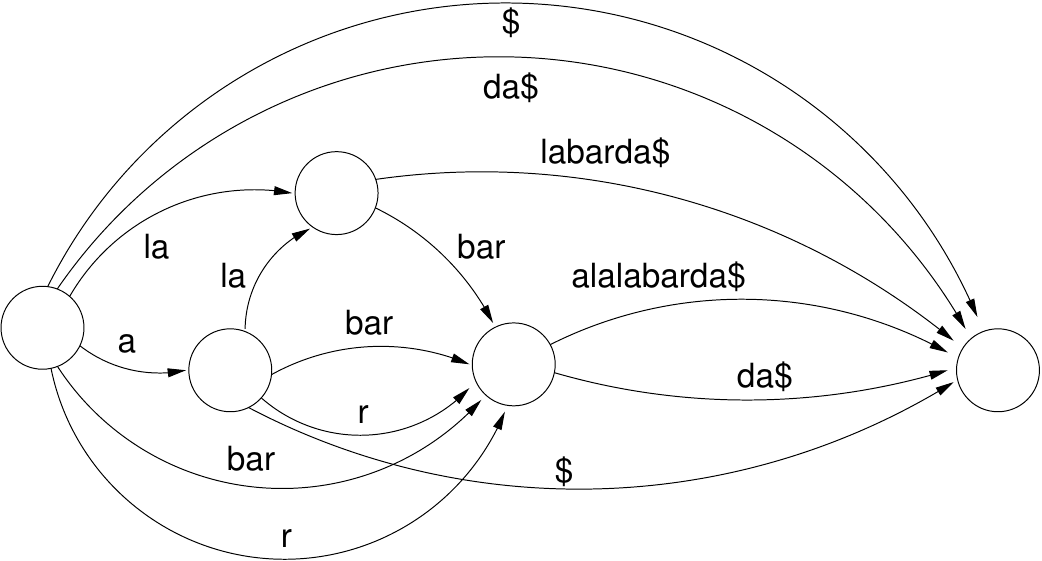}
\end{center}
\caption{The CDAWG of string $S=\mathsf{alabaralalabarda\$}$.}
\label{fig:cdawg}
\end{figure}

\begin{example}
Figure~\ref{fig:cdawg} shows the CDAWG of $S=\mathsf{alabaralalabarda\$}$, and
Figure~\ref{fig:stree} shows its suffix tree and array. The CDAWG has $5$ nodes
and $14$ edges, so $e=19$. Note, for example, that all the suffixes starting
with $\mathsf{r}$ are preceded by $\mathsf{a}$, all the suffixes starting with
$\mathsf{ar}$ are preceded by $\mathsf{b}$, and so on until $\mathsf{alabar}$.
This causes identical subtrees at the loci of all those substrings, 
$\mathsf{r}$, $\mathsf{ar}$, $\mathsf{bar}$, $\mathsf{abar}$, $\mathsf{labar}$,
and $\mathsf{alabar}$. All those loci become the single CDAWG node that is 
reachable from the root by those strings. This also relates with $r$: the loci
correspond to suffix array ranges $A[16\dd 17]$, $A[8\dd 9]$, $A[10\dd 11]$, 
$A[3\dd 4]$, $A[13\dd 14]$, and $A[5\dd 6]$. All but the last such intervals, 
consequently, fall inside BWT runs with symbols
$\mathsf{a}$, $\mathsf{b}$, $\mathsf{a}$, $\mathsf{l}$, and $\mathsf{a}$. 
There are other larger identical subtrees, like those rooted by the loci of 
$\mathsf{la}$ and $\mathsf{ala}$, corresponding to intervals $A[13\dd 15]$ and 
$A[5\dd 8]$, the first of which is within a run, 
$S^{bwt}[13\dd15]=\mathsf{aaa}$. Finally, the run 
$S^{bwt}[13\dd 17]=\mathsf{aaaaa}$ corresponds to two consecutive equal 
subtrees, with the suffix array intervals $A[13\dd 17]$ and $A[5\dd 9]$.
\end{example}

This is the weakest repetitiveness measure among those we study. It always
holds that $e = \Omega(\max(z,r))$ \cite{BCGPR15} and $e = \Omega(g)$ 
\cite{BC17}. Worse, on some string families (as simple as $\mathsf{a}^{n-1}
\mathsf{\$}$) $e$ can be $\Theta(n)$ times larger than $r$ or $z$ 
\cite{BCGPR15} and $\Theta(n/\log n)$ times larger than $g$ \cite{BC17}. 
The CDAWG is, on the other hand, well
suited for pattern searching, as we show in Section~\ref{sec:suffix-indexing}.

\subsection{String Attractors: Measure $\gamma$} \label{sec:gamma}

\citeN{KP18} proposed a new measure of repetitiveness that takes a different
approach: It is a direct measure on the string $S$ instead of the result of
a specific compression method. Their goal was to unify the existing measures 
into a cleaner and more abstract characterization of the string. An 
{\em attractor} of $S$ is a set $\Gamma$ of positions in $S$ such that any
substring $S[i\dd j]$ must have a copy including an element of $\Gamma$.
The substrings of a repetitive string should be covered with small attractors.
The measure is
then $\gamma=\gamma(S)$, the smallest size of an attractor $\Gamma$ of $S$.
This measure is obviously invariant to string reversals,
$\gamma(S)=\gamma(S^{rev})$, but it is not monotonic when we append
symbols to the string \cite{MRRRS20}.

\begin{example}
An attractor of string $S = \mathsf{alabaralalabarda\$}$
is $\Gamma = \{ 4,6,7,8,15,17 \}$. We know that this is the smallest possible
attractor, $\gamma(S)=6$,
because it coincides with the alphabet size $\sigma$, and it must obviously 
hold that $\gamma \ge \sigma$.
\end{example}

In general, it is NP-complete to find the smallest attractor size for $S$ 
\cite{KP18}, but in exchange they show that $\gamma = O(\min(b,c,z,z_{no},r,
g_{rl},g))$.\footnote{For $c$ they consider internal collage systems, recall
Section~\ref{sec:collage}.} Note that,
with current knowledge, it would be sufficient to prove that $\gamma=O(b)$, 
because $b$ asymptotically lower-bounds all those measures, as well as $v$.
Indeed, we can easily see that $\gamma \le b$: given a bidirectional macro 
scheme, take its explicit symbol positions as the attractor $\Gamma$. Every 
substring $S[i\dd j]$ not containing an explicit symbol (i.e., a position of
$\Gamma$) is inside a phrase and thus it occurs somewhere else, in particular
at $S[f(i)\dd f(j)]$. If this new substring does not contain an explicit 
position, we continue with $S[f^2(i)\dd f^2(j)]$, and so on. In a valid macro
scheme we must eventually succeed; therefore $\Gamma$ is a valid attractor.

\begin{example}
Our example attractor $\Gamma = \{ 4,6,7,8,15,17 \}$ is derived in this way 
from the bidirectional macro scheme of Figure~\ref{fig:bms}.
\end{example}

That is, $\gamma$ is a lower bound to all the other repetitiveness measures. 
Yet, we do not know if it is reachable, that is, if we can represent $S$ 
within space $O(\gamma)$. Instead, \citeN{KP18} show that
$O(\gamma\log(n/\gamma))$ space suffices not only to encode $S$ but also to
provide logarithmic-time access to any $S[i]$ 
(Section~\ref{sec:other-blocktrees}).

\citeN{CEKNP19} also show how to support indexed searches within
$O(\gamma\log(n/\gamma))$ space; see Section~\ref{sec:searchXY}. They
actually build a particular run-length grammar of that size, thus implying the
bound $g_{rl} = O(\gamma\log(n/\gamma))$. A stronger bound $g =
O(\gamma\log(n/\gamma))$ is very recent.\footnote{T. Kociumaka, personal
communication.}

A relevant step towards understanding the reachability of $\gamma$ is the
recent finding that $\gamma = o(b)$ on Thue-Morse words \cite{BFIKMN21}. This 
shows that, if $\gamma$ is to be reached, it will not be via copy-paste 
mechanisms, of which $b$ is arguably the smallest possible size.
 To definitely show that not all strings can be represented
in $O(\gamma)$ space, we should find a string family of common measure 
$\gamma$ and of size $n^{\omega(\gamma)}$. 

\subsection{String Complexity: Measure $\delta$} \label{sec:delta}

A more recent measure of repetitiveness for a string $S$, $\delta=\delta(S)$,
is built on top of the concept of {\em string complexity}, that is, the number
$S(k)$ of distinct substrings of length $k$.
\citeN{RRRS13} define $\delta = \max \{ S(k)/k, 1 \le k \le n \}$. 
It is not hard to see that $\delta(S) \le \gamma(S)$ for every string $S$
\cite{CEKNP19}: Since every substring of length $k$ in $S$ has a copy 
including some of its $\gamma$ attractor elements, there can be only $k\gamma$
distinct substrings, that is, $S(k) \le k\gamma$ for all $k$.
\citeN{CEKNP19} also show how $\delta$ can be computed in linear time.%
\footnote{A lightweight version is obtained by (1) computing the $LCP$
array, (2) initialize $S[k] \leftarrow 0$ for all $k$, (3) increment $S[LCP[i]]$
for all $i$, (4) accumulate $S[k] \leftarrow S[k]-1+S[k-1]$
for all $k$, and finally (5) $\delta \leftarrow \max(S[k]/k)$ over all $k$.} 
Further, $\delta$ is clearly monotonic and invariant upon reversals.

\begin{example}
For our string $S = \mathsf{alabaralalabarda\$}$ we have
$S(1)=6$, $S(2)=9$, $S(3)=10$, $S(4)=S(5)=S(6)=11$, and $S(k)=17-k+1$ for
$k > 6$ (i.e., all the substrings of length over 6 are different); therefore 
$\delta(S)=6$.
\end{example}

\citeN{KNP20,KNP21} show that, for every $\delta$, there are string families whose
members all have measure $\delta$ and where
$\gamma = \Omega(\delta \log (n/\delta))$. Although $\delta$ is then strictly
stronger than $\gamma$ as a compressibility measure, they also show that it
is possible not only to represent $S$ within $O(\delta\log(n/\delta))$ space,
but also to efficiently access any symbol $S[i]$ (see
Section~\ref{sec:other-blocktrees}) and support indexed searches on $S$
within that space. Indeed, this
space is optimal as a function of $\delta$: for every $2 \le \delta \le
n$ there are string families that need $\Omega(\delta\log(n/\delta))$ space
to be represented.
This means that we know that $o(\delta\log n)$ space is unreachable in general,
whereas it is unknown if $o(\gamma\log n)$ space can always be reached.

\citeN{RRRS13} prove that $z = O(\delta\log(n/\delta))$, and it can also be
proved that $g_{rl} = O(\delta\log(n/\delta))$ by building a run-length
grammar of that size \cite{KNP21}.
The same cannot be said about $g$: \citeN{KNP21}
prove that for every $n$ and $2\le\delta\le n$ there are string families where
$g = \Omega(\delta \log^2 (n/\delta) / \log\log (n/\delta))$.
This establishes another separation between $g_{rl}$ and $g$.
Recently, the only upper bound on $r$ in terms of another repetitiveness
measure was obtained: $r = O(\delta \log(n/\delta)\log\delta)$ \cite{KK19};
this also shows that $r(S)$ and $r(S^{rev})$ can differ only by a factor up
to $O(\log^2 n)$, because $\delta$ is invariant upon reversals. Even more
recently, \citeN{KS22} proved that $z_e = O(\delta\log^2(n/\delta))$.

\subsection{L-systems and NU-systems: Measures $\ell$ and $\nu$}

\citeN{NU21} recently presented new reachable measures of repetitiveness.
The first one, $\ell$, is the size of the smallest L-system generating a 
string. L-systems consist of taking a small morphism (i.e., a mapping from 
strings to strings) and iterating it from some initial string, for a given 
number of rounds; the output is a prefix of the final string with possibly a 
final character mapping. L-systems are akin to deterministic Lindenmayer 
systems \cite{Lin68a,Lin68b}, which are like context-free grammars without 
terminals.

\begin{example}
The Thue-Morse family of strings $T_k$ is obtained as $T_k=\phi^k(0)$, 
where the morphism $\phi$ maps $\phi(0) = 01$ and $\phi(1) = 10$. That is,
$T_0 = 0$, $T_1 = 01$, $T_2 = 0110$, $T_3 = 01101001$, and so on. This 
family has $\gamma=O(1)$ and $b=\Theta(\log n)$ \cite{BFIKMN21}. An L-system
generating $T_k$ has constant size, thus it holds $\ell=O(1)$ as well. This
is then a reachable representation of size $o(b)$ generating a string family.
\end{example}

Iterating morphisms then provides a different method for generating repetitive 
strings (e.g., they have measure $\gamma = O(\log n)$ under mild asumptions 
\cite{Sha20}). As seen in the example, this mechanism can provide reachable
representations that outperform any copy-paste method (i.e., macro schemes).

Indeed, \citeN{NU22} exhibit a string family where 
$\delta=\Theta(\ell\sqrt{n})$, that is, $\ell$ can be smaller than $\delta$ 
by a large factor. This is in contrast with the fact that most other measures
are within a polylogarithmic factor of each other. Note that, since $\ell$ is 
reachable, it also holds $\delta=o(\ell)$ on other string families, thus both 
measures are incomparable. This incomparability is wider, as there are
families where $g_{rl} = o(\ell)$, $z_e=o(\ell)$, and $r = o(\ell)$ \cite{NU22}.
This suggests that the regularities exploited by $\ell$ are 
orthogonal to those of copy-paste mechanisms. The only known upper bound for
$\ell$ is $O(g)$ \cite{NU21}.

\citeN{NU21} also define another measure, $\nu$ (``nu''), as the size of 
the smallest so-called NU-system generating a given string. NU-systems are
grammar-like mechanisms that combine general copy-paste mechanisms like macro 
schemes and compositions of morphisms (i.e., L-systems). 

NU-systems include rules similar to those of L-systems, but they also allow
extracting substrings from the expansions of other nonterminals. In this sense,
they are similar to collage systems, except that nonterminals are not forced 
to reference only earlier ones. This is shown to be sufficient to emulate 
bidirectional macro schemes, and thus $\nu = O(b)$ always holds, apart from
$\nu = O(\ell)$. Both kinds of rules are combined in a consistent and powerful 
manner so that it is computable to produce the generated string. The result is
more than the union of the parts: there are string families where
$\nu=o(\ell)$ and $\nu=o(b)$ \cite{NU22}.

These reachable measures
challenge the idea of $\delta$ as a lower bound on repetitive strings, if one 
accepts iteration of morphisms as a valid mechanisms for representing them. It 
is also unknown whether $\nu = O(\gamma)$ always holds; in general we lack 
mechanisms to lower-bound $\nu$. 

\subsection{Relations}

\begin{figure}[t]
\begin{center}
\includegraphics[width=\textwidth]{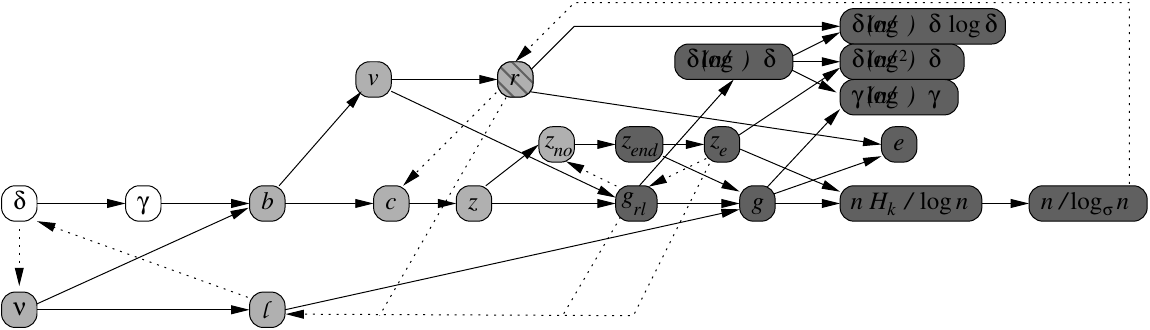}
\end{center}
\caption{Relations between the compressibility measures. A solid arrow from 
$X$ to $Y$ means that $X=O(Y)$ for all string families. For all solid and dotted
arrows, there are string families where $X=o(Y)$, with the
exceptions of $c \rightarrow z$ and $z_{no} \rightarrow z_{end} \rightarrow
z_e$. Grayed measures $X$ mean that we can encode
every string in $O(X)$ space; darker gray means that we can also provide
efficient access and indexed searches within $O(X)$ space; for $r$ we can only
provide indexed searches.}
\label{fig:measures}
\end{figure}

Figure~\ref{fig:measures} summarizes what is known about the repetitiveness
measures we have covered. Those in gray are reachable, and those in dark gray
support efficient access and indexing\footnote{Efficient access usually means
in $O(\log n)$ time per symbol, though this is stretched up to $O(\log^5 n)$
time for $z_{end}$ and $z_e$.}, as we show in the next sections. An
intriguing case is $r$, which allows for efficient indexing but not access, 
as far as we know.

Note that we do not know if $\gamma$ should be grayed or not, whereas we do
know that $\delta$ must not be grayed. The smallest grayed measure is $\nu$,
which is, by definition, the best space we can obtain via copying substrings 
and iterating morphisms.

As explained, $\delta\log(n/\delta)$ is a reachable measure
that is asymptotically optimal as a function of $\delta$.
That is, if we
separate the set of all strings into subsets $\mathcal{S}_\delta$ where the
strings have measure $\delta$, then inside each set there are families that
need $\Theta(\delta\log(n/\delta))$ space to be represented. The measure
$\delta$ is then optimal in that coarse sense. Still, we know that $b$ and
$\nu$ are string-wise better than $\delta\log(n/\delta)$ and sometimes 
asymptotically smaller; therefore they are more refined measures.

In this line, the definitive question is: {\em What is the smallest reachable 
and computable measure of repetitiveness?} Just like Shannon's entropy is a
lower bound when we decide to exploit only frequencies, bidirectional macro
schemes are a lower bound when we decide to exploit only string copies, as are
NU-systems when we include string morphisms. Still, there may be some useful 
compressibility measure between $\nu$ and the bottom 
line of the (uncomputable) Kolmogorov complexity. 

Other equally fascinating questions can be asked about accessing and indexing
strings: {\em What is the smallest reachable measure under which we can access
and/or index the strings efficiently?} Right now, $O(z_{end})$ is the best known
limit for efficient access; it is unknown if one can access $S[i]$ efficiently
within $O(z_{no})$ or $O(r)$ space. Indexing can also be supported within
$O(z_{end})$ space, but also in $O(r)$; we do not know if this is possible
within $O(z_{no})$ or $O(v)$ space. We explore this in the upcoming sections.

\paragraph*{In practice}

Some direct \cite{BCGPR15,GNP19,NPO19,RCNF20} and less 
direct \cite{MNSV09,KN13,CFMPN16,BCGPR17} experiments suggest that in typical 
repetitive texts it holds that $b < z \approx v < z_e < g < r < e$, where ``$<$'' 
denotes a clear difference in magnitude.

Since those
experiments have been made on different texts and it is hard to combine them,
and because no experiments on $\delta$ have been published, 
Table~\ref{tab:measures} compares some measures that can be computed in
polynomial time on a sample of repetitive collections obtained from the Repetitive
Corpus of Pizza\&Chili\footnote{\tt http://pizzachili.dcc.uchile.cl/repcorpus/real}. 
We include an upper bound on $g$ obtained by a heuristically balanced RePair algorithm,%
\footnote{{\tt https://www.dcc.uchile.cl/gnavarro/software/repair.tgz}, directory {\tt bal/}. To further reduce the grammar, we removed rules defining nonterminals that were
used only once.} which consistently outperforms other grammar-based compressors,
including those that offer theoretical guarantees of approximation to $g$.
We also build the CDAWG on the DNA collections, where the
code we have allows it\footnote{{\tt https://github.com/mathieuraffinot/locate-cdawg}, which
works only on the alphabet $\Sigma = \{ \mathsf{A}, \mathsf{C}, \mathsf{G}, \mathsf{T}\}$. For
\textsf{escherichia} we converted the other symbols to \textsf{A}, and verified that this would
have a negligible impact on the other measures.}.
The table suggests $z \approx v \approx 1.5{-}2.5\cdot\delta$, 
$g \approx 3{-}6\cdot\delta$, 
$r \approx 7{-}11\cdot\delta$, and
$e \approx 32{-}35\cdot\delta$. 

\begin{table}[t]
\begin{center}
\scriptsize
\begin{tabular}{l@{~}|@{~}r@{~}|@{~}r@{~}|@{~}r@{~}|@{~}r@{~}|@{~}r@{~}|@{~}r@{~}|@{~}r}
File			& \multicolumn{1}{@{~}c|@{~}}{$n$}	& \multicolumn{1}{c|@{~}}{$\lfloor\delta\rfloor$}
& \multicolumn{1}{c|@{~}}{$z$} 	& \multicolumn{1}{c|@{~}}{$v$}	& \multicolumn{1}{c|@{~}}{$g$}	& \multicolumn{1}{c|@{~}}{$r$} & \multicolumn{1}{c}{$e$} \\
\hline
\textsf{cere}		& 461,286,644	& 1,003,280	& 1,700,630	& 1,649,448	& 4,069,452	& 11,574,640	& 35,760,304 \\
\textsf{escherichia}	& 112,689,515 	& 1,337,977	& 2,078,512	& 2,014,012	& 4,342,874	& 15,044,487	& 43,356,169 \\
\hline
\textsf{einstein}	& 467,626,544 	&    42,884	&    89,467	&    97,442	&   212,902	&    290,238	& \\
\textsf{worldleaders}	&  46,968,181 	&    68,651	&   175,740	&   179,696	&   399,667	&    573,487	& \\
\hline
\textsf{coreutils}   	& 205,281,778 	&   636,101	& 1,446,468	& 1,439,918	& 2,409,429	&  4,684,460	& \\
\textsf{kernel}   	& 257,961,616 	&   405,643	&   793,915	&   794,058	& 1,374,651	&  2,791,367	& \\
\end{tabular}
\end{center}
\caption{Several repetitiveness measures computed on a sample of the Pizza\&Chili 
repetitive corpus. We chose two DNA, two natural language, and two source code 
files.}
\label{tab:measures}
\end{table}

\section{Accessing the Compressed Text and Computing Fingerprints}
\label{sec:access}

The first step beyond mere compression, and towards compressed indexing, is to
provide direct access to the compressed string without having to fully 
decompress it. We wish to extract arbitrary substrings $S[i\dd j]$ in a time
that depends on $n$ only polylogarithmically. Further, some indexes also need 
to efficiently compute Karp-Rabin fingerprints (Section~\ref{sec:kr}) of 
arbitrary substrings. 

In this section
we cover the techniques and data structures that are used to provide these 
functionalities, depending on the underlying compression method. In all cases,
any substring $S[i\dd j]$ can be computed in time $O(j-i+\log n)$ or less,
whereas the fingerprint of any $S[i\dd j]$ can be computed in time $O(\log n)$
or less. Section~\ref{sec:access-grammars} shows how this is done in $O(g)$ and
even $O(g_{rl})$ space by enriching (run-length) context-free grammars, whereas
Section~\ref{sec:blocktrees} shows how to do this in space $O(z\log(n/z))$, and
even $O(\delta\log(n/\delta))$, by using so-called block trees. Some indexes 
require more restricted forms of access, which those data structures can 
provide in less time. Section~\ref{sec:bookmarking} shows another speedup
technique called bookmarking.

The experiments \cite{B+19} show that practical access data structures built 
on block trees take about the same space as those built on balanced grammars 
(created with RePair \cite{LM00}), but block trees grow faster as soon as the 
repetitiveness decreases. On the other hand, access on block trees is over
an order of magnitude faster than on grammars.

A recent technique we are not describing in detail \cite{KS22} provides 
$O(\log^5 n)$-time access to strings encoded in LZ-End format (i.e., for
measures $z_{end}$ and $z_e$). While it works with high probability only, 
there are on average only $O(1)$ positions that cannot be
retrieved. This can be made worst-case by using a Las Vegas construction,
which iterates until some $O(1)$-bound on the number of failing positions is
reached (in most cases the first attempt will succeed). Those failing
positions are then handled via a perfect hash function, for example.

\subsection{Enhanced Grammars} \label{sec:access-grammars}

If the compressed string is represented with a context-free grammar of 
size $g$ or a run-length grammar of size $g_{rl}$, we can enrich the 
nonterminals with information associated with the length of the string they 
expand to, so as to provide efficient access within space $O(g)$ or $O(g_{rl})$,
respectively. 

For a simple start, let $A \rightarrow X_1 \cdots X_k$. Then, we store 
$\ell_0=0$, $\ell_1=\ell_0+|X_1|$, $\ell_2=\ell_1+|X_2|$, $\ldots$, $\ell_k = 
\ell_{k-1}+|X_k|$ associated with $A$. To extract the $i$th symbol of $exp(A)$,
we look for the predecessor of $i$ in those values, finding $j$ such that 
$\ell_{j-1} < i \le \ell_j$, and then seek to obtain the $i'$th symbol of 
$X_j$, with $i' = i-\ell_{j-1}$. Since predecessors can be computed in time
$O(\log\log_w n)$ \cite{BN14,NRacmcs20}, on a grammar of height $h$ we can extract
any $S[i]$ in time $O(h\log\log_w n)$, which is $O(\log n\log\log_w n)$ if the 
grammar is balanced. If the right-hand sides of the rules are of constant length, 
then the predecessors take constant time and the extraction time drops to
$O(\log n)$, as with the simple method described in Section~\ref{sec:grammar}.

\citeN{BLRSRW15} showed how this simple idea can be extended to extract any
$S[i]$ in time $O(\log n)$ from arbitrary grammars, not necessarily balanced.
They extract a {\em heavy path} \cite{ST83} from the parse tree of $S$. 
A heavy path starts
at the root $A \rightarrow X_1\cdots X_k$ and continues by the child $X_j$ with
the longest expansion, that is, with maximum $|X_j|$ (breaking ties in some
deterministic way), until reaching a leaf.
We store the heavy path separately and remove all its nodes and edges from the 
parse tree, which gets disconnected and becomes a forest. We then repeat the 
process from each root of the forest until all the nodes are in the extracted 
heavy paths.

Consider the path going through a node labeled $B$ in the parse tree, whose
last element is the terminal $exp(B)[t_B]$. We associate with $B$ its start and 
end values relative to $t_B$, $s_B = 1-t_B$ and $e_B = |B|-t_B$, respectively. 
Note that these values will be the same wherever $B$ appears in the parse tree, 
because the heavy path starting from $B$ will be identical. Further, if $C$
follows $B$ in the heavy path, then $exp(C)[t_C]$ is the same symbol 
$exp(B)[t_B]$. For a heavy path rooted at $A$, the values $s_B$ of the nodes 
we traverse downwards to the leaf, then the zero, and then the values $e_B$ of 
the nodes we traverse upwards to $A$ again, form an increasing sequence of 
positions, $P_A$. The search for $S[i]$ then proceeds as follows. We search 
for the predecessor of $i-t_A$ in the sequence $P_A$ associated with the root 
symbol $A$. Say that $B$ is followed by $C$ downwards in the path and their 
starting positions are $s_B \le i-t_A < s_C$, or their ending positions are 
$e_C < i-t_A \le e_B$. Then the search for $S[i]$ must continue as the search 
for $i' = i - t_A + t_B$  inside $B$, because $i$ is inside $exp(B)$ but not 
inside $exp(C)$. With another predecessor search for 
$i'$ on the starting positions $\ell_j$ of the children of $B$, we find the 
child $B_j$ by which our search continues, with $i'' = i' - \ell_{j-1}$. Note 
that $B_j$ is the root of another heavy path, and therefore we can proceed 
recursively.

\begin{figure}[t]
\begin{center}
\includegraphics[width=12cm]{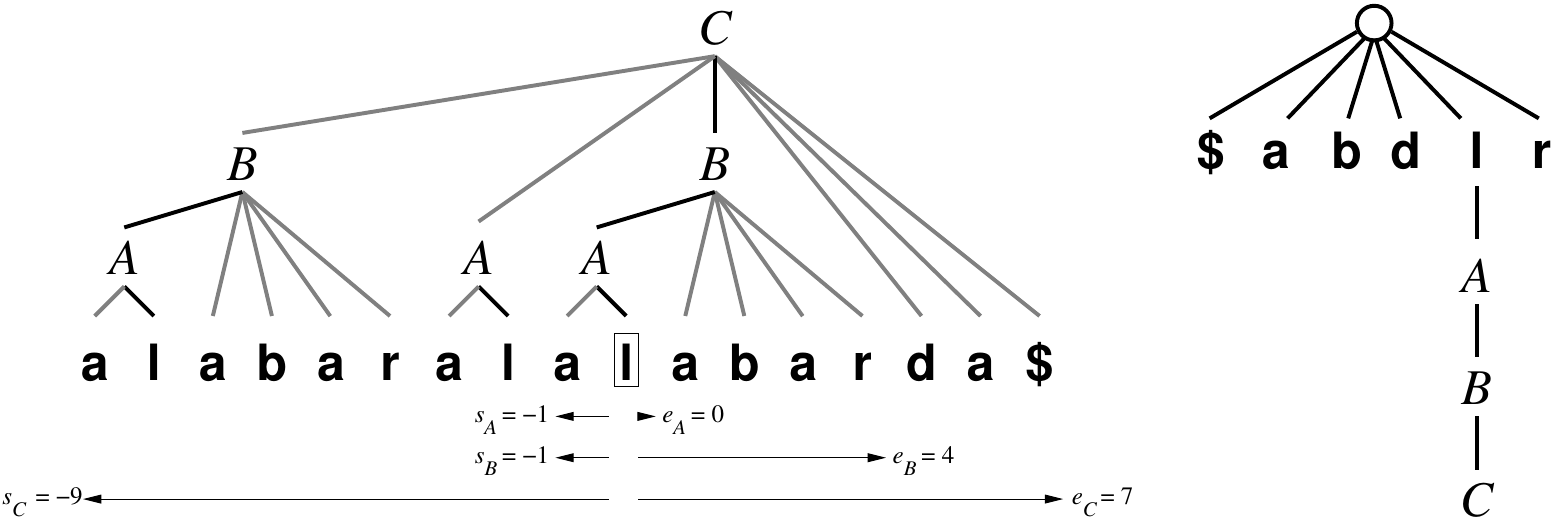}
\end{center}
\caption{On the left, the heavy path decomposition of the parse tree of 
Figure~\ref{fig:g}. Heavy edges are in black and light edges are in gray.
For the heavy path that starts on the root, we box the last element and show
how the values $s_X$ and $e_X$ of the intermediate nodes are computed.
On the right, the trie storing all the heavy paths.}
\label{fig:gaccess}
\end{figure}

\begin{example}
Figure~\ref{fig:gaccess} (left) shows an example for the grammar we used in 
Figure~\ref{fig:g} on our string $S=\mathsf{alabaralalabarda\$}$. The first
heavy path, extracted from the root, is $C \rightarrow B \rightarrow A
\rightarrow \mathsf{l}$ (breaking ties arbitrarily). From the other $B$ of
the parse tree, which becomes a tree root after we remove the edges of
the first heavy path, we extract another heavy path: $B \rightarrow A 
\rightarrow \mathsf{l}$. The remaining $A$ produces the final heavy path,
$A \rightarrow \mathsf{l}$. All the other paths have only one node.
Note that $t_C = 10$ and $t_B = t_A = 2$, that is, $exp(C)[10]=exp(B)[2]
=exp(A)[2]=\mathsf{l}$ is the last element in the heavy path. 
Therefore, $s_C = -9$, $e_C = 7$,
$s_B = -1$, $e_B = 4$, $s_A = -1$, $e_A = 0$. The $\ell$ values for $C$
are $0, 6, 8, 14, 15, 16, 17$. To find $S[12]$, we determine that $8 < 12 \le
14$, thus we have to descend by $B$, which follows the heavy path. We then
search for $12-t_C = 2$ in the sequence $s_C,s_B,s_A,0,e_A,e_B,e_C =
-9, -1, -1, 0, 0, 4, 7$ to find
$0 < 2 \le 4$, meaning that we fall between $e_A=0$ and $e_B=4$. We thus
follow the search from $B$, for the new position $12-t_C+t_B = 4$. Since the
$\ell$ values associated with $B$ are $0,2,3,4,5,6$, we descend by a light
edge towards the $\mathsf{b}$, which is $S[12]$.
\end{example}

The important property is that, if $X_j$ follows $A$ in the heavy path, then 
all the other children $X_{j'}$ of $A$ satisfy $|X_{j'}| \le |A|/2$, because
otherwise $X_{j'}$ would have followed $A$ in the heavy path. Therefore, every
time we traverse a light edge to
switch to another heavy path, the length of the expansion of the
nonterminal is halved. As a consequence, we cannot switch more than $\log n$
times to another heavy path in our traversal from the root to the leaf that
holds $S[i]$. Since we perform two predecessor searches to find 
the next heavy path, the total extraction cost is $O(\log n\log\log_w n)$, even 
if the grammar is unbalanced.

\citeN{BLRSRW15} remove the $O(\log\log_w n)$ factor by using a different
predecessor search data structure that, if $i$ falls between positions 
$p_{j-1}$ and $p_j$ inside a universe of $u$ positions, then the search takes 
time $O(\log (u/(p_j-p_{j-1})))$. This makes the successive searches on heavy
paths and children telescope to $O(\log n)$.

The other problem is that the parse tree has $\Theta(n)$ nodes, and thus we
cannot afford storing all the heavy paths. Fortunately, this is not necessary:
If $X_j$ is the child of $A$ with the largest $|X_j|$, then $X_j$ will follow 
$A$ in every heavy path where $A$ appears. We can then store all the heavy 
paths in a trie where $X_j$ is the {\em parent} of $A$. Heavy paths are then 
read as upward paths in this trie, which has exactly one node per nonterminal
and per terminal, the latter being children of the trie root. 
The trie then represents all the heavy paths within $O(g)$ space.
\citeN{BLRSRW15} show how an $O(g)$-space data structure on the trie provides 
the desired predecessor searches on upward trie paths.

\begin{example}
Figure~\ref{fig:gaccess} (right) shows the trie associated with our example
parse tree. Every time $B$ appears in the parse tree, the heavy path
continues by $A$, so $A$ is the parent of $B$ in the trie. 
\end{example}

\subsubsection{Extracting substrings}
\label{sec:extract-grammar}

\citeN{BLRSRW15} also show how to extract $S[i\dd j]$
in time $O(j-i+\log n)$. They find the path towards $i$ and the path towards $j$.
These coincide up to some node in the parse tree, from which they descend by
different children. From there, all the subtrees to the right of the 
path towards $i$ (from the deepest to the shallowest), and then all the 
subtrees to the left of the path towards $j$ (from the shallowest to the
deepest), are fully traversed in order to obtain the $j-i+1$ symbols of 
$S[i\dd j]$ in optimal time (because they output all the leaves of the traversed 
trees, and internal nodes have at least two children).

With a bit more of sophistication, \citeN{BPT15} obtain RAM-optimal time on the 
substring length, $O((j-i)/\log_\sigma n + \log n)$, among other tradeoffs. We 
note that the $O(\log n)$ additive overhead is almost optimal: any structure
using $g^{O(1)}$ space requires $\Omega(\log^{1-\epsilon} n)$ time to access a 
symbol, for any constant $\epsilon>0$ \cite{VY13}.

\subsubsection{Karp-Rabin fingerprints} \label{sec:kr-grammar}

An easy way to obtain the Karp-Rabin fingerprint of any $S[i\dd j]$ is to
obtain $\kappa(S[1\dd i-1])$ and $\kappa(S[1\dd j])$, and then operate them
as sketched in Section~\ref{sec:kr}. To 
compute the fingerprint of a prefix of $S$, \citeN{BGCSVV17} store for each 
$A \rightarrow X_1 \cdots X_k$ the fingerprints $\kappa_1 = \kappa(exp(X_1))$,
$\kappa_2 = \kappa(exp(X_1)\cdot exp(X_2))$, $\ldots$, $\kappa_k = 
\kappa(exp(X_1) \cdots exp(X_k))$. Further, for each node $B$ in a heavy path 
ending at a leaf $exp(B)[t_B]$, they store $\kappa(exp(B)[1\dd t_B-1])$.
Thus, if we have to leave at $B$ a heavy path that starts in $A$, the 
fingerprint of the prefix of $exp(A)$ that precedes $exp(B)$ is obtained by 
combining $\kappa(exp(A)[1\dd t_A-1])$ and $\kappa(exp(B)[1\dd t_B-1])$.
In our path towards extracting $S[i]$, we can 
then compose the fingerprints so as to obtain $\kappa(S[1\dd i-1])$ at the 
same time. Any fingerprint
$\kappa(S[i\dd j])$ can therefore be computed in $O(\log n)$ time.

\begin{example}
In the same example of Figure~\ref{fig:gaccess}, say we want to compute
$\kappa(S[1\dd 11])$. We start with the heavy path that starts at $C$, which
we leave at $B$. For the heavy path, we have precomputed
$\kappa(exp(C)[1\dd t_C-1])
= \kappa(\mathsf{alabarala})$ for $C$ and $\kappa(exp(B)[1\dd t_B-1]) =
\kappa(\mathsf{a})$ for $B$. By operating them, we obtain the fingerprint of
the prefix of $exp(C)$ that precedes $exp(B)$, $\kappa(\mathsf{alabaral})$.
We now descend by the second child of $B$. We have also precomputed the 
fingerprints of the prefixes of $exp(B)$ corresponding to its children,
in particular the first one, $\kappa(exp(A))=\kappa(\mathsf{al})$. By
composing both fingerprints, we have $\kappa(\mathsf{alabaralal})$ as desired.
\end{example}

\subsubsection{Extracting rule prefixes and suffixes in real time}
\label{sec:extract-prefixes}

The typical search algorithm of compressed indexes 
(Section~\ref{sec:parsed-indexing}) does not need to extract 
arbitrary substrings, but only to expand prefixes or suffixes of nonterminals.
\citeN{GKPS05} showed how one can extract prefixes or suffixes of any
$exp(A)$ in real time, that is, $O(1)$ per additional symbol. They build a 
trie similar to that used to store all the heavy paths, but this time they
store leftmost paths (for prefixes) or rightmost paths (for suffixes). That
is, if $A \rightarrow X_1 \cdots X_k$, then $X_1$ is the parent of $A$ in
the trie of leftmost paths and $X_k$ is the parent of $A$ in the trie of 
rightmost paths.

Let us consider leftmost paths; rightmost paths are analogous. To extract the 
first symbol of $exp(A)$, we go to the root of the trie, descend to the child
in the path to node $A$, and output its corresponding terminal, $a$. This takes
constant time with level ancestor queries \cite{BF04}. Let $B \rightarrow a B_2
\cdots B_s$ be the child of $a$ in the path to $A$ (again, found with level
ancestor queries from $A$).
The next symbols are then extracted recursively from $B_2, \ldots, B_s$. Once
those are exhausted, we continue with the child of $B$ in the path to $A$,
$C \rightarrow B C_2 \cdots C_t$, and extract $C_2,\ldots,C_t$, and so on,
until we extract all the desired characters.

\subsubsection{Run-length grammars} \label{sec:runlenght-access}

\citeN{CEKNP19} (App.~A) showed how the results above can be obtained on
run-length context-free grammars as well, relatively easily, by regarding
the rule $A \rightarrow X^t$ as $A \rightarrow X \cdots X$ and managing to
use only $O(1)$ words of precomputed data in order to simulate the desired
operations.

\subsubsection{All context-free grammars can be balanced}

Recently, \citeN{GJL19} proved that every context-free grammar of size
$g$ can be converted into another of size $O(g)$, right-hand sides of size 2, and
height $O(\log n)$. While the conversion seems nontrivial at first sight, once 
it is carried out we need only very simple information associated with nonterminals to
extract any $S[i\dd j]$ in time $O(j-i+\log n)$ and to
compute any fingerprint $\kappa(S[i\dd j])$ in time $O(\log n)$. It is not 
known, however, if run-length grammars can be balanced in the same way.

\subsection{Block Trees and Variants} \label{sec:blocktrees}

Block trees \cite{BGGKOPT15} are in principle built knowing the size $z$ of
the Lempel-Ziv parse of $S[1\dd n]$. Built with a parameter $\tau$, they 
provide a way to access any $S[i]$ in time $O(\log_\tau(n/z))$ with a data 
structure of size $O(z\tau\log_\tau(n/z))$. For example, with $\tau=O(1)$, 
the time is $O(\log(n/z))$ and the space is $O(z\log(n/z))$. Recall that 
several heuristics build grammars of size $g = O(z\log(n/z))$, and thus block 
trees are not asymptotically smaller than structures based on grammars, but 
they can be asymptotically faster.

The block tree is of height $\log_\tau(n/z)$. The root has $z$ children, $u_1,
\ldots,u_z$, which logically divide $S$ into blocks of length $n/z$, $S =
S_{u_1} \cdots S_{u_z}$. Each such node $v=u_j$ has $\tau$ children, 
$v_1,\ldots,v_\tau$,
which divide its block $S_v$ into equal parts, $S_v = S_{v_1}\cdots S_{v_\tau}$.
The nodes $v_i$ have, in turn, $\tau$ children that subdivide their block, and 
so on. After slightly less than $\log_\tau(n/z)$ levels, the blocks are of 
length $\log_\sigma n$, and can be stored explicitly using $\log n$ bits, that 
is, in constant space.

Some of the nodes $v$ can be removed because their block $S_v$ appears earlier
in $S$. The precise mechanism is as follows: every consecutive pair of nodes 
$v_1,v_2$ where the concatenation $S_{v_1} \cdot S_{v_2}$ does not appear 
earlier is {\em marked} (that is, we mark $v_1$ and $v_2$). After this, 
every unmarked node $v$ has an earlier occurrence, so instead of creating 
its $\tau$ children, we replace $v$ by a leftward pointer to the first 
occurrence of $S_v$ in $S$. This first occurrence spans in general two
consecutive nodes $v_1,v_2$ at the same level of $v$, and these exist and are 
marked by construction. We then make $v$ a leaf pointing to $v_1,v_2$, also 
recording the offset where $S_v$ occurs inside $S_{v_1} \cdot S_{v_2}$.

To extract $S[i]$, we determine the top-level node $v=u_j$ where $i$ falls and 
then extract its corresponding symbol. In general, to extract $S_v[i]$, 
there are three cases. (1) If $S_v$ is stored explicitly (i.e., $v$ is a node 
in the last level), we access $S_v[i]$ directly. (2) If $v$ has $\tau$ children,
we determine the corresponding child $v'$ of $v$ and the corresponding offset 
$i'$ inside $S_{v'}$, and descend to the next level looking for $S_{v'}[i']$.
(3) If $v$ points to a pair of nodes $v_1,v_2$ to the left, at the same level,
then $S_v$ occurs inside $S_{v_1} \cdot S_{v_2}$. With the offset information, 
we translate the query $S_v[i]$ into a query inside $S_{v_1}$ or inside
$S_{v_2}$. Since nodes $v_1$ and $v_2$ are marked, they have children, so we
are now in case (2) and can descend to the next level. Overall, we do $O(1)$ 
work per level of the block tree, for
a total access time of $O(\log_\tau(n/z))$.

To see that the space of this structure is $O(z\tau\log_\tau(n/z))$, it suffices
to show that there are $O(z)$ marked nodes per level: we charge $O(\tau)$ space
to the marked nodes in a level to account for the space of all the nodes in the
next level.
Note that, in a given level, there are only $z$ blocks containing Lempel-Ziv
phrase boundaries. Every pair of nodes $v_1,v_2$ without phrase boundaries in
$S_{v_1}\cdot S_{v_2}$ has an earlier occurrence because it is inside a 
Lempel-Ziv phrase. Thus, a node $v$ containing a phrase boundary in $S_v$ may
be marked and force its preceding and following nodes to be marked as well, 
but all the other nodes are unmarked. 
In conclusion, there can be at most $3z$ marked nodes per level.

Still, the construction is conservative, possibly preserving internal nodes 
$v$ such that $S_v$ occurs earlier, and no other node points inside $S_v$ nor 
inside 
the block of a descendant of $v$. Such nodes are identified and converted into
leaves with a final postorder traversal \cite{B+19}.

\begin{figure}[t]
\begin{center}
\includegraphics[width=7cm]{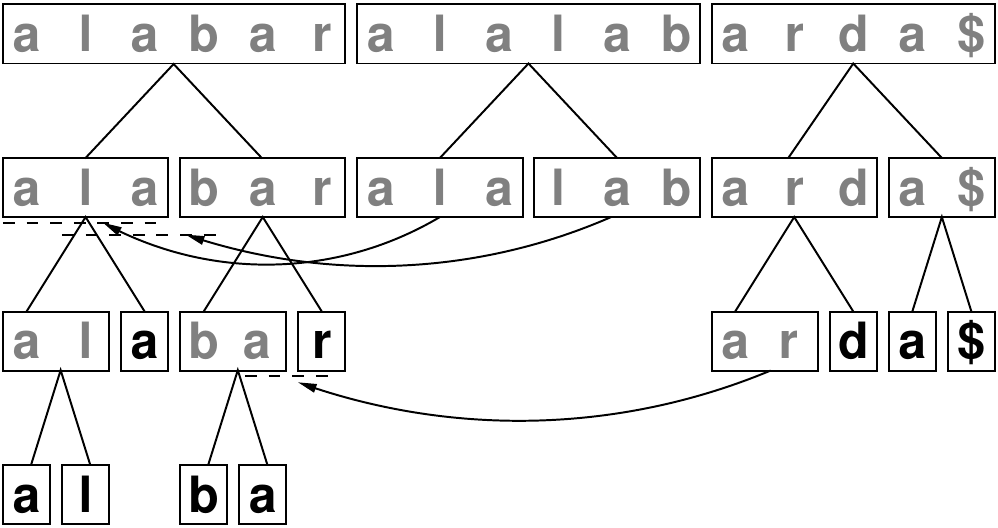}
\end{center}
\caption{A block tree for $S=\mathsf{alabaralalabarda\$}$, starting with $3$
nodes, with arity $\tau=2$, and stopping on substrings of length $1$. Grayed 
characters are conceptual and not stored; only the black ones and the pointer
structure of the tree are represented. Horizontal pointers are drawn with
curved arrows and the source areas are shown with dashed lines.}
\label{fig:bt}
\end{figure}

\begin{example}
Figure~\ref{fig:bt} shows a block tree for $S=\mathsf{alabaralalabarda\$}$
(we should start with $z(S)=11$ blocks, but $3$ is better to exemplify).
To access $S[11]$ we descend by the second block ($S_{u_2}=\mathsf{alalab}$), 
accessing $S_{u_2}[5]$. Since the block has children, we descend by the second 
($\mathsf{lab}$), aiming for its second position. But this block has no 
children because its content is replaced by an earlier occurrence of 
$\mathsf{lab}$. The quest for the second position of $\mathsf{lab}$ then
becomes the quest for the third position of $\mathsf{ala}$, within the first
block of the second level. Since this block has children, we descend to its
second child ($\mathsf{a}$), aiming for its first position. This block is
explicit, so we obtain $S[11]=\mathsf{a}$.
\end{example}

\subsubsection{Extracting substrings} \label{sec:extract-blocktree}

By storing the first and last $\log_\sigma n$ symbols of every block $S_v$,
a chunk of $S$ of that length can also be extracted in time 
$O(\log_\tau(n/z))$: we traverse the tree as for a single symbol until
the paths for the distinct symbols of the chunk diverge. At this point, the
chunk spans more than one block, and thus its content can be assembled from 
the prefixes and suffixes stored for the involved blocks.
Therefore, we can extract any 
$S[i\dd j]$ by chunks of $\log_\sigma n$ symbols, in time 
$O((1+(j-i)/\log_\sigma n)\log_\tau(n/z))$.

\subsubsection{Karp-Rabin fingerprints} \label{sec:kr-blocktree}

\citeN{NP18} show, on a slight block tree variant, that fingerprints of any
substring $S[i\dd j]$ can be computed in time $O(\log_\tau(n/z))$ by storing
some precomputed fingerprints: (1) for every top-level node $u_j$,
$\kappa(S_{u_1}\cdots S_{u_j})$; (2) for every internal node $v$ with children 
$v_1,\ldots, v_\tau$, $\kappa(S_{v_1}\cdots S_{v_j})$ for all $j$; and (3) for 
every leaf $v$ pointing leftwards to the occurrence of $S_v$ inside 
$S_{v_1}\cdot S_{v_2}$, $\kappa (S_v \cap S_{v_1})$. Then, by using the
composition operations of Section~\ref{sec:kr}, the computation of 
any prefix fingerprint $\kappa(S[1\dd i])$ is translated into the computation 
of some $\kappa(S_{u_j}[1\dd i'])$, and the computation of any
$\kappa(S_v[1\dd i'])$ is translated into the computation of some 
$\kappa(S_{v'}[1\dd i''])$ at the same level (at most once per level) or 
at the next level.

\subsubsection{Other variants} \label{sec:other-blocktrees}

The block tree concept is not as tightly coupled to Lempel-Ziv
parsing as it might seem. \citeN{KP18} build a similar structure on top of an
attractor $\Gamma$ of $S$, of minimal size $\gamma \le z$. Their structure
uses $O(\gamma\log(n/\gamma)) \subseteq O(z\log(n/z))$ space and extracts any 
$S[i\dd j]$ in time $O((1+(j-i)/\log_\sigma n)\log(n/\gamma))$.
Unlike the block tree, their structure divides $S$ 
irregularly, defining blocks as the areas between consecutive attractor 
positions. We prefer to describe the so-called $\Gamma$-tree \cite{NP18},
which is more similar to a block tree and more suitable for indexing.

The $\Gamma$-tree starts with $\gamma$ top-level nodes (i.e., in level $0$), 
each representing a block of length $n/\gamma$ of $S$. The nodes of level $l$
represent blocks of length $b_l=n/(\gamma \cdot 2^l)$. At each level $l$, every
node whose block is at distance less than $b_l$ from an attractor position is
marked. Marked nodes point to their two children in the next level, whereas
unmarked nodes $v$ become leaves pointing to a pair of nodes $v_1,v_2$ at 
the same level, where $S_v$ occurs inside $S_{v_1} \cdot S_{v_2}$ and the
occurrence contains an attractor position. Because of our marking rules, 
nodes $v_1,v_2$ exist and are marked.

It is easy to see that the $\Gamma$-tree has height $\log(n/\gamma)$, at most 
$3\gamma$ marked nodes per level, and that it requires $O(\gamma\log(n/\gamma))$
space. This space is better than the classical block tree because $\gamma \le
z$. The $\Gamma$-tree can retrieve any $S[i]$ in time $O(\log(n/\gamma))$ and
can be enhanced to match the substring extraction time of \citeN{KP18}. As
mentioned, they can also compute Karp-Rabin fingerprints of substrings of $S$
in time $O(\log(n/\gamma))$.

Recently, \citeN{KNP20,KNP21} showed that the original block tree is easily tuned to
use $O(\delta \tau \log_\tau(n/\delta)) \subseteq O(z\tau\log_\tau(n/z))$ space
(recall that $\delta \le \gamma \le z$). The only change needed is to start 
with $\delta$ 
top-level blocks. It can then be seen that there are only $O(\delta)$ marked
blocks per level (though the argument is more complex than the previous ones).
The tree height is $O(\log_\tau(n/\delta))$, higher than the block tree. 
However, from $z = O(\delta\log(n/\delta))$, they obtain that
$\log(n/\delta) = O(\log(n/g))$, and therefore the difference in query times
is not asymptotically relevant.

\subsection{Bookmarking} \label{sec:bookmarking}

\citeN{GGKNP12} combine grammars with Lempel-Ziv parsing to speed up string 
extraction over (Lempel-Ziv) phrase prefixes and suffixes, and \citeN{GGKNP14} 
extend the result to fingerprinting. We present the ideas in simplified form 
and on top of the stronger concept of attractors.

Assume we have split $S$ somehow into $t$ phrases, and let
$\Gamma$ be an attractor on $S$, of size $\gamma$. Using a structure
from Sections~\ref{sec:access-grammars} or \ref{sec:blocktrees}, we provide
access to any substring $S[i\dd i+\ell]$ in time $O(\ell + \log n)$, and 
computation of its Karp-Rabin fingerprint in time $O(\log n)$. Bookmarking 
enables, within $O((t+\gamma)\log\log n)$ additional
space, the extraction of phrase prefixes and suffixes in time 
$O(\ell)$ and their fingerprint computation in time $O(\log\ell)$.

Let us first handle extraction. We consider only the case $\ell \le \log n$,
since otherwise $O(\ell + \log n)$ is already $O(\ell)$. We build a string
$S'[1\dd n']$ by collecting all the symbols of $S$ that are at distance at 
most $\log n$ from an attractor position. It then holds that $n' = 
O(\gamma\log n)$, and every phrase prefix or suffix of $S$ of length up to
$\log n$ appears in $S'$, because it has a copy in $S$ that includes a 
position of $\Gamma$. By storing a pointer from each of the $t$ phrase
prefixes or suffixes to a copy in $S'$, using $O(t)$ space, we can 
focus on extracting substrings from $S'$.

An attractor $\Gamma'$ on $S'$ can be obtained by projecting the positions
of $\Gamma$. Further, if the area between two attractor positions in
$\Gamma$ is longer than $2\log n$, its prefix and suffix of length $\log n$
are concatenated in $S'$. In that case we add the middle position to
$\Gamma'$, to cover the possibly novel substrings. Then, $\Gamma'$ has
a position every (at most) $\log n$ symbols of $S'$, and it is an attractor of 
size $\gamma' \le 2\gamma$ for $S'$.

\begin{example}
Consider the attractor $\Gamma = \{ 4,6,7,8,15,17 \}$ of
$S=\mathsf{ala\caja{b}a\caja{r}\caja{a}\caja{l}alabar\caja{d}a\caja{\$}}$
(the boxed symbols are the attractor positions).
Let us replace $\log n$ by $2$ for the sake of the example. It then holds that
$S'=\mathsf{labaralalarda\$}$. The attractor we build for $\Gamma'$ includes the
positions $\{ 3,5,6,7,12,14 \}$, projected from $\Gamma$. In addition, since
some middle symbols of the area $S[9\dd 14]=\mathsf{alabar}$ are removed and it
becomes $S'[8\dd 11]=\mathsf{alar}$, we add one more attractor in the middle, 
to obtain $\Gamma' = \{ 3,5,6,7,10,12,14 \}$, that is,
$S'=\mathsf{la\caja{b}a\caja{r}\caja{a}\caja{l}al\caja{a}r\caja{d}a\caja{\$}}$. 
\end{example}

As mentioned at the end of Section~\ref{sec:gamma}, we can build a run-length
grammar of size $O(\gamma'\log(n'/\gamma')) = O(\gamma\log\log n)$ on $S'$
(\citeN{CEKNP19} do this without findng the attractor, which would be NP-hard).
This grammar is, in addition, locally balanced, that is, every nonterminal 
whose parse tree node has $l$ leaves is of height $O(\log l)$.

\begin{figure}[t]
\begin{center}
\includegraphics[width=6cm]{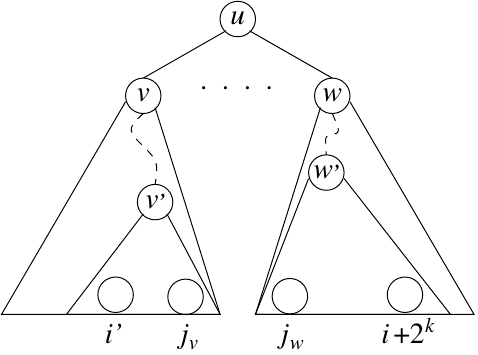}
\end{center}
\caption{Illustration of the bookmarking technique.}
\label{fig:bookmark}
\end{figure}

Assume we want to extract a substring $S'[i'\dd i'+2^k]$ for some fixed $i'$ and
$0 \le k \le \log\log n$. We may store the lowest common ancestor $u$ in the 
parse tree of the $i'$th and $(i'+2^k)$th leaves. Let $v$ and $w$ be the
children of $u$ that are ancestors of those two leaves, respectively. Let
$j_v$ be the rank of the rightmost leaf of $v$ and $j_w$ that of the leftmost 
leaf of $w$. Thus, we have $i' \le j_v < j_w \le i'+2^k$. This implies that,
if $v'$ and $w'$ are, respectively, the lowest common ancestors of the leaves
with rank $i'$ and $j_v$, and with rank $j_w$ and $i'+2^k$, then $v'$ descends
from $v$ and $w'$ descends from $w$, and both $v'$ and $w'$ are of height
$O(\log 2^k) = O(k)$ because the grammar is locally balanced. We can then 
extract $S[i'\dd i'+2^k]$ by extracting the $j_v-i'+1$ rightmost leaves of $v'$ 
by a simple traversal from the right, in time $O(j_v-i'+k)$, then the whole 
children of $u$ that are between $v$ and $w$, in time $O(2^k)$, and finally the 
$i'+2^k-j_w+1$ leftmost leaves of $w'$, in time $O(i'+2^k-j_w+k)$, with a simple
traversal from the left. All this adds up to $O(2^k)$ work. See 
Figure~\ref{fig:bookmark}.

We store the desired information on the nodes $v'$ and $w'$ for every position
$S'[i']$ to which a phrase beginning $S[i]$ is mapped. Since this is stored for
every value $k$, we use $O(t\log\log n)$ total space. To extract $S'[i'\dd
i'+\ell]$ for some $0 \le \ell \le \log n$, we choose 
$k=\lceil \log \ell\rceil$ and extract the substring in time 
$O(2^k) = O(\ell)$. The arrangement for phrase suffixes is analogous.

Similarly, we can compute the Karp-Rabin signature of $S'[i'\dd i'+\ell]$ by
storing the signature for every nonterminal, and combine the signatures of
(1) the $O(k) = O(\log\ell)$ subtrees that cover the area between $S'[i']$ and 
$S'[j_v]$, (2) the children of $u$ between $v$ and $w$ (if any), and (3) the 
$O(k) = O(\log\ell)$ subtrees that cover the area between $S'[j_w]$ and 
$S'[i'+\ell]$ (if any). (If $i'+\ell < j_w$, we only combine the $O(\log\ell)$ 
subtrees that cover the area between $S'[i']$ and $S[i'+\ell]$.)
This can be done in $O(\log\ell)$ time, recall Section~\ref{sec:kr-grammar}.

\section{Parsing-Based Indexing} \label{sec:parsed-indexing}

In this section we describe a common technique underlying a large class of
indexes for repetitive string collections. The key idea, already devised by 
\citeN{KU96}, builds on the parsing induced by a compression method, which
divides
$S[1\dd n]$ into $p$ phrases, $S = S_1 \cdots S_p$. The parsing is used to 
classify the occurrences of any pattern $P[1\dd m]$ into two types:
\begin{itemize}
\item The {\em primary} occurrences are those that cross a phrase boundary.
\item The {\em secondary} occurrences are those contained in a single phrase.
\end{itemize}

The main idea of parsing-based indexes is to first detect the primary 
occurrences with
a structure using $O(p)$ space, and then obtain the secondary ones from those,
also using $O(p)$ space. The key property that the parsing must 
hold is that it must allow finding the secondary occurrences from the primary 
ones within $O(p)$ space.

\subsection{Geometric Structure to Track Primary Occurrences}
\label{sec:grid}

Every primary occurrence of $P$ in $S$ can be uniquely described by 
$\langle i,j\rangle$, indicating:
\begin{enumerate}
\item The leftmost phrase $S_i$ it intersects.
\item The position $j$ of $P$ that aligns at the end of that phrase.
\end{enumerate}

A primary occurrence $\langle i,j \rangle$ then implies that 
\begin{itemize}
\item $P[1\dd j]$ is a {\em suffix} of $S_i$, and 
\item $P[j+1\dd m]$ is a {\em prefix} of $S_{i+1}\cdots S_p$.
\end{itemize}

The idea is then to create two sets of strings:
\begin{itemize}
\item $\mathcal{X}$ is the set of all the {\em reversed} phrase contents,
$X_i = S_i^{rev}$, for $1 \le i < p$, and
\item $\mathcal{Y}$ is the set of all the suffixes $Y_i = S_{i+1}\cdots S_p$, 
for $1 \le i < p$.
\end{itemize}

If, for a given $j$, $P[1\dd j]^{rev}$ is a prefix of $X_i$ (i.e., $P[1\dd j]$
is a suffix of $S_i$) and $P[j+1\dd m]$ is a prefix of $Y_i$, then 
$\langle i,j \rangle$ is a primary occurrence of $P$ in $S$. To find
them all, we lexicographically sort the strings in $\mathcal{X}$ and 
$\mathcal{Y}$, and set up a bidimensional grid of size $p \times p$. The grid
has exactly $p$ points, one per row and per column: if, for some $i$, the 
$x$th element of $\mathcal{X}$ in lexicographic order is $X_i$ and the $y$th
element of $\mathcal{Y}$ in lexicographic order is $Y_i$, then there is a point
at $(x,y)$ in the grid, which we label $i$.

The primary occurrences of $P$ are then found with the following procedure:
\begin{itemize}
\item For each $1 \le j < m$
\begin{enumerate}
	\item Find the lexicographic range $[s_x,e_x]$ of $P[1\dd j]^{rev}$ in 
$\mathcal{X}$.
	\item Find the lexicographic range $[s_y,e_y]$ of $P[j+1\dd m]$ in 
$\mathcal{Y}$.
	\item Retrieve all the grid points $(x,y) \in [s_x,e_x] \times
[s_y,e_y]$.
	\item For each retrieved point $(x,y)$ labeled $i$, report the
primary occurrence $\langle i,j \rangle$.
\end{enumerate}
\end{itemize}

It is then sufficient to associate the end position $p(i)=|S_1 \cdots S_i|$ 
with each phrase $S_i$, to know that the primary occurrence $\langle i,j 
\rangle$ must be reported at position $S[p(i)-j+1 \dd p(i)-j+m]$. Or we can
simply store $p(i)$ instead of $i$ in the grid.

\begin{figure}[t]
\begin{center}
\includegraphics[width=8cm]{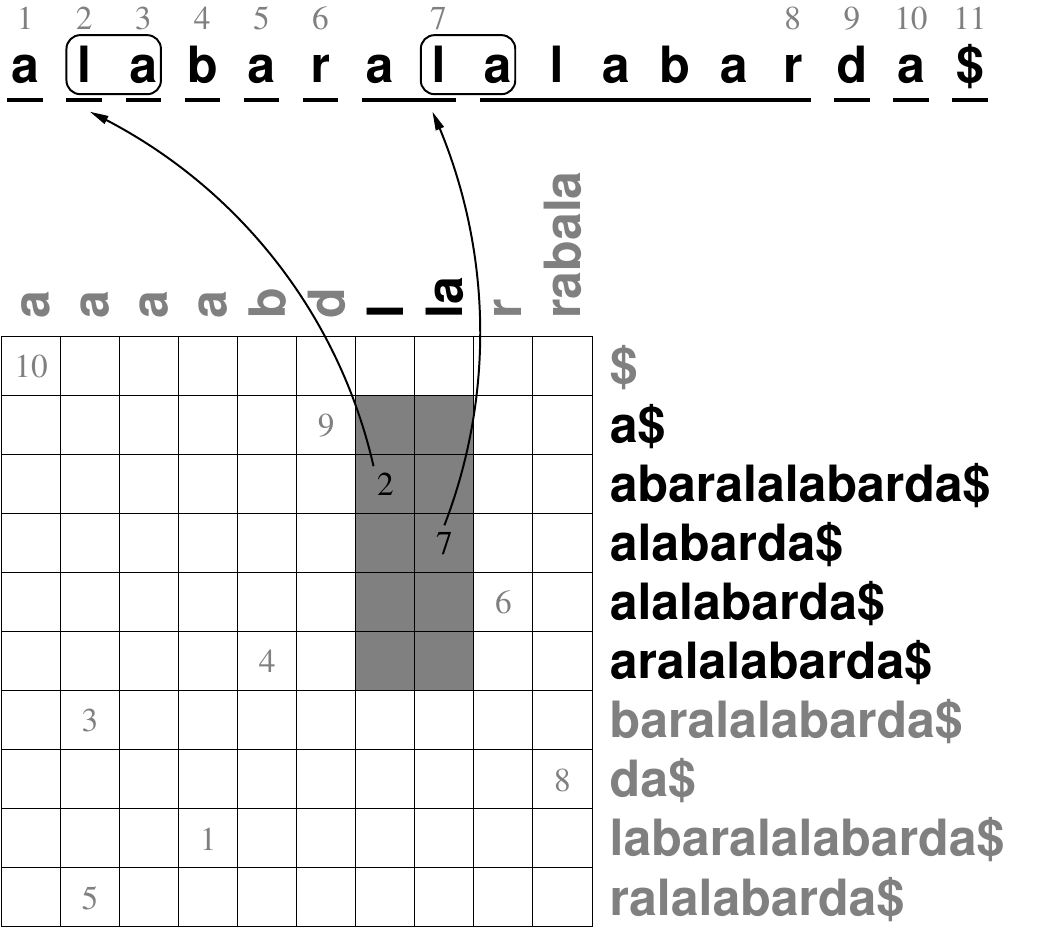}
\end{center}
\caption{A parse for $S=\mathsf{alabaralalabarda\$}$ and the corresponding grid.
We show the search process to find the primary occurrences of $P=\mathsf{la}$.}
\label{fig:grid}
\end{figure}

\begin{example}
Figure~\ref{fig:grid} shows the grid built on a parsing of 
$S=\mathsf{alabaralalabarda\$}$.
Every reversed phrase appears on top, as 
an $x$-coordinate, and every suffix appears on the right, as a $y$-coordinate.
Both sets of strings are lexicographically sorted and the points in the grid
connect phrases ($x$) with their following suffix ($y$). Instead of points we
draw the label, which is the number of the phrase in the $x$-coordinate.

A search for $P=\mathsf{la}$ finds its primary occurrences by searching 
$\mathcal{X}$ for $P[1]^{rev}=\mathsf{l}$, which yields the range $[x_s,x_e]
= [7,8]$, and searching $\mathcal{Y}$ for $P[2]=\mathsf{a}$, which gives
the range $[y_s,y_e]=[2,6]$. The search for the (grayed) zone $[7,8] \times
[2,6]$ returns two points, with labels $2$ and $7$, meaning that $P[1]$ aligns
at the end of those phrase numbers, precisely at positions $S[2]$ and $S[8]$.
\end{example}

Note that, by definition, there are no primary occurrences when $|P|=1$.
Still, it will be convenient to find all the occurrences of $P$ that lie at the
end of a phrase. To do this, we carry out the same steps above for
$j=1$, in the understanding that the lexicographic range on $\mathcal{Y}$
is $[s_y,e_y]=[1,p]$.

The challenges are then (1) how to find the intervals in $\mathcal{X}$ and
$\mathcal{Y}$, and (2) how to find the points in the grid range.

\subsubsection{Finding the intervals in $\mathcal{X}$ and $\mathcal{Y}$}
\label{sec:searchXY}

A simple solution is to perform a binary search on the sets, which requires
$O(\log p)$ comparisons of strings. The desired prefixes of $X_i$ or 
$Y_i$ to compare with, of length at most $m$, must be extracted from a 
compressed representation of $S$: we represent $X_i$ and $Y_i$ just with 
the position where they appear in $S$. By using any of the techniques 
in Section~\ref{sec:access}, we extract them in 
time $f_e = O(m/\log_\sigma n+\log n)$ 
(Section~\ref{sec:extract-grammar}) or $f_e = 
O((1+m/\log_\sigma n)\log(n/g))$ (Section~\ref{sec:extract-blocktree}).
Since this is repeated for every $1 \le j < m$, all the intervals are found in
time $O(f_e\; m \log p)$, which is in $O((m+\log n)m\log n)$ with the first
tradeoff. This complexity can be reduced to $O(m^2 \log n)$ on grammars, by
exploiting the fact that the phrases are defined as the leaves of the grammar
tree, and therefore we always need to extract prefixes or suffixes of
nonterminal expansions (Section~\ref{sec:extract-prefixes}). The same time
can be obtained with $O((p+\gamma)\log\log n)$ additional space by using 
bookmarking (Section~\ref{sec:bookmarking}). In all cases, however, the 
complexity stays quadratic in $m$: we need to search for $m-1$ prefixes/suffixes
of $P$ of length up to $m$.

The quadratic term can be removed 
by using a {\em batched search} for all the suffixes $P[j+1\dd m]$ together
(or all the suffixes $P[1\dd j]^{rev}$). The technique is based on compact 
tries and Karp-Rabin fingerprints \cite{BPPV10,GGKNP14,PhBiCPM17,GNP18,CEKNP19}.
The idea is to represent the sets $\mathcal{X}$ and $\mathcal{Y}$ with compact 
tries, storing fingerprints of the strings labeling selected paths, and then   
verifying the candidates to ensure they are actual matches. The fingerprints
of all the suffixes sought are computed in time $O(m)$. The trie is actually a 
version called {\em z-fast trie} \cite{BBPV09,BPPV10}, which allows searching 
for a string of length $\ell$ with $O(\log\ell)$ fingerprint comparisons. Once 
all the candidate results are found, a clever method does all the 
verifications with one single string extraction, exploiting the fact that we
are looking for various suffixes of a single pattern. If a fingerprint 
$\kappa(S[i\dd j])$
is computed in time $f_h$, then the lexicographic ranges of any $k$ suffixes 
of $P[1\dd m]$ can be found in time $O(m + k(f_h + \log m)+f_e)$. The structure
uses $O(p)$ space, and is built in $O(n\log n)$ expected time to ensure
that there are no fingerprint collisions in $S$.

As shown in Sections~\ref{sec:kr-grammar} and \ref{sec:kr-blocktree}, we can
compute Karp-Rabin fingerprints in time $O(\log n)$ or $O(\log(n/g))$ using
grammars or block trees, respectively. To search for the $k=m-1$ suffixes of $P$
(or of its reverse) we then need time $O(m \log n)$. 

This approach can be applied with block trees built on Lempel-Ziv 
($O(z\log(n/z))$ space, and even $O(\delta\log(n/\delta))$ as shown in 
Section~\ref{sec:other-blocktrees}), or built on attractors
($O(\gamma\log(n/\gamma))$ space). It can also be applied on grammars
($O(g)$ space), and even on run-length grammars (see 
Section~\ref{sec:runlenght-access}), within space
$O(g_{rl}) \subseteq O(\delta\log(n/\delta))$. Combined with bookmarking
(Section~\ref{sec:bookmarking}), the time can be reduced to $O(m\log m)$
because $f_h=O(\log m)$,
yet we need $O(z\log\log n)$ or $O(g\log\log n)$ further space.

A recent twist \cite{CE18,CEKNP19} is that a specific type of grammar, called
{\em locally consistent grammar},\footnote{Built via various rounds of the 
better-known locally consistent parsings \cite{CV86,SV95,MSU97,BES06}.} 
can speed up the searches because there 
are only $k=O(\log m)$ cuts of $P$ that deserve consideration. In a locally
consistent grammar, the subtrees of the parse tree expanding to two identical
substrings $S[i\dd j] = S[i' \dd j']$ are identical except for $O(1)$ nodes
in each level. \citeN{CEKNP19} show that locally balanced and locally consistent
grammars of size $O(\gamma\log(n/\gamma))$ can be built without the need to 
compute $\gamma$ (which would be NP-hard). Further, we can obtain $f_e = O(m)$
time with grammars because, as explained, we extract only rule prefixes and 
suffixes when searching $\mathcal{X}$ or $\mathcal{Y}$ 
(Section~\ref{sec:extract-prefixes}). They also show how to compute Karp-Rabin 
fingerprints in time $O(\log^2 m)$ (without the extra space bookmarking uses). 
The time to find all the relevant intervals then decreases to 
$O(m + k(f_h + \log m)+f_e) \subseteq O(m)$.

\subsubsection{Finding the points in the two-dimensional range}
\label{sec:grid-implem}

This is a well-studied geometric problem \cite{CLP11}. We can represent $p$
points on a $p \times p$ grid within $O(p)$ space, so that we can report all 
the $t$ points within any given two-dimensional range in time 
$O((1+t)\log^\epsilon p)$, for any constant $\epsilon>0$. By using slightly
more space, $O(p\log\log p)$, the time drops to $O((1+t)\log\log p)$, and
if we use $O(p\log^\epsilon p)$ space, the time drops to $O(\log\log p + t)$,
thus enabling constant time per occurrence reported.

If we look for the $k=m-1$ prefixes and suffixes of $P$ with $O(p)$ space, the 
total search
time is $O(m\log^\epsilon p) \subseteq O(m\log n)$, plus $O(\log^\epsilon p)$
per primary occurrence. If we reduce $k$ to $O(\log m)$, the search time
drops to $O(\log m\log^\epsilon p)$, plus $O(\log^\epsilon p)$
per primary occurrence. \citeN{CEKNP19} reduce this $O(\log m\log^\epsilon p)$ 
additive term to just $O(\log^\epsilon \gamma)$ by dealing with short patterns 
separately.

\subsection{Tracking Secondary Occurrences}

The parsing method must allow us infer all the secondary occurrences from
the primary ones. The precise method for propagating primary to secondary 
occurrences depends on the underlying technique. 

\subsubsection{Lempel-Ziv parsing and macro schemes} 
\label{sec:lz-second}

The idea of primary and secondary occurrences was first devised for the 
Lempel-Ziv parsing \cite{FT95,KU96}, of size $p=z$. Note that the leftmost 
occurrence
of any pattern $P$ cannot be secondary, because then it would be inside 
a phrase that would occur earlier in $S$. Secondary occurrences can then be 
obtained by finding all the phrase sources that cover each primary occurrence. 
Each such source produces a secondary occurrence at the phrase that copies the
source. In turn, one must find all the phrase sources that cover these secondary
occurrences to find further secondary occurrences, and so on. All the 
occurrences are found in that way.

A simple technique \cite{KN13} is to maintain all the sources $[b_k,e_k]$ of 
the $z$ phrases of $S$ in arrays $B[1\dd z]$ (holding all $b_k$s) and 
$E[1\dd z]$ (holding all $e_k$s), both sorted by increasing endpoint $e_k$. 
Given an occurrence $S[i\dd j]$, a successor search on $E$ finds 
(in $O(\log\log_w n)$ time \cite{BN14}) the smallest endpoint $e_k \ge j$, and 
therefore all the sources 
in $E[k\dd z]$ end at or after $S[j]$; those in $E[1\dd k-1]$ cannot cover 
$S[i\dd j]$ because they end before $j$. We then want to retrieve
the values $B[l] \le i$ with $k \le l \le z$, that is, the sources that
in addition start no later than $i$.

A technique to retrieve each source covering $S[i\dd j]$ in constant time is
as follows. We build a Range Minimum Query (RMQ) data structure 
\cite{BFCPSS05,FH11} on $B$, which uses $O(z)$ bits and returns, in constant
time, the position of the minimum value in any range $B[k\dd k']$. We first query for
$B[k \dd z]$. Let the minimum be at $B[l]$. If $B[l] > i$, then this source
does not cover $S[i\dd j]$, and no other source does because this is the 
one starting the earliest. We can therefore stop. If, instead, $B[l] \le i$,
then we have a secondary occurrence in the target of $[b_l,e_l]$. We must
report that occurrence and recursively look for other sources covering it.
In addition, we must recursively look for sources that start early enough
in $B[k \dd l-1]$ and $B[l+1 \dd z]$. Since we get an occurrence each time
we find a suitable value of $B$ in the current range, and stop as soon as 
there are no further values, it is easy to see that 
we obtain each secondary occurrence in constant time.

\begin{figure}[t]
\begin{center}
\includegraphics[width=8cm]{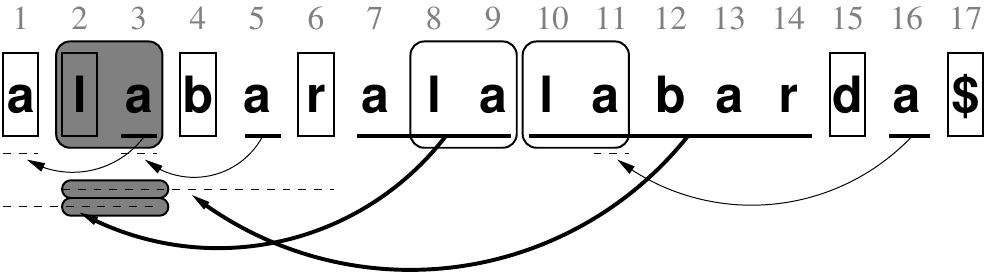}
\end{center}
\caption{Finding the secondary occurrences of $P=\mathsf{la}$ on the 
Lempel-Ziv parse of Figure~\ref{fig:lzparse}. The occurrences are marked with
rounded boxes. The primary occurrence, which is grayed, is projected from 
sources to targets to obtain the secondary ones.}
\label{fig:lz-second}
\end{figure}

\begin{example}
Figure~\ref{fig:lz-second} shows the search for $P=\mathsf{la}$ on the
Lempel-Ziv parse of Figure~\ref{fig:lzparse}. There is only one primary 
occurrence at $S[2\dd 3]$. The sources $[b_k,e_k]$ are, in increasing
order of $e_k$, $[1,1], [1,3], [3,3], [2,6], [11,11]$, so we have the arrays
$B = \langle 1,1,3,2,11 \rangle$ and $E = \langle 1,3,3,6,11 \rangle$. A
successor search for $3$ in $E$ shows that $E[2\dd 5]$ contains all the
sources that finish at or after position $3$. We now use RMQs on $B$ to find 
those that start at or before position $2$. The first candidate is $RMQ(2,5)
= 2$, which identifies the valid source $[B[2],E[2]] = [1,3]$ covering our
primary occurrence. The target of this source is $S[7\dd 9]$, which contains
a secondary occurrence at $S[8\dd 9]$ (the offset of the occurrence within the
target is the same as within the source). There is no other source covering
$S[8\dd 9]$, so that secondary occurrence does not propagate further. We
continue with our RMQs, now on the remaining interval $B[3\dd 5]$. The query
$RMQ(3,5)=4$ yields the source $[B[4],E[4]] = [2,6]$, which also covers the
primary occurrence. Its target, $S[10\dd 14]$, then contains a secondary
occurrence at the same offset as in the source, in $S[10\dd 11]$. Again,
no source covers this secondary occurrence. Continuing,
we must check the intervals $B[3\dd 3]$ and $B[5\dd 5]$. But
since both are larger than $2$, they do not cover the primary occurrence
and we are done.

\end{example}

Thus, if we use a Lempel-Ziv parse, we require $O(z)$ additional space to
track the secondary occurrences. With $occ$ occurrences in total,
the time is $O(occ \log\log_w n)$, dominated by the successor searches.
Note that the scheme works well also under bidirectional macro schemes,
because it does not need that the targets be to the right of the sources.
Thus, the space can be reduced to $O(b)$.

\subsection{Block Trees}

The sequence of leaves in any of the block tree variants we described 
partitions $S$ into a sequence of $p$ phrases (where $p$ can be as small
as $O(\delta\log(n/\delta))$, see Section~\ref{sec:other-blocktrees}).
Each such phrase is either explicit (if it is at the last level) or it
has another occurrence inside an internal node of the same level.

This parsing also permits applying the scheme of primary and secondary
occurrences, if we use leaves of length $1$ \cite{NP18}. It is not hard to see 
that every secondary occurrence $S[i\dd j]$, with $i<j$, has a copy that 
crosses a phrase boundary: $S[i\dd j]$ is inside a block $S_v$ that is a 
leaf, thus it points to another occurrence of $S_v$ inside $S_{v_1} \cdot
S_{v_2}$. If the copy $S[i'\dd j']$ spans both $S_{v_1}$ and $S_{v_2}$, 
then it is a primary
occurrence. Otherwise it falls inside $S_{v_1}$ or $S_{v_2}$, and then
it is also inside a block of the next block tree level. At this next
level, $S[i'\dd j']$ may fall inside a block $S_{v'}$ that is a leaf, and
thus it points towards another occurrence of $S_{v'}$. In this case,
$S[i'\dd j']$ is also a secondary occurrence and we will discover $S[i\dd j]$
from it; in turn $S[i'\dd j']$ will be discovered from its pointed occurrence
$S[i''\dd j'']$, and so on. Instead, $S[i'\dd j']$ may fall inside an internal
block $S_{v'} = S_{v'_1} \cdot S_{v'_2}$. If $S[i'\dd j']$ spans both $S_{v'_1}$
and $S_{v'_2}$,
then it is a primary occurrence, otherwise it appears inside a block of the
next level, and so on. We continue this process until we find an occurrence of 
$S[i\dd j]$ that crosses a block boundary, and thus it is 
primary. Our original occurrence $S[i\dd j]$ is then found from the primary
occurrence, through a chain of zero or more intermediate secondary occurrences.

\begin{example}
Consider the parsing $S=\mathsf{a|l|a|b|a|r|ala|lab|ar|d|a|\$}$ induced by the
block tree of Figure~\ref{fig:bt}, where the phrases of length 1 are the 
leaves of the last level. Then $P=\mathsf{al}$ has two primary occurrences,
at $S[1\dd 2]$ and $S[9\dd 10]$. The sources of blocks are $[1,3]$, $[2,4]$,
and $[5,6]$. Therefore the source $[1,3]$ covers the first primary occurrence,
which then has a secondary occurrence at the target, in $S[7,8]$.
\end{example}

The process and data structures are then almost identical to those for the
Lempel-Ziv parsing. We collect the sources of all the leaves at all the
levels in arrays $B$ and $E$, as for Lempel-Ziv, and use them to find, 
directly or transitively, all the secondary occurrences.
In some variants, such as block trees built on attractors
(see Section~\ref{sec:other-blocktrees}), the sources can be before or after
the target in $S$, as in bidirectional macro schemes, and the scheme works
equally well.

\subsection{Grammars}
\label{sec:g-second}

In the case of context-free grammars \cite{CNspire12} (and also run-length 
grammars), the partition of $S$ induced by the leaves of the grammar tree 
induces a suitable
parsing: a secondary occurrence $S[i\dd j]$ inside a leaf labeled by 
nonterminal $A$ has another occurrence $S[i'\dd j']$ below the occurrence of 
$A$ as an internal node of the grammar tree. If $S[i'\dd j']$ is inside a
leaf labeled $B$ (with $|B| < |A|$), then there is another occurrence 
$S[i''\dd j'']$ below the internal node labeled $B$, and so on. Eventually,
we find a copy crossing a phrase boundary, and this is our primary occurrence.

The hierarchical structure of the grammar tree enables a simplified process to
find the occurrences. For each internal node $v$ representing the rule $A 
\rightarrow X_1 \cdots X_k$, and for each $1 \le i < k$, we insert 
$exp(X_i)^{rev}$ into $\mathcal{X}$ and $exp(X_{i+1})\cdots exp(X_k)$ into
$\mathcal{Y}$, associating the corresponding grid point to node $v$
with offset $|exp(X_1)\cdots exp(X_i)|$. This ensures that every cutting
point between consecutive grammar tree leaves $X$ and $Y$ is included and
associated with the lowest common ancestor node $A=lca(X,Y)$ that covers both 
leaves. By associating the part of $exp(A)$ that follows $exp(X)$, instead of
the full suffix, one ensures at construction time that $A$ is the lowest 
nonterminal that covers the primary occurrence covering $X$ and $Y$.

Once we establish that $P$ occurs inside $exp(A)$ at position $j$, we must 
track $j$ upwards in the grammar tree, adjusting it at each step, until 
locating the occurrence in the start symbol, which gives the position where $P$
occurs in $S$. To support this upward traversal we store, in each grammar tree 
node $A$ with parent $C$, the offset of $exp(A)$ inside $exp(C)$. This is
added to $j$ when we climb from $A$ to $C$.

In addition, every other occurrence of $A$ in the grammar tree contains a 
secondary occurrence of $P$, with the same offset $j$. Note that all those
other occurrences of $A$ are leaves in the grammar tree. 
Each node labeled $A$ has then 
a pointer to the next grammar tree node labeled $A$, forming a linked list that
must be traversed to find all the secondary occurrences (in any desired order; 
the only restriction is that the list must start at the only internal node 
labeled $A$). Further, if $C$ is the parent of $A$, then any other occurrence
of $C$ in the grammar tree (which is necessarily a leaf as well) also contains a
new secondary occurrence of $P$.

The process then starts at each primary occurrence $A$ and recursively moves
to (1) the parent of $A$ (adjusting the offset $j$), and (2) the next node
labeled $A$. The recursive calls end when we reach the grammar tree root in
step (1), which occurs once per distinct secondary occurrence of $P$, and
when there is no next node to consider in step (2).

\begin{figure}[t]
\begin{center}
\includegraphics[width=8cm]{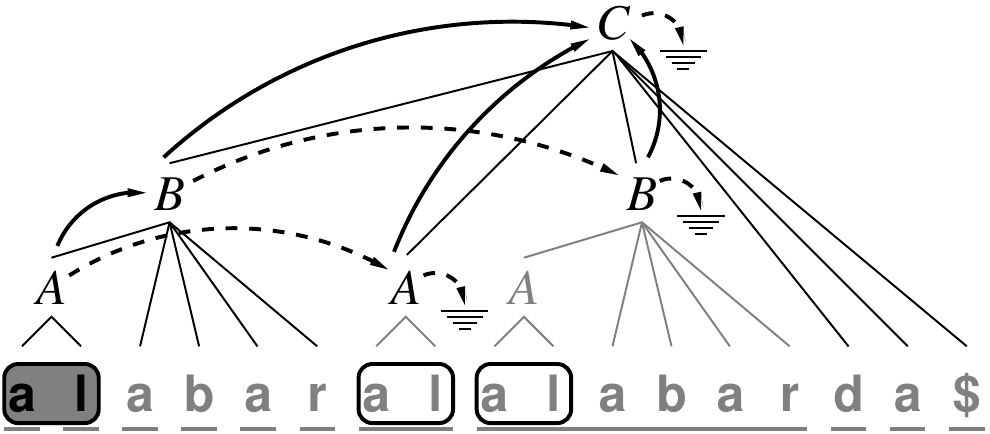}
\end{center}
\caption{Finding the secondary occurrences of $P=\mathsf{al}$ on the 
grammar-induced parse of Figure~\ref{fig:g} (the removed edges of the parse
tree are grayed). The occurrences are marked with rounded boxes; the primary 
one is grayed. The bold solid arrows translate each occurrence towards the root,
and the bold dashed arrows towards the next occurrence of the same nonterminal.}
\label{fig:g-second}
\end{figure}

\begin{example}
Figure~\ref{fig:g-second} shows how the only primary occurrence of 
$P=\mathsf{al}$ in $S=\mathsf{alabaralalabarda\$}$, using the parse of
Figure~\ref{fig:g} (and Figure~\ref{fig:grid}), are propagated using the grammar
tree. The primary occurrence, $S[1\dd 2]$ spans the first two leaves, and the
pointer of the grid sends us to the internal node labeled $A$, which is the
lowest common ancestor of those two leaves, with offset $1$ (indeed, 
$exp(A)=\mathsf{al}$). To find its position in $S$, we go up to $B$, the parent
of $A$, where the offset is still $1$ because the offset of $A$ within $B$ is
$0$ ($exp(B)=\mathsf{alabar}$). Finally, we
reach $C$, the parent of $B$ and the tree root, where the offset is still $1$
and thus we report the primary occurrence $S[1\dd 2]$.

The secondary occurrences are found by recursively following the dashed arrows
towards the other occurrences of the intermediate nonterminals. From the
internal node $A$ we reach the only other occurrence of $A$ in the grammar
tree (which is a leaf; remind that there is only one internal node per
label). This leaf has offset $6$ within its parent $C$, so the offset within
$C$ is $1+6=7$. We then move to 
$C$ and report a secondary occurrence at $S[7\dd 8]$. The list of the $A$s
ends there. Similarly, when we arrive at the internal node $B$, we follow the
dashed arrow towards the only other occurrence of $B$ in the grammar tree.
This has offset $8$ within its parent $C$, so when we move up to $C$ we report
the secondary occurrence $S[9\dd 10]$.

Note that the bold arrows, solid and dashed, form a binary tree rooted at the
primary occurrence. The leaves that are left children are list ends and those
that are right children are secondary occurrences. Thus the total number of
nodes (and the total cost of the tracking) is proportional to the
number of occurrences reported.
\end{example}

\citeN{CNspire12} show that the cost of the traversal indeed amortizes to 
constant time per secondary occurrence if we ensure that every nonterminal $A$ 
occurs at least twice in the grammar tree (as in our example). Nonterminals 
appearing only once can be
easily removed from the grammar. If we cannot modify the grammar, we can
instead make each node $A$ point not to its parent, but to its lowest 
ancestor that appears at least twice in the grammar tree (or to the root,
if no such ancestor exists) \cite{CEKNP19}. This ensures that we report the
$occ_s$ secondary occurrences in time $O(occ_p + occ_s)$.

\citeN{CEKNP19} show how the process is adapted to handle the special nodes
induced by the rules $A \rightarrow X^k$ of run-length grammars.

\subsection{Resulting Tradeoffs}

By considering the cost to find the primary and secondary occurrences, and
sticking to the best possible space in each case, it turns out that we can 
find all the $occ$ occurrences of $P[1\dd m]$ in $S[1\dd n]$ either:
\begin{itemize}
\item In time $O(m\log n + occ\log^\epsilon n)$, within
space $O(g_{rl}) \subseteq O(\delta\log(n/\delta))$.
\item In time $O(m + (occ+1)\log^\epsilon n)$, within
space $O(\gamma\log(n/\gamma))$.
\end{itemize}

The first result is obtained by using a run-length grammar to define the parse
of $S$ (thus the grid is of size $g_{rl} \times g_{rl}$), to access nonterminal
prefixes and suffixes and compute fingerprints, and to track the secondary
occurrences. Note that finding the smallest grammar is NP-hard, but there are
ways to build run-length grammars of size $O(\delta\log(n/\delta))$, recall the
end of Section~\ref{sec:delta}. The second result uses the improvement based 
on locally consistent parsing (end of Section~\ref{sec:searchXY}); recall that 
we do not need to compute $\gamma$ (which is NP-hard too) in order to obtain it.

\subsubsection{Using more space}


We can combine the slightly larger grid representations 
(Section~\ref{sec:grid-implem}) with bookmarking in
order to obtain improved times for the first result. We can use a bidirectional 
macro scheme to define the phrases, so that the grid is of size $b \times b$, 
and use the geometric data structure of size $O(b\log\log b)$ that reports the 
$occ$ points in time $O((1+occ)\log\log b)$. We then use a run-length grammar 
to provide direct access to $S$, and enrich it with bookmarking 
(Section~\ref{sec:bookmarking}) to provide substring extraction and Karp-Rabin 
hashes (to find the ranges in $\mathcal{X}$ and $\mathcal{Y}$) in 
time $f_e=O(m)$ and $f_h=O(\log m)$, respectively, at phrase boundaries.
This adds $O((b+\gamma)\log\log n) = O(b\log\log n)$ space. The time to
find the $m-1$ ranges in $\mathcal{X}$ and $\mathcal{Y}$ is then
$O(m + m(f_h+\log m)+f_e) = O(m\log m)$. The $m-1$ geometric searches take time
$O(m\log\log b + occ \log\log b)$, and the secondary occurrences are reported
in time $O(occ \log\log_w n)$ (Section~\ref{sec:lz-second}). 
\citeN{GGKNP14} get rid of the $O(m\log\log b)$ term by dealing
separately with short patterns (see their Section~4.2; it adapts to our
combination of structures without any change).

Note again that it is NP-hard to find the smallest bidirectional macro scheme,
but we can build suboptimal ones from the Lempel-Ziv parse, the lex-parse, or
the BWT runs, for example (Sections~\ref{sec:lz}, \ref{sec:bwt}, and
\ref{sec:lex-parse}). Recall also that there are heuristics to build 
bidirectional macro schemes smaller than $z$ \cite{NT19,RCNF20}.

The larger grid representation, of size $O(p \log^\epsilon p)$ for $p$ points,
reports primary occurrences in constant time, but to maintain that constant time
for secondary occurrences we need that the parse comes from a (run-length) 
grammar (Section~\ref{sec:g-second}). We must therefore use a grid of 
$g_{rl} \times g_{rl}$. The grammar
already extracts phrase (i.e., nonterminal) prefixes and suffixes in constant
time, yet bookmarking is still useful to compute fingerprints in $O(\log m)$
time. We can then search:
\begin{itemize}
\item In time $O(m\log m + occ\log\log n)$, within
space $O(g_{rl} + b\log\log n)$.
\item In time $O(m\log m + occ)$, within
space $O(g_{rl}\log^\epsilon n)$.
\end{itemize}

Finally, using larger grids directly on the result that uses 
$O(\gamma\log(n/\gamma))$ space yields the first optimal-time index
\cite{CEKNP19}. We can search:
\begin{itemize}
\item In time $O((m+occ)\log\log n)$, within
space $O(\gamma\log(n/\gamma)\log\log n)$.
\item In time $O(m+occ)$, within
space $O(\gamma\log(n/\gamma)\log^\epsilon n)$.
\end{itemize}

\subsubsection{History} \label{sec:history}

The generic technique we have described encompasses a large number of indexes
found in the literature. As said, \citeN{KU96} pioneered the idea of primary 
and secondary occurrences based on Lempel-Ziv for indexing. Their index is not 
properly a compressed index because it stores $S$ in 
plain form, and uses $O(z)$ additional space to store the grid and a mechanism 
of stratified lists of source areas to find the secondary occurrences.

\begin{figure}[t]
\begin{center}
\includegraphics[width=12cm]{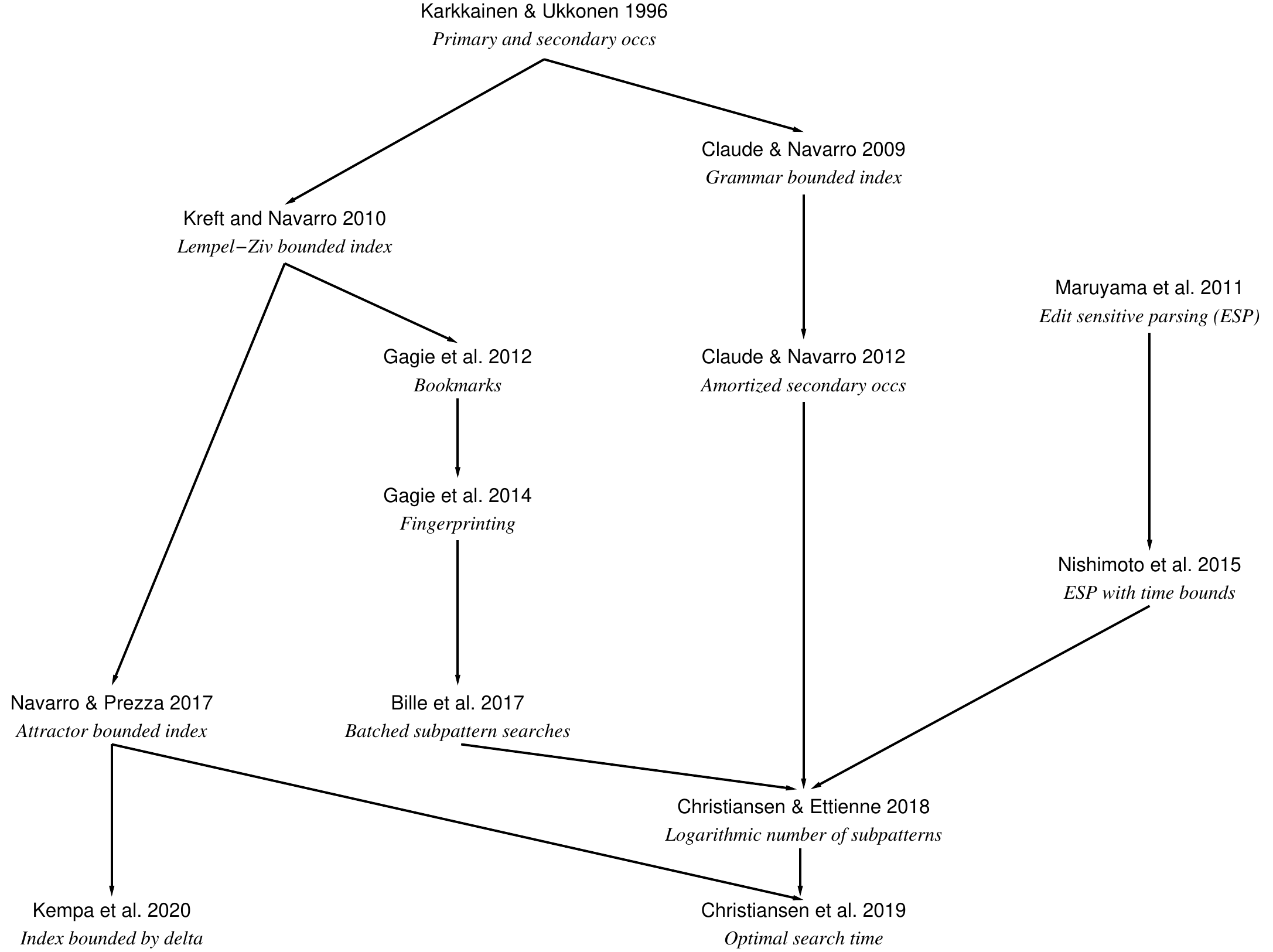}
\end{center}
\caption{Diagram of the influences and main ideas about parsing-based indexing.}
\label{fig:history}
\end{figure}

Figure~\ref{fig:history} shows a diagram with the main ideas that appeared 
along time and the influences between contributions. The Appendix gives a
detailed account. The best results to date are those we have given explicitly,
plus some intermediate tradeoffs given by \citeN{CEKNP19} (see their Table I).

\section{Suffix-Based Indexing} \label{sec:suffix-indexing}

Suffix arrays and suffix trees (Section~\ref{sec:stree}) are data structures 
designed to support indexed searches. They are of size $O(n)$, but large in 
practice. We next describe how their search algorithms translate into structures
of size $O(r)$ or $O(e)$, which are related to the regularities induced by 
repetitiveness on suffix arrays and trees.

\subsection{Based on the BWT} \label{sec:bwt-index}

The suffix array search based on the BWT dates back to \citeN{FM00,FM05}, who
showed that, with appropriate data structures, $S^{bwt}$ is sufficient to 
simulate a suffix array search and find the range $A[sp\dd ep]$ of the suffixes
that start with a search pattern $P$. Their method, called {\em backward 
search}, consecutively finds the interval $A[sp_i\dd ep_i]$ of the suffixes
starting with $P[i\dd m]$, by starting with $[sp_{m+1}\dd ep_{m+1}] = [1\dd n]$
and then computing, for $i=m$ to $i=1$,
\begin{eqnarray*}
	sp_i & = & C[P[i]] + rank_{P[i]}(S^{bwt},sp_{i+1}-1)+1, \\
	ep_i & = & C[P[i]] + rank_{P[i]}(S^{bwt},ep_{i+1}), 
\end{eqnarray*}
where $C[c]$ is the number of occurrences in $S$ of symbols lexicographically 
smaller than $c$, and $rank_c(S^{bwt},j)$ is the number of occurrences of $c$
in $S^{bwt}[1\dd j]$.\footnote{If $sp_i > ep_i$, then $P$ does not 
occur in $S$ and we must not continue the backward search.} 
Further, if $A[j] = i$, that is, the lexicographically
$j$th smallest suffix of $S$ is $S[i\dd]$, then $A[j']=i-1$ for $c=S^{bwt}[j]$
and
\[  j' ~=~ LF(j) ~=~ C[c] + rank_c(S^{bwt},j),
\]
which is called an {\em LF-step} from $j$. By performing LF-steps on the BWT
of $S$, we virtually traverse $S$ backwards. Operation $rank$ on $S^{bwt}$ can
be implemented in $n\log\sigma + o(n\log\sigma)$ bits (and even in statistically
compressed space) and in the optimal time $O(\log\log_w\sigma)$ \cite{BN14},
which yields $O(m\log\log_w\sigma)$ time for backward searching of $P$.

To understand the rationale of the backward search formula, let us start with
the backward step. Recall from Section~\ref{sec:bwt} that $c=S^{bwt}[j]$ is the 
symbol preceding the suffix $A[j]$, $c=S^{bwt}[j]=S[A[j]-1]$. The function 
$j'=LF(j)$ computes where is in $A$ the suffix that points to $A[j]-1$. First, 
all the $C[c]$ suffixes that start with symbols less than $c$ precede $A[j']$. 
Second, the suffixes $S[A[j]-1\dd]$ are stably 
sorted by $\langle S[A[j]-1], A[j] \rangle = \langle S^{bwt}[j], A[j] \rangle$,
that is, by their first symbol and breaking ties with the rank of the suffix 
that follows. Therefore, $LF(j)$ adds the number $C[c]$ of suffixes starting
with symbols less than $c$ and the number $rank_c(S^{bwt},j)$ of suffixes that
start with $c$ up to the one we want to translate, $A[j]$.

\begin{example}
Consider $S = \mathsf{alabaralalabarda\$}$, with
$S^{bwt} = \mathsf{adll\$lrbbaaraaaaa}$, in Figure~\ref{fig:runs}.
From $A[14]=10$ and $S^{bwt}[14]=\mathsf{a}$ (which correspond to the suffix
$S[10\dd]=\mathsf{labarda\$}$), we compute $LF(14)=C[\mathsf{a}]+
rank_\mathsf{a}(S^{bwt},14) = 1 + 5 = 6$. Indeed $A[6]=9=A[14]-1$, 
corresponding to the suffix $\mathsf{alabarda\$}$.
\end{example}

Let us now consider the backward search steps. Note that we know that the 
suffixes in $A[sp_{i+1}\dd ep_{i+1}]$ start with $P[i+1\dd m]$. The range
$A[sp_i\dd ep_i]$ lists the suffixes that start with $P[i\dd m]$, that is, they
start with $P[i]$ and then continue with a suffix in $A[sp_{i+1}\dd ep_{i+1}]$.
We then want to capture all the suffixes in $A[sp_{i+1}\dd ep_{i+1}]$ that are 
preceded by $c=P[i]$ and map them to their corresponding position in $A$. Since
they will be mapped to a range, the backward search formula is a way to perform
all those LF-steps in one shot.

\begin{example}
Consider again $S = \mathsf{alabaralalabarda\$}$, with
$S^{bwt} = \mathsf{adll\$lrbbaaraaaaa}$, in Figure~\ref{fig:runs}.
To search for $P=\mathsf{la}$, we start with the range $A[sp_3\dd ep_3]=[1\dd
17]$. The first backward step, for $P[2]=\mathsf{a}$, gives $sp_2 = 
C[\mathsf{a}]+rank_\mathsf{a}(S^{bwt},0)+1 = 1+1 = 2$ and $ep_2 = 
C[\mathsf{a}]+rank_\mathsf{a}(S^{bwt},17) = 1+8 = 9$. Indeed,
$A[2\dd 9]$ is the range of all the suffixes starting with $P[2\dd 2]=
\mathsf{a}$. The second and final backward step, for $P[1]=\mathsf{l}$, gives
$sp_1 = C[\mathsf{l}]+rank_\mathsf{l}(S^{bwt},1)+1 = 12+1 = 13$ and $ep_2 = 
C[\mathsf{l}]+rank_\mathsf{l}(S^{bwt},9) = 12+3 = 15$. Indeed, $A[13\dd 15]$
is the range of the suffixes starting with $P=\mathsf{la}$, and thus the 
occurrences of $P$ are at $A[13] = 2$, $A[14]=10$, and $A[15]=8$.
Note that, if we knew that the suffixes in $A[2\dd 9]$ preceded by 
$\mathsf{l}$ were at positions $3$, $4$, and $6$, and we had computed
$LF(3)$, $LF(4)$, and $LF(6)$, we would also have obtained the interval
$A[13\dd 15]$.
\end{example}

\citeN{FM05} and \citeN{FMMN07} show how to represent $S^{bwt}$ within 
$n{\cal H}_k(S)+o(n\log\sigma)$ bits of space, that is, asymptotically within the 
$k$th order empirical entropy of $S$, while supporting pattern searches in time 
$O(m\log\sigma + occ \log^{1+\epsilon} n)$ for any constant $\epsilon>0$.
These concepts are well covered in other surveys \cite{NM06}, so we will not
develop them further here; we will jump directly to how to implement them when
$S$ is highly repetitive.

\subsubsection{Finding the interval} \label{sec:rindex-count}

\citeN{MN05} showed how to compute $rank$ on $S^{bwt}$ when it is represented 
in run-length form (i.e., as a sequence of $r$ runs). We present the results in 
a more recent setup \cite{MNSV09,GNP19} that ensures $O(r)$ space. The positions
that start runs in $S^{bwt}$ are stored in a predecessor data structure that
also tells the number of the corresponding runs. A string $S'[1\dd r]$ stores the 
symbol corresponding to each run of $S^{bwt}$, in the same order of $S^{bwt}$.
The run lengths are also stored in another array, $R[1\dd r]$, but they are
stably sorted 
lexicographically by the associated symbol. More precisely, if $R[t]$ is
associated with symbol $c$, it stores the cumulative length of the runs 
associated with $c$ 
in $R[1\dd t]$. Finally, $C'[c]$ tells the number of runs of symbols $d$ for 
all $d < c$. Then, to compute $rank_c(S^{bwt},j)$, we:
\begin{enumerate}
\item Find the predecessor $j'$ of $j$, so that we know that $j$ belongs to the 
$k$th run in $S^{bwt}$, which starts at position $j' \le j$.
\item Determine that the symbol of the current run is $c' = S'[k]$.
\item Compute $p = rank_{c}(S',k-1)$ to determine that there are $p$ runs
of $c$ {\em before} the current run.
\item The position of the run $k-1$ in $R$ is $C'[c]+p$: $R$ lists the $C'[c]$ 
runs of symbols less than $c$, and then the $p$ runs of $c$ preceding our run
$k$ (because $R$ is stably sorted, upon ties it retains the order of
the runs in $S^{bwt}$).
\item We then know that $rank_c(S^{bwt},j'-1) = R[C'[c]+p]$.
\item This is the final answer if $c\not=c'$. If $c=c'$, then $j$ is within a 
run of $c$s and thus we must add $j-j'+1$ to the answer.
\end{enumerate}

\begin{example}
The BWT of $S=\mathsf{alabaralalabarda\$}$ has $r(S)=10$ runs (recall
Figure~\ref{fig:runs}), $S^{bwt} = \mathsf{a|d|ll|\$|l|r|bb|aa|r|aaaaa}$.
The predecessor data structure then contains the run start positions,
$\langle 1,2,3,5,6,7,8,10,12,13 \rangle$. The string of distinct run symbols
is $S'[1\dd 10]=\mathsf{adl\$lrbara}$. Stably sorting the runs 
$\langle 1\dd 10\rangle$ by symbol we obtain $\langle 4,1,8,10,7,2,3,5,6,9
\rangle$ (e.g., we first list the $4$th run because its symbol is the smallest,
$\mathsf{\$}$, then we list the $3$ positions of $\mathsf{a}$ in $S'$, 
$1,8,10$, and so
on), and therefore $R=\langle 1,1,3,8,2,1,2,3,1,2\rangle$ (e.g., $R[2\dd 4]=
\langle 1,3,8\rangle$ because the runs of $\mathsf{a}$ are of lengths $1$, $2$,
and $5$, which cumulate to $1$, $3$, and $8$). Finally, $C'[\mathsf{\$}]=0$, 
$C'[\mathsf{a}]=1$, $C'[\mathsf{b}]=4$, $C'[\mathsf{d}]=5$, $C'[\mathsf{l}]=6$,
and $C'[\mathsf{r}]=8$ precede the positions where the runs of each symbol start
in $R$.

To compute $rank_\mathsf{a}(S^{bwt},15)$ we find the predecessor $j'=13$ of $j=15$ and 
from the same structure learn that it is the run number $k=10$. We then know
that it is a run of $\mathsf{a}$s because $S'[10]=\mathsf{a}$. We then find
out that there are $p=2$ runs of $\mathsf{a}$s preceding it because
$rank_\mathsf{a}(S',9)=2$. Further, there are $C'[\mathsf{a}]=1$ runs of
symbols smaller than $\mathsf{a}$ in $S^{bwt}$. This means that the runs of
$\mathsf{a}$s start in $R$ after position $C'[\mathsf{a}]=1$, and that the run
$k-1=9$ is, precisely, at $C'[\mathsf{a}]+p=3$. With $R[3]=3$ we learn that 
there
are $3$ $\mathsf{a}$s in $S^{bwt}[1\dd 12]$. Finally, since we are counting
$\mathsf{a}$s and $j$ is in a run of $\mathsf{a}$s, we must add the 
$j-j'+1 = 15-13+1 = 3$ $\mathsf{a}$s in our current run. The final answer is
then $rank_\mathsf{a}(S^{bwt},15) = 3+3=6$.
\end{example}

The cost of the above procedure is dominated by the time to find the predecessor
of $j$, and the time to compute $rank$ on $S'[1\dd r]$. Using only structures 
of size $O(r)$ \cite{GNP19}, the predecessor can be computed in time 
$O(\log\log_w(n/r))$ if there are $r$ elements in a universe $[1\dd n]$, and
$rank$ on $S'$ can be computed in time $O(\log\log_w\sigma)$. This yields a total
time of $O(m\log\log_w(\sigma+n/r))$ to determine the range $A[sp\dd ep]$ using 
backward search for $P$, in space $O(r)$ \cite{GNP19}.
Recently, \citeN{NT20} showed that, adding some artificial cuts to the
runs, it is possible to avoid the $O(\log\log_w(n/r))$-time predecessor
searches by always maintaining the run to which the position $j$
belongs in constant time. This reduces the time to
$O(m\log\log_w\sigma)$, still within $O(r)$ space.

\subsubsection{Locating the occurrences} \label{sec:rindex-locate}

Once we have determined the interval $[sp\dd ep]$ where the answers lie in the
suffix array, we must output the positions $A[sp],\ldots,A[ep]$ to complete the
query. We do not have, however, the suffix array in explicit form. The classical
procedure \cite{FM05,MNSV09} is to choose a sampling step $t$ and sample the 
suffix array entries that point to all the positions of the form 
$S[i \cdot t + 1]$, for all $0 \le i < n/t$. Then, if $A[j]$ is not sampled,
we compute $LF(j)$ and see if $A[LF(j)]$ is not sampled, and so on. Since we
sample $S$ regularly, some $A[LF^s(j)]$ must be sampled for $0 \le s < t$.
Since function $LF$ implicitly moves us one position backward in $S$, it holds
that $A[j] = A[LF^s(j)]+s$. Therefore, within $O(n/t)$ extra space, we can
report each of the occurrence positions in time $O(t\log\log_w(n/r))$ (the 
LF-steps do not require $O(\log\log_w\sigma)$ time for the $rank$ on $S'$ because
its queries are of the form $rank_{S'[i]}(S',i)$, for which we can just store 
the answers to the $r$ distinct queries).

Though this procedure is reasonable for statistical compression, the extra
space $O(n/t)$ is usually much larger than $r$, unless we accept a significantly
high time $O((n/r)\log\log_w(n/r))$ to report each occurrence. This had been a 
challenge for BWT-based indexes on repetitive collections
until recently \cite{GNP19}, where a way to efficiently locate the
occurrences within $O(r)$ space was devised.

\citeN{GNP19} solve the problem of reporting $A[sp\dd ep]$ in two steps, in
their so-called r-index (we
present the simplified version of \citeN{BGI20}). First, they show that the
backward search can be modified so that, at the end, we know the value of
$A[ep]$ (in turn, this simplifies a previous solution \cite{PP18}). Second,
they show how to find $A[j-1]$ given the value of $A[j]$.

The first part is not difficult. When we start with $[sp\dd ep]=[1\dd n]$, we
just need the value of $A[n]$ stored. Now, assume we know $[sp_{i+1}\dd
ep_{i+1}]$ and $A[ep_{i+1}]$, and compute $[sp_i\dd ep_i]$ using the backward
search formula. If the last suffix, $A[ep_{i+1}]$, is preceded by $P[i]$ 
(i.e., if $S^{bwt}[ep_{i+1}]=P[i]$), then the last suffix of $A[sp_i\dd ep_i]$ 
will be precisely $A[ep_i] = A[LF(ep_{i+1})] = A[ep_{i+1}]-1$, and thus we know
it. Otherwise, we must find the last occurrence of $P[i]$ in 
$S^{bwt}[sp_{i+1}\dd ep_{i+1}]$, because this is the one that will be mapped
to $A[ep_i]$.
This can be done by storing an array $L[1\dd r]$ parallel to $R$, so that
$L[t]$ is the value of $A$ for the last entry of the run $R[t]$ refers to.
Once we determine $p$ using the backward search step described above, we have
that $A[ep_i] = L[C'[c]+p]-1$.

\begin{example}
For $S=\mathsf{alabaralalabarda\$}$ and $S^{bwt} = \mathsf{adll\$lrbbaaraaaaa}$,
we have $L=\langle 1,17,12,14,13,16,11,9,7,15 \rangle$. For example, $L[3]$ 
refers to the end of the $2$nd run of $\mathsf{a}$s, as seen in the previous 
example for $R[3]$. This ends at $S^{bwt}[11]$, and $A[11]=12=L[3]$. In the
backward search for $P=\mathsf{la}$, we start with $[sp_3\dd ep_3] = [1\dd 17]$,
and know that $A[17] = 14$. The backward search computation then yields 
$[sp_2\dd ep_2] = [2\dd 9]$. Since $S^{bwt}[17]=
\mathsf{a}=P[2]$, we deduce that $A[9]=14-1 = 13$.
A new backward step yields $[sp_1\dd ep_1] = [13\dd 15]$. Since $S^{bwt}[9]
=\mathsf{b} \not= P[1]$, we must consult $L$. The desired position of $L$ is
obtained with the same process to find $rank_\mathsf{l}(S^{bwt},9)$: $k=7$,
$p=rank_\mathsf{l}(S',6)=2$, $t=C'[\mathsf{l}]+p=6+2=8$, from where we obtain 
that $A[15]=L[8]-1=9-1=8$.
\end{example}

For the second part, finding $A[j-1]$ from $A[j]$, let us define $d =
A[j-1]-A[j]$, and assume both positions are in the same run, that is,
$S^{bwt}[j-1]=S^{bwt}[j]=c$ for some $c$. By the LF-step formula, it
is not hard to see that $LF(j-1) = LF(j)-1$, and thus $A[LF(j-1)]-A[LF(j)] =
(A[j-1]-1)-(A[j]-1)=d=A[LF(j)-1]-A[LF(j)]$.\footnote{With the possible exception of $A[j-1]$ or
$A[j]$ being $1$, but in this case the BWT symbol is $\mathsf{\$}$, and thus
they cannot be in the same run.} This means that, as we perform LF-steps from
$j$ and $j-1$ and both stay in the same run, their difference $d$ stays the
same. After performing $s$ LF-steps, for some $s$, $S^{bwt}[j'] = S^{bwt}[LF^s(j)]$
will be a run head and $S^{bwt}[j'-1] = S^{bwt}[LF^s(j-1)]$ will belong to
another run. If we store $d=A[j'-1]-A[j']$ for the run head $j'$, we can
compute $A[j-1] = A[j]+d$.

The key to find the proper run head is to note that $A[j'] = A[j]-s$ is
the only position of a run head mapped to $S$ in $A[j]-s,\ldots,A[j]$.
We then store another predecessor data structure with the positions $A[j']$ in
$S$ that correspond to run heads in $S^{bwt}$, $S^{bwt}[j'-1] \not=
S^{bwt}[j']$. To the position $t=A[j']$ we associate $d(t)=A[j'-1]-A[j']$.
To compute $A[j-1]$ from $A[j]$, we simply find the predecessor $t=A[j']$ of
$A[j]$ and then know that $A[j-1] = A[j]+d(t)$.\footnote{\citeN{GNP19} store
$A[j'-1]$ instead of $d(t)$,
and thus add $s=A[j]-t$ to return $A[j-1]=A[j'-1]+s$.}

\begin{figure}[t]
\begin{center}
\includegraphics[width=10cm]{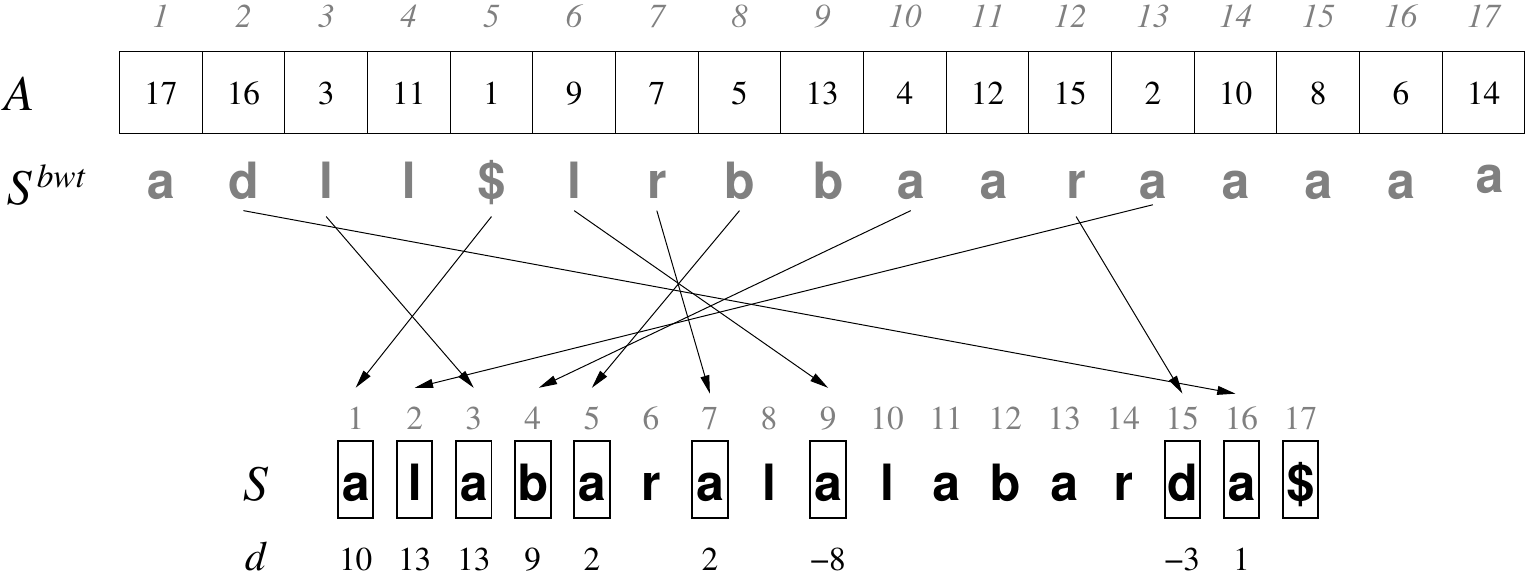}
\end{center}
\caption{The sampling on $S=\mathsf{alabaralalabarda\$}$ induced by the runs of
its BWT with the purpose of locating occurrences. We omit the sample of $A[1]
=17$ because no pattern can occur at $S[17]$.}
\label{fig:rindex}
\end{figure}

\begin{example}
Figure~\ref{fig:rindex} shows the run beginnings projected to $S$, and the
associated values $d$. 
Once we find the interval $A[13\dd 15]$ for $P=\mathsf{la}$ in the previous
example, and since we know that $A[15]=8$, we can compute $A[14]$ as follows.
The boxed predecessor of $S[8]$ is $S[7]$. Since $A[7]=7$, we stored $d(7)=
A[6]-A[7]=2$ associated with $S[7]$, and thus we know that $A[14]=A[15]+d(7)=
10$. Now, the boxed predecessor of $S[10]$ is $S[9]$. Since $A[6]=9$, we
stored $d(9)=A[5]-A[6]=-8$ associated with $S[9]$, and thus we know that $A[13]
=A[14]+d(9)=2$.
\end{example}

Each new position is then found with a predecessor search, yielding total
search time $O(m \log\log_w(\sigma+n/r) + occ \log\log_w(n/r))$, and $O(r)$ space
\cite{GNP19}. \citeN{NT20} manage to compute these predecessors in constant
time as well, thus reducing the search time to
of $O(m\log\log_w\sigma + occ)$ in $O(r)$ space.

This index was implemented and shown to be 1--2 orders of 
magnitude faster than parsing-based indexes, though up to twice as large 
\cite{GNP19}. When the collections are very repetitive, its size is still 
small enough, but the index (as well as measure $r$) degrades faster than $z$ 
or $g$ when repetitiveness starts to decrease.

Note that the index does not provide direct access to $S$ within
$O(r)$ space, only within $O(r\log(n/r))$ space (and $O(\ell+\log(n/r))$ time), 
and this is provided through a run-length grammar of that size and height
$O(\log(n/r))$. What is more interesting is that they also build grammars of 
size $O(r\log(n/r))$ that provide access in time $O(\log(n/r))$ to any cell
of $A$ or $A^{-1}$.

A previous attempt to provide fast searches on top of the BWT 
\cite{BCGPR15} combines it with Lempel-Ziv parsing: it uses 
$O(z(S)+r(S)+r(S^{rev}))$ space and searches in time
$O(m(\log\log n + \log z) + occ \log^\epsilon z)$.
A careful implementation \cite{BCGPR17} shows to be relevant,
for example it uses about 3 times more space and is faster than 
the index of \citeN{KN13}.

\subsubsection{Optimal search time} 

\citeN{Kem19} generalizes the concept of BWT runs to $s$-runs, where the $s$
symbols preceding each suffix $A[j]$ must coincide. He shows that, if $S$
has $r$ runs, then it has $O(rs)$ $s$-runs. \citeN{GNP19} use this idea to
define a new string $S^* = S^0 \cdot S^1 \cdots S^{s-1}$, where $S^k$ is formed
by discarding the first $k$ symbols of $S$ and then packing it into
``metasymbols'' of length $s$. The metasymbols are lexicographically compared
in the same way as the string that composes them. They show that the suffix
array interval for
$P$ in $S$ and for $P^0$ in $S^*$ are the same. Since the length of $P^0$ is
$m' = m/s$, one can choose $s=\log\log_w(\sigma+n/r)$ in order to represent
$S^*$ within $O(rs)  = O(r\log\log_w(\sigma+n/r))$ space and find the interval
$A[sp\dd ep]$ of $P$ in $S$ by searching for $P^0$ in $S^*$, in time
$O(m'\log\log_w(\sigma^s+n/r)) = O(m)$.

In turn, the occurrences are located also in chunks of $s$, by storing in
$d(t)$ not only the information on $A[j'-1]$, but also on $A[j'-2],\ldots,
A[j'-s]$, in space $O(rs)$ as well. Thus, we invest a predecessor
search, time $O(\log\log_w(n/r))$, but in exchange retrieve $s$
occurrences. The resulting time is $O(m + \log\log_w(n/r) + occ)$, which is
converted into the optimal $O(m+occ)$ by handling short patterns separately.

With the new technique of \citeN{NT20}, we can obtain $O(m+occ)$ time
within $O(r\log\log_w\sigma)$ space, because it is sufficient to set
$s=\log\log_w\sigma$ to find the interval $A[sp\dd ep]$ in time $O(m)$ and
the occurrences are already located in time $O(occ)$. Note that this space
is $O(r)$ if $\sigma = O(\mathrm{poly}\,w)$.

RAM-optimal search time is also possible with this index, within
$O(r w \log_\sigma\log_w n)$ space \cite{GNP19}.
Interestingly, RAM-optimal search time had been obtained only in the
classical scenario, using $O(n)$ words of space.

\subsection{Based on the CDAWG} \label{sec:cdawg-search}

In principle, searching the CDAWG as easy as searching a suffix tree
\cite{CH97}: since any suffix can be read from the root node, we simply move 
from the root using $P$ until finding its locus node (like on the suffix tree, 
if we end in the middle of an edge, we move to its target node).

A first problem is that we do not store the strings that label the edges of the
CDAWG. Instead, we may store only the first symbols and the lengths of those 
strings, as done for suffix trees in Section~\ref{sec:stree}. Once we reach 
the locus node, we must verify that all the skipped symbols coincide with $P$ 
\cite{CH97,BCspire17}. The problem is that the string $S$ is not directly
available for verification. Since $e=\Omega(g)$, however, we can in principle 
build a grammar of size $O(e)$ so that we can extract a substring of $S$ of 
length $m$, and thus verify the skipped symbols, in time $O(m+\log n)$;
recall Section~\ref{sec:access-grammars}.

To determine which position of $S$ to extract, we use the property that all
the strings arriving at a CDAWG node are suffixes of one another
\cite[Lem.~1]{BBHMCE87}. Thus, we store the final position in $S$ of the 
longest string arriving at each node from the root. If the longest string 
arriving at the locus node $v$ ends at position $t(v)$, and 
we skipped the last $l$ symbols of the last edge, then $P$ should be equal to 
$S[t(v)-l-m+1\dd t(v)-l]$, so we extract that substring and compare it with $P$.
If they coincide, then every distinct path from $v$ to the final node, of total
length $L$, represents an occurrence at $S[n-L-l-m+1\dd n-L-l]$. Since every
node has at least two outgoing edges, we spend $O(1)$ amortized time per 
occurrence reported \cite{BBHMCE87}. 
The total search time is then $O(m+\log n + occ)$.

\begin{example}
Let us search for $P=\mathsf{la}$ in the CDAWG of Figure~\ref{fig:cdawg}. We
leave the root by the arrow whose string starts with $\mathsf{l}$, and arrive
at the target node $v$ with $l=0$ (because the edge consumes the $m=2$ symbols 
of $P$). The node $v$ can be associated with position $t(v)=3$, which ends an
occurrence of the longest string arriving at $v$, $\mathsf{ala}$. We then 
extract $S[t(v)-l-m+1\dd t(v)-l] = S[2\dd 3]=\mathsf{la}$ and verify that the
skipped symbols match $P$. We then traverse all the forward paths from $v$, 
reaching the final node in three ways, with total lengths $L=8,14,6$. 
Therefore, $P$ occurs in $S$ at positions $n-L-l-m+1 = 8,2,10$.
\end{example}

Another alternative, exploiting the fact that $e = \Omega(r)$, is to enrich
the CDAWG with the BWT-based index of size $O(r)$: as shown in
Section~\ref{sec:rindex-count}, we can determine in time $O(m\log\log_w\sigma)$
whether $P$ occurs in $S$ or not, that is, if $sp \le ep$ in the interval
$A[sp\dd ep]$ we compute. If $P$ occurs in $S$, we do not need the grammar to
extract and verify the skipped symbols; we can just proceed to output all
the occurrences \cite{BCGPR15}. The total time is then $O(m\log\log_w\sigma + occ)$.
This variant is carefully implemented by \citeN{BCGPR17}, who show that the
structure is about two orders of magnitude faster than the index of
\citeN{KN13}, though it uses an order of magnitude more space.

It is even possible to reach optimal $O(m+occ)$ time with the CDAWG, by
exploiting the fact that the CDAWG induces a particular grammar of size $O(e)$
where there is a distinct nonterminal per string labeling a CDAWG edge
\cite{BC17}. Since we need to extract a prefix of the string leading to the
locus of $P$, and this is the concatenation of several edges plus a prefix of
the last edge, the technique of Section~\ref{sec:extract-prefixes} allows us
to extract the string to verify in time $O(m)$. Variants of this idea are 
given by \citeN{BCspire17} (using $O(e)$ space) and \citeN{TGFIA17}
(using $O(e(S)+e(S^{rev}))$ space).

\section{Other Queries and Models} \label{sec:others}

In this section we briefly cover the techniques to handle other type of 
queries, apart from the fundamental one of finding all the occurrences of a 
pattern in a string. We also consider indexes based on more ad-hoc scenarios
where repetitiveness arises.

\subsection{Counting}

A query that in principle is simpler than locating the $occ$ occurrences of a 
pattern $P$ in $S$ is that of counting them, that is, determining $occ$. This
query is useful in Data Mining and Information Retrieval, for example, where 
one might want to know how frequent or relevant $P$ is in $S$.

In suffix-based indexes, counting is generally a step that predeces locating:
for the latter, we find the interval $A[sp\dd ep]$ of the suffixes starting 
with $P$ and then we output $A[sp],\ldots,A[ep]$. For counting, we just have
to output $occ = ep-sp+1$. As shown in Section~\ref{sec:rindex-count}, this can be easily done in $O(r)$ space and
$O(m\log n)$ time \cite{MNSV09}, or even $O(m\log\log_w n)$ time with stronger
predecessor data structures \cite{GNP19}. It is also possible to count in
$O(m)$ time and $O(r\log\log_w(\sigma+n/r))$ space, and in the RAM-optimal 
$O(\lceil m\log(\sigma)/w \rceil)$ time and $O(rw\log_\sigma\log_w n)$ space
\cite{GNP19}.

Similary, we can count in $O(m)$ time and $O(e)$ space using CDAWGs
(Section~\ref{sec:cdawg-search}). By storing
in each CDAWG node $v$ the number of paths that lead from the node to the final
state, we simply find the locus $v$ of $P$ in $O(m)$ time and report this 
number.

It is less obvious how to count in phrase-based indexes, as these are
oriented to reporting all the occurrences. \citeN{Nav18} observes that, in a 
grammar-based index, the number of occurrences (one primary and many 
secondary) triggered by each point of the grid (Section~\ref{sec:grid}) depends
only on the point (Section~\ref{sec:g-second}), and thus one can associate 
that number of the point itself. Note that this is not the case of the grid 
associated with, say, the Lempel-Ziv parse, where each source may or not 
contain the primary occurrence depending on the alignment of $P$
(Section~\ref{sec:lz-second}).

Counting on a grammar-based index then reduces to summing the number of 
occurrences of all the points inside the grid ranges obtained from each 
partition of $P$. With appropriate geometric data structures \cite{Cha88}, 
\citeN{Nav18} obtains $O(m^2 + m\log^{2+\epsilon} n)$ counting time in $O(g)$ 
space. This time can be reduced to $O(m\log^{2+\epsilon} n)$ by searching for
all the partitions of $P$ in batch (Section~\ref{sec:searchXY}) and computing
fingerprints on the grammar (Section~\ref{sec:kr-grammar}).

\citeN{CEKNP19} further reduce this time by using their locally consistent 
grammar, which is of size $O(\gamma\log(n/\gamma))$ and requires one to test 
only $O(\log m)$ partitions of $P$ (recall the end of 
Section~\ref{sec:searchXY}). Such a lower number of partitions of $P$ leads
to counting in time $O(m+\log^{2+\epsilon} n)$, once again without the need to 
find the smallest attractor. This is more complex than before because theirs
is a run-length grammar, and the run-length rules challenge the observation
that the number of occurrences depends only on the points. Among other 
tradeoffs, they show how to count in optimal time
$O(m)$ using $O(\gamma\log(n/\gamma)\log n)$ space.


\subsection{Suffix Trees} \label{sec:strees}

Compressed suffix-based indexes can be enhanced in order to support full
suffix-tree functionality (recall Section~\ref{sec:e}). Suffix trees enable
a large number of complex analyses on the strings, and are particularly 
popular in stringology \cite{CR02} and bioinformatics \cite{Gus97}.

\citeN{Sad07} pioneered the compressed suffix trees, defining a basic set
of primitives for traversing the tree and showing how to implement them with
just $6n$ bits on top of a compressed suffix array. The required primitives
are:
\begin{itemize}
\item Access to any cell of the suffix array, $A[i]$.
\item Access to any cell of the inverse suffix array, $A^{-1}[j]$.
\item Access to any cell of the {\em longest common prefix array}, $LCP[i]$
is the length of the longest common prefix of $S[A[i]\dd]$ and $S[A[i-1]\dd]$.
\item Either:
\begin{itemize}
	\item The navigable topology of the suffix tree \cite{Sad07}, or
	\item Support for range minimum queries (RMQs, 
Section~\ref{sec:lz-second}) on the $LCP$ array \cite{FMN09}.
\end{itemize}
\end{itemize}

The idea that the {\em differential} suffix array, $DA[i]=A[i]-A[i-1]$, is
highly compressible on repetitive sequences can be traced back to \citeN{GN07},
and it is related with the BWT runs: if $S^{bwt}[i\dd i+\ell]$ is a run, then
the LF-step formula (Section~\ref{sec:bwt-index}) implies that 
$LF(i+k) = LF(i)+k$ for all $0 \le k \le \ell$, and therefore
$DA[LF(i)+k] = A[LF(i)+k] - A[LF(i)+k-1] = A[LF(i+k)] - A[LF(i+k-1)] = 
(A[i+k]-1) - (A[i+k-1]-1) = A[i+k]-A[i+k-1]=DA[i+k]$. That is, 
$DA[i+1\dd i+\ell] = DA[LF(i)+1 \dd LF(i)+\ell]$ is a repetition in $DA$.

\begin{example}
In Figure~\ref{fig:runs}, the run $S^{bwt}[13\dd 17]=\mathsf{aaaaa}$ implies
that $LF(13+k)=LF(13)+k=5+k$ for $0 \le k \le 4$. Therefore, $A[5+k] = 
A[13+k]-1$ for $0 \le k \le 4$: $A[5\dd 9] = \langle 1,9,7,5,13\rangle$ and
$A[13\dd 17] = \langle 2,10,8,6,14\rangle$. This translates into a copy in
$DA[2\dd 17] = \langle -1,-13,8,-10,8,-2,-2,8,-9,8,3,-13,8,-2,-2,8\rangle$:
we have $DA[14\dd17]=DA[6\dd 9]$.
\end{example}

The same happens with the differential versions of arrays $A^{-1}$ and $LCP$,
and also with representations of the tree topology, which also has large
identical subtrees. It is also interesting that the array $PLCP[j] = 
LCP[A^{-1}[j]]$, can be represented in $O(r)$ space \cite{FMN09}. All these 
regularities inspired several practical compressed suffix trees for highly 
repetitive strings, where some of those components were represented using 
grammars or block trees, and even with the original run-length BWT
\cite{MNSV09,ACN13,NO16,CN19}. Though implemented and shown to be practical, 
only recently \cite{GNP19} it was shown that one can build run-length
grammars of size $O(r\log(n/r))$ to represent those differential arrays and
support most of the suffix tree primitives in time $O(\log(n/r))$.

CDAWGs are also natural candidates to implement suffix trees. \citeN{BCGP16}
combine run-length compressed BWTs with CDAWGs to implement several primitives
in $O(1)$ or $O(\log\log n)$ time, within space $O(e(S)+e(S^{rev}))$.
\citeN{BC17} use heavy path decomposition of the suffix tree to expand the
set of supported primitives, in times from $O(1)$ to $O(\log n)$ and within
the same asymptotic space. 

There are also developments to solve particular problems that, although
can be solved with suffix trees, admit more efficient solutions. Some
examples are finding the longest common substring between $P$ and $S$ (LCS)
in space $O(z\log n)$ \cite{GGN13,AHGT18}, or finding all the maximal substrings
of $P$ appearing in $S$ (maximal exact matches, or MEMs) and the longest
substrings in $S$ starting at each position of $P$ (matching statistics)
in space $O(r)$ \cite{BGI20}.

\subsection{Persistent Strings}

Various repetitive collections consist of a sequence of versions of a base
document, where each version differs from the previous one in a few
``edits'' (symbol insertions, deletions, or substitutions). Branching in the
versions is also possible, so that we have a tree of versions where each
version differs from its parent in a few edits. This can be addressed with
{\em persistent data structures} \cite{DSST89}, which allow updates
but can be accessed as they were at any point in the past. Partial persistence
allows updates only in the current version, and thus can model version
sequences; full persistence allows updating any version and enables
version trees.

The basic results of \citeN{DSST89} allow one to represent a fully-persistent
balanced binary
search tree of $t$ leaves in $O(t)$ space, and then apply each update
persistently by creating only one new path from the root to the affected node.
The pointers sprouting from this path lead to the previous version's subtrees,
and the root of the new path becomes the root of the new version. In this 
way, performing $s$ updates take $O(s\log (t+s))$ additional space, and the 
operations are carried out in time $O(\log(t+s))$. The total space can be 
reduced to $O(t+s)$ with more careful binary tree implementations
\cite{DSST89,ST86}.

If we use such a fully-persistent balanced tree to represent the symbols of
a base document at its leaves, then we can represent the edit operations as
updates to the tree. Maintaining the number of leaves that descend from any
node leads us to any desired string position.
A base document of length $t$ on which $s$ edit
operations are carried out along all its versions is then represented in
$O(t+s)$ space, and it supports edits and access to individual
symbols in time $O(\log(t+s))$.

Recently, \citeN{BG20} showed how to improve the time of persistent strings to
$O(\log (t+s)/\log\log (t+s))$, on a model where they start with the empty 
string (so one can build the first string via $n'$ insertions). Further, they
can extract a substring of length $\ell$ in $O(\ell)$ additional time.
They use
$O(t+s)$ space and show that their time is optimal for any space in 
$O((t+s)\,\mathrm{polylog}\, (t+s))$. 


This result, which makes use of the particular structure of the repetitiveness,
is not directly obtained with the general techniques we have covered. For
example, it is easy to see that, processing this versioned collection from the
beginning, we obtain $z_{no} = O(t/\log_\sigma t + s)$ Lempel-Ziv phrases, even
without overlaps. It is unknown, however, if one can provide direct access
within $O(z_{no})$ space in general. We can provide logarithmic access
time by using space that is in general larger than $O(t+s)$. For example, we
can use the block trees of Section~\ref{sec:blocktrees}, which would use space
$O(\delta\log(n/\delta)) \subseteq O((t/\log_\sigma t+s)\log(s\log_\sigma t))$,
because $n \le (t+s) s$ and $\delta \le z_{no}$. We can also use grammars.
A grammar of size $O(t/\log_\sigma t + s\log(t+s))$ can be
obtained as follows \cite{Nav18}: First, we build a nonterminal $A_0$ expanding
to the original text (the grammar is of size $O(t/\log_\sigma t)$ up to here);
then, for each new version $i$ that applies $s_i$ edits to version $j$,
produce the nonterminal $A_i$ expanding to version $i$ by copying the parse
tree of $A_j$ and modifying the paths to those edits. The grammar trees
of all those nonterminals $A_i$ add up to $O(s\log(t+s))$ nodes.

\subsection{Document Retrieval} \label{sec:docret}

While basic pattern matching is at the core of a number of data retrieval 
activities, typical Information Retrieval queries operate at the {\em document}
level, more than at the individual level of occurrences. That is, we have a 
collection of $\$$-terminated strings $S_1,\ldots,S_d$, and concatenate them 
all into a single string $S[1\dd n] = S_1 \cdots S_d$. Given a search pattern
$P[1\dd m]$, we are interested in three basic queries:
\begin{description}
\item[{\rm\em Document listing:}] List all the $docc$ documents where $P$ appears. 
\item[{\rm\em Document counting:}] Compute $docc$, the number of documents where $P$
appears.
\item[{\rm\em Top-$k$ retrieval:}] List the $k$ documents where $P$ appears more 
prominently, which is typically defined as having the most occurrences.
\end{description}

While there has been some activity in supporting document retrieval queries on
general string collections, even considering compressed indexes \cite{Nav14},
the developments for highly repetitive string collections is very incipient.
Most of the developments are of practical nature. 

\subsubsection{Document listing} \label{sec:doclist}

The first document listing index \cite{Mut02} obtained optimal $O(m+docc)$ time
and $O(n)$ space. They use the {\em document array} $D[1\dd n]$, where $D[i]$ 
is the document where $A[i]$ belongs and $A$ is the suffix array of $S$. 
They use the suffix tree of $S$ to find the interval $A[sp\dd ep]$, and thus 
the task is to output the distinct values in $D[sp\dd ep]$.

\citeN{CM13} proposed the first document listing index for repetitive string
collections. They enrich a typical grammar-based index as we have described in
Section~\ref{sec:parsed-indexing} with an {\em inverted list} that stores, for
each nonterminal, the documents where its expansion appears. They then find
the primary occurrences, collect all the nonterminals involved in secondary
occurrences and, instead of listing all the occurrence positions, they merge
all the inverted lists of the involved nonterminals. To reduce space, the set
of inverted lists is also grammar-compressed, so each list to merge must be
expanded. They do not provide worst-case time or space bounds.

\citeN{GHKKNPS17} propose two techniques, ILCP and PDL. In the former, they
create an array $ILCP[1\dd n]$ that interleaves the local $LCP$ arrays 
(Section~\ref{sec:strees}) of the documents $S_i$, according to the documents
pointed from $D[1\dd n]$, that is, if $D[i]=d$ and $d$ appears $k$ times in
$D[1\dd i]$, then $ILCP[i]=LCP_i[k]$, where $LCP_i$ is the LCP array of 
$S_i$. They show that (1) $ILCP$ tends to have long runs of equal values on 
repetitive string collections, and (2) the algorithm of \citeN{Mut02} runs
almost verbatim on top of $ILCP$ instead of $D$. This yields a document listing
index bounded by the number $\rho$ of the $ILCP$ runs.
With the help of a compressed suffix array that finds 
$[sp\dd ep]$ in time $t_s$ and computes an arbitrary cell $A[i]$ in time $t_a$,
they can perform document listing in time $O(t_s + docc\cdot t_a)$; for 
example we can obtain $t_s=O(m)$ and $t_a=O(\log(n/r))$ within
$O(r\log(n/r))$ space. They have, however, only average-case bounds for
the size $O(\rho)$ of their index: if the collection can be regarded as a base
document of size $n'$ generated at random and the other $d-1$ documents are
copies on which one applies $s$ random edits,
then $\rho = O(n' + s\log (n'+s))$.

In PDL, \citeN{GHKKNPS17} use a pruned suffix tree, where the nodes expanding
to less than $s$ positions of $A$ are removed. In the remaining nodes, they
store the inverted lists of the documents where the node string appears. The
lists are then grammar-compressed as in \citeN{CM13}. Document listing is then
done by finding the locus of $P$. If it exists, the answer is precomputed there.
Otherwise, we have to build the answer by brute force from a range of only
$O(s)$ positions of $A$. By using a run-length compressed BWT index in addition
to the sampled suffix tree, we can then answer in time $O((m+s)\log\log n + 
docc)$. The space is $O(r)$ for the BWT-based index, plus $O(n/s)$ for the
sampled suffix tree, plus an unbounded space for the grammar-compressed lists.

\citeN{CNspire19} propose a simpler variant where the document array $D[1\dd n]$
is compressed using a balanced grammar. Note that $D$ is compressible for the
same reasons of the differential suffix array $DA$ (Section~\ref{sec:strees}).%
\footnote{It is not hard to see that $D$ has an attractor of size $r+d$ (T.
Gagie, personal communication).} 
For each nonterminal $A$, they store the (also grammar-compressed) inverted 
list of all the distinct document numbers to which $A$ expands. The 
range $D[sp\dd ep]$ is then covered by $O(\log n)$ nonterminals, whose lists
can be expanded and merged in time $O(docc \log n)$. With a run-length 
compressed BWT index, for example, they obtain time $O(m\log\log n + 
docc \log n)$. The run-length BWT takes $O(r)$ space and the grammar-compressed 
document array $D$ takes $O(r\log(n/r))$ space, but the space
of the grammar-compressed inverted lists is not bounded. 

Although some offer good time bounds for document listing, none of the previous
indexes offer worst-case space bounds. They have all been implemented, however,
and show good performance \cite{CM13,GHKKNPS17,CNspire19}. \citeN{Nav18} offers
the only index with a limited form of worst-case space bound: if we have a base 
document of length $n'$ and $s$ arbitrary edits applied on the other $d-1$ 
documents, the index is of size $O(n'\log\sigma + s \log^2 n)$. It performs
document listing in time $O(m\log n + docc \cdot m \log^{1+\epsilon} n)$.
The technique is to store the inverted lists inside the
components of the geometric data structure for range searches that we use to
find the primary occurrences.

\subsubsection{Document counting}

\citeN{Sad07b} showed that the number of distinct documents where $P$ appears
can be counted in constant time once $[sp\dd ep]$ is known, by adding a data 
structure of just $2n+o(n)$ bits on top of a suffix tree or array on $S$. While
this is a good solution in classical scenarios, spending even $\Theta(n)$ bits
is excessive for large highly repetitive string collections.

\citeN{GHKKNPS17} show that the $2n$ bits of \citeN{Sad07b} do inherit the
repetitiveness of the string collection, in different ways (runs, repetitions,
etc.) depending on the type of repetitiveness (versioning, documents that are
internally repetitive, etc.), and explore various ways to exploit it. They 
experimentally
show that the structure can be extremely small and fast in practice.
They also build a counting structure based on ILCP, which uses $O(r+\rho)$
space and counts in time $O(m\log\log n)$, but it does not perform so well in
the experiments.

\subsubsection{Top-$k$ retrieval}

PDL \cite{GHKKNPS17} can be adapted for top-$k$ retrieval, by sorting the
inverted lists in decreasing frequency order. One can then read only the first
$k$ documents from the inverted list of the locus of $P$, when it exists;
otherwise a brute-force solution over $O(s)$ suffix array cells is applied.
They compare PDL experimentally with other solutions (none of which is designed
for highly repetitive string collections) and find it to be very competitive.

Note that this idea is not directly applicable to other indexes that use 
inverted lists \cite{CM13,CNspire19} because they would have to merge various
inverted lists to find the top-$k$ candidates. It is possible, however, to 
obtain exact or approximate results by applying known pruning techniques 
\cite{BCC10,BYRN11,GHKKNPS17}.

\subsection{Heuristic Indexes}

Apart from those covered in Section~\ref{sec:measures},
other compression techniques for highly repetitive string collections have been
proposed, but they are aimed at specific situations. In this section we briefly
cover a couple of those for which compressed indexes have been developed.

\subsubsection{Relative Lempel-Ziv}

\citeN{KPZ10} proposed a variant of Lempel-Ziv specialized for genome 
collections of the same species, where every genome is sufficiently similar to
each other. The idea is to create a {\em reference} string $R$ (e.g., one of the
genomes, but it is also possible to build artificial ones 
\cite{KPZ11,KBSCZ12,GPV16,LPMW16}) 
and parse every other string $S$ in the same way of 
Lempel-Ziv, but with the sources being extracted only from $R$. Very good
compression ratios are reported on genome collections \cite{DDN15}. Further, 
if we retain direct access to any substring of $R$, we also have efficient 
direct access to any substring $S[i\dd j]$ of every other string $S$ 
\cite{DG11,FGGP14,CFGPS16}.

The simplicity of random access to the strings also impacts on indexing.
The parse-based indexing we described in Section~\ref{sec:parsed-indexing}
adapts very well to Relative Lempel-Ziv compression \cite{GGKNP12,DJSS14,NS19}.
The reference $R$ can be indexed within statistical entropy bounds \cite{NM06},
so that we first locate the occurrences of $P$ in it. The occurrences inside
phrases in the other strings $S$ are obtained by mapping sources in $R$ 
covering occurrences of $P$ to their targets in $S$, using the same mechanism 
of Section~\ref{sec:lz-second}. Finally, the occurrences that span more than
one phrase in other strings $S$ are found with a mechanism similar to the
use of the grid (Section~\ref{sec:grid}). 

From the existing indexes, only that of \citeN{NS19} is implemented. It uses
$O(|R|/\log_\sigma n + z)$ space (where $z$ is the size of the parse) and
searches in time $O((m+occ)\log n)$; the others \cite{GGKNP12,DJSS14} obtain
slightly better complexities. Their experiments show that an index based on 
Relative Lempel-Ziv performs well in space only when all
the documents are very similar to each other. For example, this is not the 
case of versioned document collections, where each document is very similar
to its close versions but possibly very different from far versions.

Relative Lempel-Ziv has also been used to compress the differential suffix
array $DA$, offering more space but less time than previous approaches
\cite{PZ20}.

\subsubsection{Alignments}

Another way to exploit having a reference $R$ and many other strings similar 
to it is to build a classical or entropy-compressed index for $R$ and a
``relative'' index for every other string $S$. The rationale is that the
similarity between the two strings should translate into similarities in the
index data structures. 

\citeN{BGGMS14} use this idea on BWT-based indexes 
(Section~\ref{sec:bwt-index}). By working on the symbols that are not in the 
longest common subsequence of the BWTs of $R$ and $S$, they simulate the 
BWT-based index of each string $S$. This is expanded into full
suffix-tree functionality \cite{FGNPS18} by adding an $LCP$ array for $S$ 
that is compressed using Relative Lempel-Ziv with respect to the $LCP$ array
of $R$, and managing to efficiently compute RMQs on it.

\citetwoN{NKPKKMP16}{NKMPLLMP18} use a different approach, also based on the 
alignments of $R$ and the strings $S$. To build the (BWT-based) index, they 
separate the regions that are common to all sequences, and the uncommon regions
(all the rest). The main 
vulnerability of this approach is that a single change in one sequence destroys
an otherwise common region. Still, their implementation is shown to 
outperform a generic index \cite{MNSV09} on genome collections, in space and 
time. A similarly inspired structure providing suffix tree functionality was 
also proposed \cite{NPCHIMP13}, though not implemented. They do not give clear 
space bounds, but show how to search for patterns in optimal time 
$O(m+occ)$, and conjecture that most suffix tree operations can be supported.

Several other alignment-based indexes have appeared, generally specific of 
pan-genomic applications. We only mention a few of them
\cite{M+17,VNVPM18,G+18}.

\section{Current Challenges} \label{sec:concl}

In the final section of this survey, we consider the most important open 
challenges in this area: (1) obtaining practical implementations, and
(2) being able to build the indexes for very large text collections.

\subsection{Practicality and Implementations}

There is usually a long way between a theoretical finding and its practical
deployment. Many decisions that are made for the sake of obtaining good
worst-case complexities, or for simplicity of presentation, are not good in
practice. Algorithm engineering is the process of modifying a theoretically
appealing idea into a competitive implementation, involving knowledge 
of the detailed cost model of computers (caching, prefetching, multithreading, 
etc.). Further, big-$O$ space figures ignore constants, which must be carefully
considered to obtain competitive space for the indexes. In practice variables 
like $z$, $g$, $r$, etc.\ are a hundredth or a thousandth of $n$, and therefore 
using space like $10z$ bytes may yield a large index in practice.

While competitive implementations have been developed for indexes based on
Lempel-Ziv \cite{KN13,FGHP13,CFMPN16,FKP18}, 
grammars \cite{MNKS13,TTS14,CFMPN16,CNP20},
the BWT \cite{MNSV09,BCGPR17,GNP19,KMBGLM20}, and CDAWGs \cite{BCGPR17}, the most recent
and promising theoretical developments \cite{BEGV18,NP18,CEKNP19,KNP20} are 
yet to be implemented and tested. It is unknown how much these 
improvements will impact in practice.

\begin{figure}[t]
\begin{center}
\includegraphics[width=7cm]{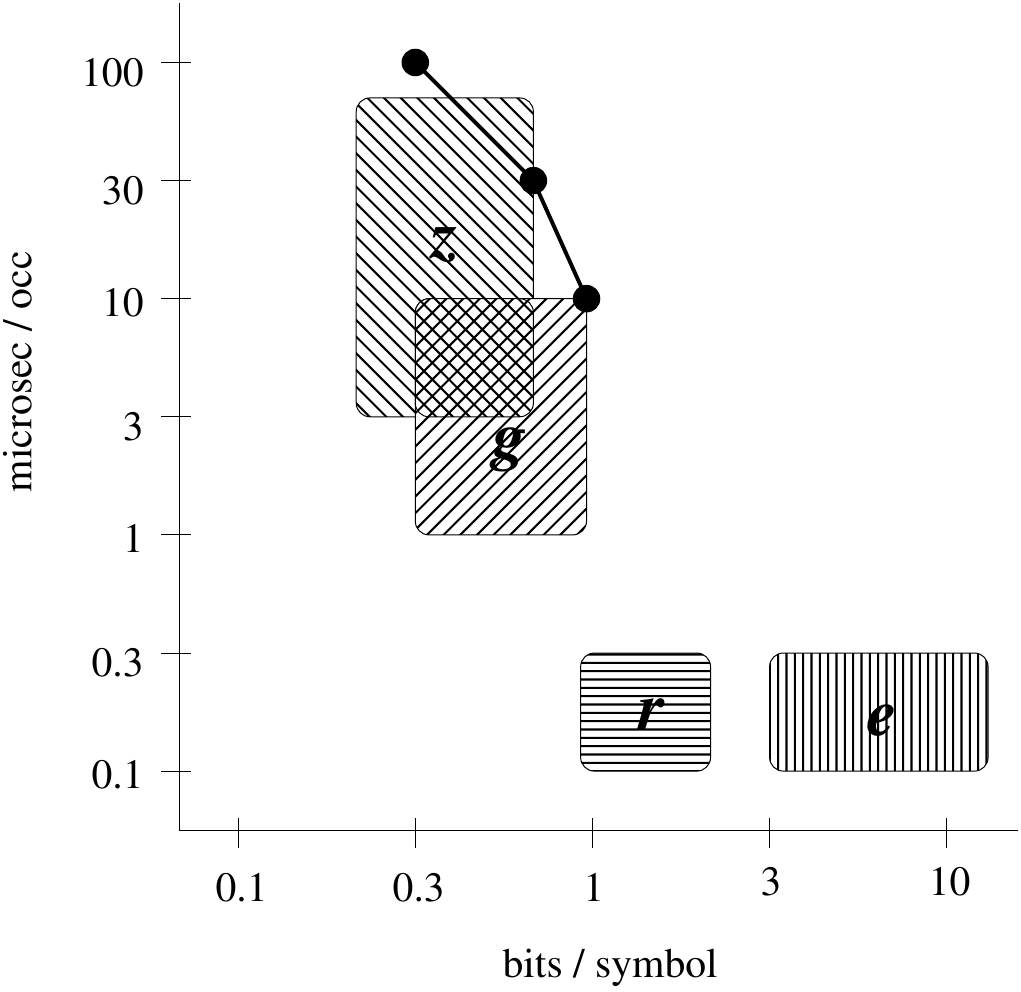}
\end{center}
\caption{Space/time tradeoffs of the indexes building on 
different repetitiveness measures. Both axes are logarithmic:
bits per symbol ($x$) and search time per occurrence in microseconds
($y$).}
\label{fig:spacetime}
\end{figure}

Figure~\ref{fig:spacetime} shows, in very broad terms, the space/time tradeoffs
obtained by the implementations built on the different repetitiveness concepts.
It is made by taking the most representative values obtained across the 
different experiments of the publications mentioned above, discarding too
repetitive and not repetitive enough collections. The black dots represent the 
run-length BWT built on regular sampling \cite{MNSV09}, which has been a 
baseline for most comparisons. Though $r$ seems to dominate $e$, the latter
(represented by CDAWG indexes) is implemented only on DNA, where it is
nearly twice as fast as $r$ (represented by the r-index).

\subsection{Construction and Dynamism}

An important obstacle for the practical adoption of the indexes we covered is
how to build them on huge datasets. Once built, the indexes are orders of
magnitude smaller than the input and one hopes to handle them in main memory.
However, the initial step of computing the parsing, the run-length BWT, or 
another small representation of a very large text collection, even if it can be
generally performed in the optimal $O(n)$ time, usually requires $O(n)$ main 
memory space with a significant constant. There are various approaches that aim
to reduce those main memory requirements and/or read the text in streaming mode,
but some are still incipient.

\paragraph*{Burrows-Wheeler Transform} The BWT is easily obtained from the
suffix array, which in turn can be built in $O(n)$ time and space 
\cite{KSPP05,KA05,KSB06}. However, the constant associated with the space is
large. \citeN{KSB06} allow using $O(nv)$ time and $O(n/\sqrt{v})$ space for
a parameter $v$, but they still need to store the $n\log n + n\log\sigma$ bits 
of the suffix array and the text. External-memory suffix array construction
requires optimal $O(Sort(n))$ I/Os and time \cite{FFM00,KSB06}.

There are various algorithms to build the BWT directly using compact space or
in external memory \cite{Kar07,HLSSY07,OS09,HSS09,FGM12,BZGO13,MNN17,KKstoc19,BCKM20,FNN20},
but they do not produce it in run-length compressed form. Recently,
\citeN{Kem19} showed how to build the run-length BWT in $O(n/\log_\sigma n +
r\,\mathrm{polylog}~n)$ time and working space. 

With a dynamic representation of sequences that supports insertions of symbols
\cite{MN15} one can build the run-length encoded BWT incrementally by
traversing the text in reversed form; the LF-mapping formula given in 
Section~\ref{sec:bwt-index} shows where to insert the next text symbol. This
idea is used by \citeN{PP18} to build the run-length compressed BWT directly, 
in streaming mode, in $O(n\log r)$ time and within $O(r)$ main memory space.
\citeN{OSTIS18} improve their practical performace by a factor of $50$ (using
just twice the space). 

\citeN{Sir16} presents a practical technique to build very large BWTs 
(i.e., for terabytes of data) in run-length compressed form.
It splits the collection into subcollections, builds the individual BWTs, and
then merges them into a global one. 

\citeN{BGKLMM19} propose a practical method called
``prefix free parsing'', which first parses the text using a rolling hash
(a Karp-Rabin-like hash that depends on the last $\ell$ symbols read; a 
phrase ends whenever the hash modulo a parameter value is zero). The result 
is a dictionary
of phrases and a sequence of phrase identifiers; both are generally much
smaller than $n$ when the text is repetitive. They then build the BWT from
those elements. Their experiments show that they can build BWTs of very large
collections in reasonable time.
\citeN{KMBGLM20} show how to add in this construction the sampling used by
the run-length BWT \cite{GNP19}; recall Section~\ref{sec:rindex-locate}.

\paragraph*{Lempel-Ziv parsing} While it has been known for decades how to 
obtain this parse in $O(n)$ time \cite{RPE81,SS82}, these algorithms
use $O(n)$ space (i.e., $O(n\log n)$ bits) with a significant constant. A long line of research 
\cite{CPS08,OG11,KP13,KKP13,GB13,GB14,KKPdcc14,YIBIT14,FIK15,KKP16,KS16,BP16,MNN17,FIKS18,Kem19} 
has focused on using little space, reducing the constant associated 
with the $O(n\log n)$ bits and even reaching $O(n\log\sigma)$ bits
\cite{KKP13,BP16,MNN17,Kem19} and statistically compressed space \cite{PP15}.
This is still unaware of repetitiveness, however.

Interestingly, the only known method to build the Lempel-Ziv parsing in
repetition-aware space ($O(z+r)$) is to build the run-length BWT first and
then derive the Lempel-Ziv parse from it \cite{PP18,OSTIS18,BGI20}. These
methods use up to 3 orders of magnitude less space (and are just
2--8 times slower) than the previous approaches. Another
interesting development \cite{KKPdcc14} uses external memory: with a RAM of
size $M$, it performs $O(n^2/M)$ I/Os and requires $2n$ bytes of disk working
space. Despite this quadratic complexity, their experiments show that this 
technique can handle, in practice, much larger texts than previous
approaches. 

Other approaches aim at approximating the Lempel-Ziv parse.
\citeN{FGGK15} build an $(1+\epsilon)$-approximation in $O((n/\epsilon)\log n)$
time and $O(z)$ space. 
\citeN{VKNP19} use Relative Lempel-Ziv as a building block.

\paragraph*{Grammar construction} RePair \cite{LM00} is the heuristic that
obtains the best grammars in practice. While it computes the grammar in 
$O(n)$ time and space, the constant associated with the space is significant
and prevents using it on large texts. Attempts to reduce this space have paid a 
significant price in time \cite{BGP17,SOGTIS19,KIFTSG19}. A recent heuristic
\cite{GIMNST19} obtains space close to that of RePair using a semi-streaming
algorithm, but it degrades quickly as the repetitiveness decreases.
Various other grammar construction algorithms, for example \citeN{Sak05}
and \citeN{Jez15,Jez16}, build a balanced grammar that approximates the 
smallest grammar within an $O(\log n)$ factor by performing a logarithmic 
number of passes on the text (which halves at each pass), and could be amenable
to a semi-streamed construction.
The same is true of recent grammar constructions \cite{NLGARN18,DN21} inspired
in suffix array construction algorithms.

An important line of research in this regard are the online grammar construction
algorithms \cite{MST12,MTSS13,TIS17}. In OLCA \cite{MST12}, the authors build
a grammar in $O(n)$ time by reading the text in streaming mode. They obtain
an $O(\log^2 n)$ approximation to the smallest grammar using $O(g \log^2 n)$
space. In SOLCA \cite{MTSS13} they reduce the space to $O(g)$.
FOLCA \cite{TIS17} improves the space to $g\log g + o(g\log g)$ bits; the
authors also prove that the grammar built by SOLCA and FOLCA is an 
$O(\log n \log^* n)$ approximation. Their experiments show that the grammar 
is built very efficiently in time and main memory space, though the resulting 
grammar is 4--5 times larger than the one generated by RePair. 

\paragraph*{Dynamism}

A related challenge is dynamism, that is, the ability to modify the index when
the text changes. Although a dynamic index is clearly more practical than one
that has to be rebuilt every time, this is a difficult topic on which little
progress has been made. The online construction methods for run-length BWTs
\cite{PP18,OSTIS18,BGI20} and grammars \cite{MTSS13,TIS17} naturally allow us adding
new text at the beginning or at the end of the string. A dynamic BWT
representation allows adding or removing arbitrary documents from a text
collection \cite{MN08}. Supporting arbitrary modifications to the text is much
more difficult, however.
We are only aware of two works. One \cite{NIIBT20} builds
on edit-sensitive parsing to maintain a grammar under arbitrary substring
insertions and deletions. They use $O(z\log n \log^* n)$ space and search in
time $O(m(\log\log n)^2 + \log n \log^* n(\log n + \log m \log^* n) + 
occ \log n)$ (simplified, see the precise formula in the appendix).
A substring of length $\ell$ is inserted/deleted in time
$O((\ell+\log n \log^* n)\log^2 n \log^* n)$. In practice, the search is fast
for long patterns only; \citeN{NTT18} improved their search time on short
patterns. A second work \cite{GKKLS15} uses another kind of locally-consistent
grammar to insert substrings of length $\ell$ in time $O(\ell \log n)$, split
and concatenate represented strings in time $O(\log^2 n)$, and search for
patterns in those strings in time
$O(m+ occ \log n)$; all the operations succeed with high probability.

\subsection{New Queries}

Counting and locating queries have been the fundamental object of study since
the beginning of compressed indexes \cite{FM00,GV00}. In a highly repetitive
scenario, however, their relevance can be questioned. For example, consider
a versioned collection where we search for a pattern $P$. If $P$ appears in the
common part of many documents, the index will report all those occurrences 
in every document, each time with identical context. It is not clear that the
$\Omega(occ)$ effort of producing such a large output is worthy. We finish with
a proposal of some query variants that can be more appropriate on highly 
repetitive collections; note that some are closer to document retrieval queries 
(see Section~\ref{sec:docret}):

\begin{description}
\item[{\rm\em Contextual locating:}] Instead of reporting all the positions
where $P$ occurs in $S$, list the $cocc$ distinct ``contexts'' where it appears,
for example, the distinct lines (separated by newlines), or the distinct
substrings of some length $\ell$ preceding and following $P$, and one (or
all, or the amount of) the positions where $P$ occurs in each context. 
While it is feasible to
solve this query in time proportional to $occ$, the challenge is to solve it
in time proportional to $cocc$. Grammar, BWT, and CDAWG based indexes seem
well suited for this task. A first result in this direction
\cite{Nav20contextual} obtains $O(m+cocc \log n)$ time in $O(r^*\log(n/r^*))$
space, where $r^*$ is the sum of the number of runs in the BWTs of $S$ and
$S^{rev}$.

\item[{\rm\em Range document listing:}] On a document collection formed by a 
linear list of versions (like Wikipedia or periodic publications), list the 
$nrange$ maximal ranges of versions where $P$ appears (e.g.,
``$P$ appears in the documents $45$--$57$ and $101$--$245$''). On a document
collection formed by a tree of versions (like GitHub and most versioned document
repositories), list the $nrange$ maximal subtrees such that $P$ appears in
all the subtree nodes. This makes sense because close versions should 
be more similar than far versions, and then the occurrences should cluster. 
This query is solved by document listing in time 
proportional to $docc$, the number of documents where $P$ appears 
(Section~\ref{sec:doclist}), but the challenge is to solve it in time 
proportional to $nrange$. Handling subtrees essentially boils down to handling
linear ranges if we consider preorder numbering of the documents in the tree
of versions. Document listing indexes that use inverted lists
\cite{CM13,GHKKNPS17,CNspire19,Nav18} could be adapted to handle this query.

\item[{\rm\em Granular document listing:}] On a document collection formed by a 
tree of versions, list the nodes of a certain depth where $P$ appears in
some node of their subtree (e.g., ``$P$ appears somewhere in versions 1.2.x and
4.0.x''). This query allows us, for example, to determine in which ``major 
versions'' of documents we can find $P$, without the detail of precisely in 
which minor versions it appears. Again the challenge is to perform this query
in time proportional to the size of the output, and again document listing
structures could be well suited to face it.

\item[{\rm\em Document restricted listing:}] List the occurrences of $P$ only
within a given range or subtree of documents one is interested in (``find $P$
only within the reports of $2010$--$2013$''). This query can be combined with
any of the previous ones. 
\end{description}

Those queries can be combined with the aim of retrieving the $k$ ``most
important'' results, where importance can be defined in terms of the documents
themselves (as done by search engines with Web pages) and/or in terms of the
presence of $P$ in the documents (typically favoring those where $P$ occurs
most often, as in classical Information Retrieval systems), see \citeN{Nav14}. 

In general, we expect that the techniques developed for locating pattern
occurrences on highly repetitive text collections can be used as a base to
solve these more sophisticated queries. Efficiently solving some of them can be 
challenging even without repetitiveness, however. For example, document 
restricted listing is related to the problem of ``position-restricted substring 
searching'', which is unlikely to be solvable within succinct space
\cite{HSTV12}.

\section*{Acknowledgements}

We thank Travis Gagie and Nicola Prezza for useful comments.

\bibliographystyle{acmtrans}
\bibliography{paper}

\appendix

\section{History of the Contributions to Parsing-Based Indexing}

We cover only the developments related to the repetitiveness measures we have
considered. Other parsed-based indexes, such as those building on the LZ78
compression format \cite{ZL78}, are omitted because they are not competitive
on highly repetitive text collections.

\begin{description}
\item[\sf\citeN{CN09,CNfi10}] proposed the first compressed index based on 
grammar 
compression. Given any grammar of size $g$, their index uses $O(g)$ space to
implement the grid and the tracking of occurrences over the grammar tree, but 
not yet the amortization mechanism we described. On a grammar tree of height 
$h$, the index searches in time $O(m(m+h)\log n +occ\cdot h\log n)$ and 
extracts a substring of length $\ell$ in time $O((\ell+h)\log n)$. The terms 
$O(\log n)$ can be reduced by using more advanced data structures, but the 
index was designed with practice in mind and it was actually implemented
\cite{CFMPN16},
using a RePair construction \cite{LM00} that is heuristically balanced.

\item[\sf\citeN{KN11,KN13}] proposed the first compressed index based on 
Lempel-Ziv, and the only one so far of size $O(z)$. Within this size, they
cannot provide access to $S$ with good time guarantees: each accessed symbol
must be traced through the chain of target-to-source dependencies. If the
maximum length of such a chain is $h \le z$, their search time is $O(m^2 h +
(m+occ)\log z)$. The term $\log z$ could be $\log^\epsilon z$ by using
the geometric structure we have described but, again,
 they opt for a practical version.
Binary searches in $\mathcal{X}$ and $\mathcal{Y}$ are sped up with Patricia 
trees \cite{Mor68}. A substring of length $\ell$ is extracted in time 
$O(\ell\,h)$. This is the smallest implemented index; it is rather
efficient unless the patterns are too long \cite{KN13,CFMPN16}. Interestingly,
it outperforms the previous index \cite{CNfi10} both in space (as expected)
and time (not expected).

\item[\sf\citeN{MNKS11,MNKS13},\citeN{TTS14}] 
propose another grammar index based on
``edit-sensitive parsing'', which is related to locally consistent parsing
(see the end of Section~\ref{sec:searchXY}). This ensures that the parsing of
$P$ and of any of its occurrences in $S$ will differ only by a few ($O(\log n 
\log^* n)$) symbols in the extremes of the respective parse trees, and therefore
the internal symbols are consistent. By looking for those one captures all the
$occ_c$ {\em potential} occurrences, which however can be more than the
actual occurrences. Given a grammar of size $g_e \ge g$ built using 
edit-sensitive parsing, their index takes $O(g_e)$ space and searches in time
$O(m\log\log n\log^* n + occ_c \log m \log n \log\log n\log^* n)$. Substrings of length
$\ell$ are extracted in time $O((\ell+\log n)\log\log g_e)$. Their index is implemented,
and outperforms that of \citeN{KN13} for $m \ge 100$.

\item[\sf\citeN{CNspire12},\citeN{CNP20}] improved the proposal of
\citeN{CNfi10} by introducing the amortization mechanism and also using the 
mechanism to extract phrase prefixes and suffixes in optimal time 
(Section~\ref{sec:extract-prefixes}). The result is an index of size
$O(g)$ built on any grammar of size $g$, which searches in time 
$O(m^2 \log\log_g n + (m+occ)\log n)$. Again, this index is described with
practicality in mind; they show that with larger data structures of size
$O(g)$ one can reach search time $O(m^2 + (m+occ)\log^\epsilon n)$. Any
substring of size $\ell$ can be extracted in time $O(\ell+\log n)$ with the 
mechanisms seen in Section~\ref{sec:extract-grammar}. An implementation of 
this index \cite{CNP20} outperforms the Lempel-Ziv based index \cite{KN13} 
in time, while using somewhat more space. The optimal-time extraction of 
prefixes and suffixes is shown to have no practical impact on balanced grammars.

\item[\sf\citeN{GGKNP12}] invented bookmarking to speed up substring extraction
in the structure of \citeN{KN13}. They use bookmarking on a Lempel-Ziv parse, 
of size $O(z\log\log z)$, which is added to a grammar of size $O(g)$ to provide
direct access to $S$. As a result, their index is of size $O(g+ z\log\log z)$ 
and searches in time $O(m^2 + (m+occ) \log\log n)$. Their technique is more
sophisticated than the one we present in Section~\ref{sec:bookmarking}, but it
would not improve the tradeoffs we obtained.

\item[\sf\citeN{FGHP13,FKP18}] proposed the so-called {\em hybrid indexing}.
Given a maximum pattern length $M$ that can be sought, and a suitable parse
(Lempel-Ziv, in their case) of size $z$, they form a string $S'$ of size
$< 2Mz$ by collecting the symbols at distance at most $M$ from a phrase
boundary and separating disjoint areas with $\$$s. Any primary occurrence
in $S$ is then found in $S'$, and any occurrence in $S'$ is a distinct 
occurrence in $S$. They then index $S'$ using any compact index and search
it for $P$. The occurrences of $P$ in $S'$ that cross the middle of a piece
are the primary occurrences of $P$ in $S$; the other occurrences in $S'$ are 
discarded, but these are at most $occ$. The mechanism of 
Section~\ref{sec:lz-second} to propagate primary to secondary occurrences is 
then used. Patterns longer than $M$
are searched for by cutting them into chunks of length $M$ and assembling their
occurrences. Within space $O(Mz)$, they can search in time $O((m+occ)
\log\log n)$ if $m \le M$. Though they offer no guarantees for longer patterns,
their implementation outperforms other classical 
indexes \cite{MNSV09,KN13} when $m$ is up to a few times $M$. The weak point of
this index shows up when $m$ is much smaller or much larger than the value $M$ 
chosen at index construction time. \citeN{GP15} relate this technique with
earlier more specific developments, and call {\em kernelization} the general
technique to solve string matching problems on repetitive sequences by working 
on the texts surrounding phrases.

\item[\sf\citeN{GGKNP14}] extended bookmarking to include fingerprinting as
well (Section~\ref{sec:bookmarking}, again more sophisticated than our
presentation), and invented the technique of using fingerprinting to remove 
the $O(m^2)$ term that appeared in all previous indexes. In the way they 
present their index, the space is $O(z \log n)$ and the search
time is $O(m\log m + occ \log\log n)$.\footnote{Their actual space is
$O(z(\log^* n + \log(n/z) + \log\log z))$, which they convert to
$O(z\log(n/z))$ by assuming a small enough alphabet and using 
$z = O(n/\log_\sigma n)$.}

\item[\sf\citeN{NIIBT15,NIIBT20}] propose the first dynamic compressed index 
(i.e., one can modify $S$ without rebuilding the index from scratch). It is 
based on edit-sensitive parsing, and they manage to remove the term $occ_c$ 
in the previous index \cite{TTS14} by finding stronger properties of the 
encoding of $P$ via its parse tree. Their search time is
$O(m \min(\log\log n \log\log g_e/\log\log\log n,$ $\sqrt{\log g_e/\log\log g_e} )
 + \log m \log n \log^* n (\log n + \log m \log^* n) +
occ \log n)$.

\item[\sf\citeN{PhBiCPM17,BEGV18}] improve upon the result of
\citeN{GGKNP14}. They propose for the first time the batched search
for the pattern prefixes and suffixes, recall 
Section~\ref{sec:searchXY}. They also speed up the searches by storing more
points in the grid: if we store the points $S[i],\ldots,S[i+\tau-1]$ for every
phrase starting at $S[i]$, then we need to check only one out of $\tau$ 
partitions of $P$, that is, we check $m/\tau$ partitions. This leads to various
tradeoffs, which in simplified form are:
$O(z\log(n/z)\log\log z)$ space and $O((m+occ)\log\log n)$ time,
$O(z(\log(n/z)+\log\log z))$ space and $O((m+occ)\log^\epsilon n)$ time,
$O(z(\log(n/z)+\log\log z)\log\log z)$ space and $O(m+occ\log\log n)$ time, and
$O(z(\log(n/z)+\log^\epsilon n))$ space and 
$O(m+occ\log^\epsilon n)$ time.
The last two reach for the first time linear complexity in $m$. 
They also show how to extract a substring of length $\ell$ in time
$O(\ell + \log(n/z))$.

\item[\sf\citeN{Nav17},\citeN{NP18}] build a compressed index based on block 
trees (Section~\ref{sec:blocktrees}), which are used to provide both access 
and a suitable
parse of $S$. They reuse the idea of the grid and the mechanism to propagate 
secondary occurrences. By using a block tree built on attractors \cite{NP18},
they obtain $O(\gamma\log(n/\gamma))$ space and $O(m\log n+occ\log^\epsilon n)$
search time. They called this index ``universal'' because it was the first 
one built on a general measure of compressibility (attractors) instead of on 
specific compressors like grammars or Lempel-Ziv. For example, if one builds
a bidirectional macro scheme of size $b$ (on which no index has been proposed),
one can use it as an upper bound to $\gamma$ and have a functional index
of size $O(b\log(n/b))$.

\item[\sf\citeN{CE18}] were the first to show that, using a locally consistent
parsing, only $O(\log m)$ partitions of $P$ need be considered (see the end of 
Section~\ref{sec:searchXY}). Building on
the grammar-based index of \citeN{CNspire12} and on batched pattern searches 
(Section~\ref{sec:searchXY}), they obtain an index using 
$O(z(\log(n/z)+\log\log z))$ space and $O(m+\log^\epsilon(z\log(n/z)) +
occ(\log\log n + \log^\epsilon z)) \subseteq O(m+(1+occ)\log^\epsilon n)$
time,\footnote{This corrected time is 
given in the journal version \cite{CEKNP19}.} thus offering another tradeoff
with time linear in $m$.

\item[\sf\citeN{CEKNP19}] rebuild the result of \citeN{CE18} on top of 
attractors, like \citeN{NP18}. They use a slightly different run-length
grammar, which is proved to be of size $O(\gamma\log(n/\gamma))$, and
a better mechanism to track secondary occurrences within constant amortized
time \cite{CNspire12}. Their index, of size $O(\gamma\log(n/\gamma))$,
then searches in time $O(m + \log^\epsilon \gamma + occ \log^\epsilon
(\gamma\log(n/\gamma))) \subseteq O(m + (1+occ)\log^\epsilon n)$.
By enlarging the index to size $O(\gamma\log(n/\gamma)\log^\epsilon n)$,
they reach for the first time optimal time in parsing-based indexes,
$O(m+occ)$. Several other intermediate tradeoffs are obtained too.
Interestingly, they obtain this space in terms of $\gamma$ without
the need to find the smallest attractor, which makes the index implementable
(they use measure $\delta$, Section~\ref{sec:delta}, to approximate $\gamma$).
Finally, they extend the current results on indexes
based on grammars to run-length grammars, thus reaching an index of size
$O(g_{rl})$ that searches in time $O(m\log n + occ \log^\epsilon n)$.

\item[\sf\citeN{KNP20}] prove that the original block trees \cite{BGGKOPT15}
are not only of size $O(z\log(n/z))$, but also $O(\delta\log(n/\delta))$.
They then show that the universal index of \citeN{NP18} can also be 
represented in space $O(\delta\log(n/\delta))$ and is directly implementable
within this space. The search time, $O(m\log n + 
occ \log^\epsilon n)$, is also obtained in space
$O(g_{rl})$ \cite{CEKNP19}, which as explained can be proved to be in
$O(\delta\log(n/\delta))$ \cite{KNP21}, though there is no efficient way to obtain a 
run-length grammar of the optimal size $g_{rl}$.

\item[\sf\citeN{TKNIBT20}] use the grammar-based techniques we have presented
to build an index that searches in time $O(m+\log n \log m + occ\log n)$,
by exploiting the properties of a particular grammar: they decompose $S$
recursively into Lyndon words (a Lyndon word is lexicographically smaller than
its suffixes), and then build a binary grammar that follows the decomposition,
in the hope that the same nonterminals are generated many times. They empirically
show that the resulting grammar is only 1.5--2.0 times larger than that of
RePair.
\end{description}

\end{document}